\documentclass[12pt]{article}

\usepackage[a4paper,text={16.8cm,22.4cm}]{geometry}
\usepackage{amsmath,amsfonts,slashed,amssymb,tikz,bm,psfrag,graphicx,color,dsfont}
\usepackage{multicol}

\RequirePackage[sort&compress,square,comma,numbers]{natbib}
\allowdisplaybreaks
\begin{document}

\begin{titlepage}

\begin{flushright}
\normalsize
UWTHPH 2015-30 \\
% arXiv:1505.nnnnn
% v1:
November 30, 2015
\end{flushright}

\vspace{0.1cm}
\begin{center}
\Large\bf
Perturbative Corrections to  $\Lambda_b \to \Lambda$ Form Factors from QCD Light-Cone Sum Rules
\end{center}

\vspace{0.5cm}
\begin{center}
{\bf Yu-Ming Wang$^{a,b}$ and Yue-Long Shen$^c$} \\
\vspace{0.7cm}
{\sl  ${}^a$\, Fakult\"{a}t f\"{u}r Physik, Universit\"{a}t Wien, Boltzmanngasse 5, 1090 Vienna, Austria \\
${}^b$\, Physik Department T31, Technische Universit\"at M\"unchen,
James-Franck-Stra\ss e~1, D-85748 Garching, Germany \\
${}^c$\, College of Information Science and Engineering,
Ocean University of China, Qingdao, Shandong 266100, P.R. China}
\end{center}

\vspace{0.2cm}
\begin{abstract}
We compute  radiative corrections to $\Lambda_b \to \Lambda$ from factors,  at next-to-leading logarithmic accuracy,
from QCD light-cone sum rules with $\Lambda_b$-baryon distribution amplitudes. Employing the diagrammatic approach
factorization of the vacuum-to-$\Lambda_b$-baryon correlation function is justified at leading power in $\Lambda/m_b$,
with the aid of the method of regions. Hard functions entering the factorization formulae are identical to the corresponding
matching coefficients of heavy-to-light currents from QCD onto soft-collinear effective theory. The universal jet function from
integrating out the hard-collinear fluctuations exhibits richer structures compared with the one involved in the factorization
expressions of the vacuum-to-$B$-meson correlation function. Based upon the QCD resummation improved sum rules we observe that
the perturbative corrections at ${\cal O}(\alpha_s)$ shift the $\Lambda_b \to \Lambda$ from factors at large recoil significantly
and the dominant contribution originates from the next-to-leading order jet function instead of the hard coefficient functions.
Having at hand the sum rule predictions for the $\Lambda_b \to \Lambda$ from factors we further investigate several decay observables
in the electro-weak penguin $\Lambda_b \to \Lambda \, \ell^{+} \ell^{-}$  transitions in the factorization limit (i.e., ignoring
the ``non-factorizable" hadronic effects which cannot be expressed in terms of the $\Lambda_b \to \Lambda$ from factors), including the
invariant mass distribution of the lepton pair, the forward-backward asymmetry in the dilepton system and the longitudinal polarization
fraction of the leptonic sector.

\end{abstract}
\vfil

\end{titlepage}

\section{Introduction}

Electro-weak penguin $b \to s \ell \ell$ decays are widely believed to be sensitive probes to physics
beyond the Standard Model (SM) and continuous efforts have been devoted to investigations of exclusive
$B \to K^{(\ast)} \ell^{+} \ell^{-}$ decays towards understanding  the strong interaction dynamics in QCD and
constructing  the optimized angular
observables of phenomenological interest. Unfortunately,  no evident new physics signals have been revealed
in the exclusive $B$-meson decays yet, albeit with several ``anomalies" under extensive discussions and debates.
It is therefore natural to explore the dynamics of flavour-changing neutral current induced hadronic transitions
in a complementary way.

In this respect the baryonic counter channels $\Lambda_b \to \Lambda \, \ell^{+} \ell^{-}$ can serve the purpose thanks to
the dedicated $b$-physics program at the LHC. Theory descriptions of exclusive  heavy baryon decays have been initiated
in the early days of the heavy-quark effective theory (HQET) in an attempt to understand the QCD dynamics of heavy quark decays,
and they have attracted renewed attentions recently \cite{Mannel:2011xg,Feldmann:2011xf,Wang:2011uv}
towards a better understanding of the heavy-to-light baryonic form factors at large recoil in the heavy quark limit.
Also,  there are  good arguments in favor of studying the exclusive
$\Lambda_b \to \Lambda \, \ell^{+} \ell^{-}$ decays on the phenomenological side.
First,  the polarization asymmetry of the  $\Lambda$-baryon in the decay products allows a ``clean" extraction
of the helicity structure of the weak effective Hamiltonian in the factorization limit \cite{Huang:1998ek,Chen:2001zc}.
Second, the angular distribution for the four-body decays $\Lambda_b \to \Lambda (\rightarrow N \pi) \, \ell^{+} \ell^{-}$
offers additional information on  the Wilson coefficients of effective weak operators \cite{Boer:2014kda}, due to the fact that
the cascade weak decay $\Lambda \to N \pi$ is parity violating.
Third, the systematic uncertainty entering  the computation of the $\Lambda_b \to \Lambda \, \ell^{+} \ell^{-}$ amplitude,
induced by the $\Lambda$-baryon decay width, is negligible compared with the counterpart  $B \to K^{\ast} \ell^{+} \ell^{-}$
channels.

Precision QCD calculations of the electro-weak penguin decays $\Lambda_b \to \Lambda \, \ell^{+} \ell^{-}$ are complicated
by the poorly known hadronic form factors and by the notoriously ``non-factorizable" hadronic effects defined by the non-local
matrix elements of the weak operators acting together with the QED quark currents.
The main purpose of this paper is to perform a complete analysis of 10 independent $\Lambda_b \to \Lambda$ form factors, at
${\cal O}(\alpha_s)$, from QCD light-cone sum rules (LCSR) with the $\Lambda_b$-baryon distribution amplitudes (DA)
originally developed in the context of the $B$-meson decays \cite{Khodjamirian:2005ea,Khodjamirian:2006st,DeFazio:2005dx,DeFazio:2007hw},
paving the way for the construction of  a systematic approach to the exclusive
$\Lambda_b \to \Lambda \, \ell^{+} \ell^{-}$ decays
in analogy to the mesonic counterpart case \cite{Beneke:2001at}.
As already emphasized in \cite{Wang:2015vgv} one of the primary tasks of constructing the $\Lambda_b$-baryon LCSR
is to demonstrate QCD factorization for the vacuum-to-$\Lambda_b$-baryon correlation function in the proper kinematic regime.
In the framework of soft-collinear effective theory (SCET) factorization of the correlation function defined with the ``A-type" weak current
and an interpolating current of the $\Lambda$-baryon was established at tree level in the heavy quark limit \cite{Feldmann:2011xf}.
Instead of using the SCET technique we will, following \cite{Wang:2015vgv},  adopt the method of regions \cite{Beneke:1997zp} to prove  factorization of the  vacuum-to-$\Lambda_b$-baryon correlation function at next-to-leading-order (NLO) in $\alpha_s$ diagrammatically and resum  large logarithms in the short distance functions with the renormalization-group (RG) approach in momentum space.

Soft QCD dynamics of the vacuum-to-$\Lambda_b$-baryon correlation function is parameterized by
the non-perturbative but universal wave functions of the $\Lambda_b$-baryon \cite{Ball:2008fw} which also serves as fundamental
inputs for the theory description of semileptonic  $\Lambda_b \to p \, \ell \nu$  transitions \cite{Wang:2009hra},
$\Lambda_b \to \Lambda_c \, \ell \nu$  decays \cite{Guo:2005qa}
and hadronic $\Lambda_b \to p \, \pi, p \, K$ decays \cite{Lu:2009cm}.
Despite the recent progress in understanding the renormalization property of the twist-2 $\Lambda_b$-baryon DA
\cite{Bell:2013tfa,Braun:2014npa}, modelling the higher twist DA in compatible with the perturbative QCD constraints
still demands dedicated studies. As we will observe later, it is actually the  twist-4 DA of the $\Lambda_b$-baryon
entering the QCD factorization formulae of the vacuum-to-$\Lambda_b$-baryon correction functions,
whose RG evolution equation at one loop is not explicitly known yet (though building blocks of the renormalization kernels
for the desired light ray operators  can be found in  \cite{Knodlseder:2011gc}).
Investigating renormalization scale evolution of the convolution integral of the NLO twist-4 partonic DA and the tree-level
hard kernel constitutes another non-trivial target  of this  paper.

Different QCD-based approaches were adopted in the literature to compute the $\Lambda_b \to \Lambda$ form factors in addition to the recent Lattice QCD determinations \cite{Detmold:2012vy}. A closely related approach  was applied to construct the LCSR for
$\Lambda_b \to \Lambda$ form factors at tree level from the vacuum-to-$\Lambda$-baryon correlation function \cite{Wang:2008sm}
where the $\Lambda$-baryon DA entering the sum rules were only considered at the leading conformal spin accuracy
(the non-asymptotic corrections were worked out in \cite{Liu:2014uha} now)
and the Chernyak-Zhitnitsky \cite{Chernyak:1984bm} type of the $\Lambda$-baryon interpolating current was used
(see \cite{Aliev:2010uy,Gan:2012tt} for alternative choices and \cite{Braun:2006hz}
for interesting comments on the choices of the  baryonic interpolating currents).
Another approach to compute the  $\Lambda_b \to \Lambda$ form factors based upon the transverse-momentum-dependent (TMD) factorization
 was carried out in \cite{He:2006ud} where the soft overlap contribution was assumed to be suppressed by the Sudakov factor and only the
 hard spectator interactions induced  by two-hard-collinear-gluon exchanges are taken into account.
 A comparison of the resulting form factors from two different methods tends to indicate that the heavy-to-light baryonic form factors
 at large recoil is dominated by the {\it formally sub-leading} soft gluon exchanges
instead of the leading power hard spectator contributions \footnote{Strictly speaking, separation of the soft overlap  contributions
(Feynman mechanism) and the hard-scattering effects are both factorization scale- and scheme- dependent.}.

The paper is organized as follows. In section \ref{section: tree-level LCSR} we first set up  convention
of the helicity-based parametrization of the $\Lambda_b \to \Lambda$ form factors and then discuss the choice of
the interpolating currents for the $\Lambda$-baryon and introduce the correlation functions for constructions
of the LCSR for all the independent form factors. We also present the essential ingredients for proof of QCD
factorization of the correlation functions  and derive the tree level LCSR for $\Lambda_b \to \Lambda$ form factors.
Applying the method of regions we compute the  hard coefficients and the jet functions at ${\cal O}(\alpha_s)$
entering the QCD factorization formulae in section \ref{section: NLO factorization} where we demonstrate explicitly
cancellation of the factorization-scale dependence in the correlation functions and resummation of large logarithms
in the short-distance functions is also achieved at next-to-leading-logarithmic (NLL) accuracy with the standard RG approach.
Resummation improved LCSR for the $\Lambda_b \to \Lambda$ form factors presented in section \ref{section: resummation improved LCSR}
constitute the main new results of this paper. The details of the numerical analysis of the newly derived LCSR, including
various sources of perturbative and systematic uncertainties, the $z$-series expansion and a comparison with  the Lattice
determinations at small recoil, are collected in section \ref{section: numerical analysis}.
Phenomenological applications of our results to the exclusive electro-weak penguin decays $\Lambda_b \to \Lambda \ell^{+} \ell^{-}$
 at large recoil  are discussed in the factorization limit in section  \ref{section: phenomenologies}.
Section \ref{section: conclusion} is reserved for the concluding discussion.
Appendix \ref{appendix: spectral representations} contains dispersion representations of the convolution integrals
entering expressions of the factorized correlation functions, which are essential to construct the LCSR for
the $\Lambda_b \to \Lambda$ form factors presented in section \ref{section: resummation improved LCSR}.

\section {Tree-level LCSR of  $\Lambda_b \to \Lambda$ form factors }
\label{section: tree-level LCSR}

\subsection{Helicity-based $\Lambda_b \to \Lambda$ form factors}
\label{subsection: form factor definition}

Following \cite{Feldmann:2011xf} we define $\Lambda_b \to \Lambda$ form factors in the helicity basis
which lead to rather compact expressions for angular distributions,  unitary bounds and sum rules,
and  we collect the definitions as follows
\begin{eqnarray}
\langle \Lambda(p^{\prime}, s^{\prime} )|\bar  s \, \gamma_{\mu} \, b|  \Lambda_b(p,s) \rangle
&=& \bar \Lambda(p^{\prime}, s^{\prime}) \bigg [ f_{\Lambda_b \to \Lambda}^{0}(q^2) \,
\frac{m_{\Lambda_b} - m_{\Lambda}}{q^2}  \, q_{\mu}   \nonumber \\
&& + f_{\Lambda_b \to \Lambda}^{+}(q^2) \, \frac{m_{\Lambda_b} + m_{\Lambda}}{s_{+}}
\left ( (p+p^{\prime})_{\mu}  - \frac{m_{\Lambda_b}^2 - m_{\Lambda}^2}{q^2}  \, q_{\mu} \right ) \nonumber \\
&& +  f_{\Lambda_b \to \Lambda}^{T }(q^2) \, \left (\gamma_{\mu} - \frac{2 \, m_{\Lambda}}{s_{+}} p_{\mu}
- \frac{2 \, m_{\Lambda_b}}{s_{+}} p^{\prime}_{\mu}  \right ) \bigg ] \Lambda_b (p, s) \,,
\\
\langle \Lambda(p^{\prime}, s^{\prime} )|\bar  s \, \gamma_{\mu} \gamma_5 \, b|  \Lambda_b(p,s) \rangle
&=& - \bar \Lambda(p^{\prime}, s^{\prime}) \gamma_5 \bigg [ g_{\Lambda_b \to \Lambda}^{0}(q^2) \,
\frac{m_{\Lambda_b} + m_{\Lambda}}{q^2}  \, q_{\mu}   \nonumber \\
&& + g_{\Lambda_b \to \Lambda}^{+}(q^2) \, \frac{m_{\Lambda_b} - m_{\Lambda}}{s_{-}}
\left ( (p+p^{\prime})_{\mu}  - \frac{m_{\Lambda_b}^2 - m_{\Lambda}^2}{q^2}  \, q_{\mu} \right ) \nonumber \\
&& +  g_{\Lambda_b \to \Lambda}^{T }(q^2) \, \left (\gamma_{\mu} + \frac{2 \, m_{\Lambda}}{s_{-}} p_{\mu}
- \frac{2 \, m_{\Lambda_b}}{s_{-}} p^{\prime}_{\mu}  \right ) \bigg ] \Lambda_b (p, s) \,,
\\
\langle \Lambda(p^{\prime}, s^{\prime} )|\bar  s \, i \sigma_{\mu \nu}  q^{\nu} \, b|  \Lambda_b(p,s) \rangle
&=& - \bar \Lambda(p^{\prime}, s^{\prime}) \bigg [ h_{\Lambda_b \to \Lambda}^{+}(q^2) \, \frac{q^2}{s_{+}}
\left ( (p+p^{\prime})_{\mu}  - \frac{m_{\Lambda_b}^2 - m_{\Lambda}^2}{q^2}  \, q_{\mu} \right ) \nonumber \\
&& \hspace{-1.5 cm} + (m_{\Lambda_b} + m_{\Lambda}) \, h_{\Lambda_b \to \Lambda}^{T }(q^2)
\, \left (\gamma_{\mu} - \frac{2 \, m_{\Lambda}}{s_{+}} p_{\mu}
- \frac{2 \, m_{\Lambda_b}}{s_{+}} p^{\prime}_{\mu}  \right ) \bigg ] \Lambda_b (p, s) \,,
\\
\langle \Lambda(p^{\prime}, s^{\prime} )|\bar  s \, i \sigma_{\mu \nu}  q^{\nu} \gamma_5 \, b|  \Lambda_b(p,s) \rangle
&=& - \bar \Lambda(p^{\prime}, s^{\prime})\gamma_5
\bigg [ \tilde{h}_{\Lambda_b \to \Lambda}^{+}(q^2) \, \frac{q^2}{s_{-}}
\left ( (p+p^{\prime})_{\mu}  - \frac{m_{\Lambda_b}^2 - m_{\Lambda}^2}{q^2}  \, q_{\mu} \right ) \nonumber \\
&& \hspace{-1.5 cm} + (m_{\Lambda_b} - m_{\Lambda}) \, \tilde{h}_{\Lambda_b \to \Lambda}^{T }(q^2)
\, \left (\gamma_{\mu} + \frac{2 \, m_{\Lambda}}{s_{-}} p_{\mu}
- \frac{2 \, m_{\Lambda_b}}{s_{-}} p^{\prime}_{\mu}  \right ) \bigg ] \Lambda_b (p, s) \,,
\end{eqnarray}
where $m_{\Lambda_b}$ ($s$) is the mass (spin) of the $\Lambda_b$-baryon,
$m_{\Lambda}$ ($s^{\prime}$) is the mass (spin) of the $\Lambda$-baryon and
we introduce
\begin{eqnarray}
s_{\pm}=(m_{\Lambda_b} \pm  m_{\Lambda})^2 -q^2.
\end{eqnarray}
We work in the rest frame of the $\Lambda_b$-baryon with the velocity vector $v_{\mu}=p_{\mu}/m_{\Lambda_b}$
and define a light-cone vector $\bar n_{\mu}$ parallel to the four-momentum $p^{\prime}$ of the $\Lambda$-baryon in the massless limit.
Another light-cone vector can be introduced as $n_{\mu}= 2\, v_{\mu} - \bar n_{\mu}$
with $n \cdot \bar n =2$  for the later convenience. At large hadronic recoil we write
\begin{eqnarray}
n \cdot p^{\prime} \simeq \frac{m_{\Lambda_b}^2+m_{\Lambda}^2-q^2}{m_{\Lambda_b}} = 2 E_{\Lambda} \sim {\cal O}(m_{\Lambda_b})\,.
\label{momentum npprime}
\end{eqnarray}
Exploiting the heavy quark symmetry and the collinear equations of motion yields \cite{Mannel:2011xg,Feldmann:2011xf}
\begin{eqnarray}
&& f_{\Lambda_b \to \Lambda}^{0}(q^2)\simeq f_{\Lambda_b \to \Lambda}^{+}(q^2) \simeq f_{\Lambda_b \to \Lambda}^{T}(q^2)
\simeq h_{\Lambda_b \to \Lambda}^{+}(q^2) \simeq h_{\Lambda_b \to \Lambda}^{T}(q^2) \nonumber \\
&& \simeq g_{\Lambda_b \to \Lambda}^{0}(q^2) \simeq g_{\Lambda_b \to \Lambda}^{+}(q^2) \simeq g_{\Lambda_b \to \Lambda}^{T}(q^2)
\simeq \tilde{h}_{\Lambda_b \to \Lambda}^{+}(q^2) \simeq \tilde{h}_{\Lambda_b \to \Lambda}^{T}(q^2) \,
\label{symmetry relation of FFs at tree level}
\end{eqnarray}
at large recoil, where the strong interaction dynamics of the hadronic transitions is assumed to be dominated by
the  soft gluon exchanges.   Hard spectator interactions induced by the two-hard-collinear-gluon exchanges
are shown to still respect these symmetry relations at leading power in $\Lambda/m_b$ \cite{Wang:2011uv},
where $\Lambda$ is a hadronic scale of order $\Lambda_{\rm QCD}$.
We will first confirm such form factor relations from the tree-level LCSR (see also \cite{Feldmann:2011xf}) and then
compute the symmetry-breaking effects induced by the hard fluctuations of QCD decay currents
(also known as the matching coefficients of weak currents from QCD onto SCET) and the one-loop jet function in the next section.

\subsection{Interpolating currents and correlation function}

Following the standard strategy we start with construction of the correlation function
\begin{eqnarray}
\Pi_{\mu, a}(p, q)= i \int d^4 x  \, e^{i q \cdot x} \, \langle 0 |T \{j_{\Lambda}(x), j_{\mu, a}(0) \}| \Lambda_b(p) \rangle \,,
\label{definition: correlator}
\end{eqnarray}
where the local current $j_{\Lambda}$ interpolates the $\Lambda$-baryon and
$j_{\mu, a}$ stands for the weak transition current $\bar s \, \Gamma_{\mu, a} \, b $
with the index ``$a$" indicating a certain Lorenz structure, i.e.,
\begin{eqnarray}
j_{\mu, V} =\bar s \, \gamma_{\mu} \, b\,, &  \qquad & j_{\mu, A}=\bar s \, \gamma_{\mu}\, \gamma_5 \, b\,, \nonumber \\
j_{\mu, T} =\bar s \, \sigma_{\mu \nu} \, q^{\nu}\, b\,, &  \qquad &
j_{\mu, \tilde{T}}=\bar s \, \sigma_{\mu \nu} \, q^{\nu} \, \gamma_5 \, b \,.
\end{eqnarray}
 As discussed in \cite{Khodjamirian:2011jp} the general structure of the $\Lambda$-baryon current reads
\begin{eqnarray}
j_{\Lambda}= \epsilon_{i j k} \, \left ( u_i^{\rm T} \, C \, \Gamma \, d_j \right ) \, \tilde{\Gamma} \, s_k \,,
\end{eqnarray}
where $C$ is the charge conjugation matrix and the sum runs over the color indices $i, j, k$.
Implementing the isospin constraint of the light diquark $[ud]$ system we are left with three independent choices
\begin{eqnarray}
j_{\Lambda}^{\cal{A}} = \epsilon_{i j k} \, \left ( u_i^{\rm T} \, C \, \gamma_5 \, \!\! \not n \, d_j \right )
\,  s_k  \,, \qquad
j_{\Lambda}^{\cal{P}}= \epsilon_{i j k} \, \left ( u_i^{\rm T} \, C \, \gamma_5 \, d_j \right )
\, s_k  \,,  \qquad
j_{\Lambda}^{\cal{S}} = \epsilon_{i j k} \, \left ( u_i^{\rm T} \, C \,  d_j \right )
\, \gamma_5 \, s_k  \,. \nonumber  \\
\end{eqnarray}
Projecting out the large and small components of the (hard)-collinear quark fields one can readily identify that
the two currents $j_{\Lambda}^{\cal{P}}$ and $j_{\Lambda}^{\cal{S}}$ are power suppressed compared with
the axial-vector current $j_{\Lambda}^{\cal{A}}$. Having in mind that the interpolating current should
couple strongly to the $\Lambda$-baryon in order to minimize the contamination generated by its coupling to the unwanted
hadronic states, we will only consider the axial-vector current $j_{\Lambda}^{\cal{A}}$
for construction of the correlation function.

To derive the hadronic dispersion relation of the correlation function  we need to define the coupling of
the $\Lambda$-baryon with the $j_{\Lambda}^{\cal{A}}$ current
\begin{eqnarray}
\langle 0| j_{\Lambda}^{\cal{A}} | \Lambda(p^{\prime})\rangle = f_{\Lambda}(\mu) \,
(n \cdot p^{\prime}) \, \Lambda(p^{\prime}) \,,
\end{eqnarray}
where the renormalization scale dependence of $f_{\Lambda}(\mu)$ is indicated explicitly
and the corresponding evolution equation is given by
\begin{eqnarray}
\frac{d}{d \ln \mu} \ln f_{\Lambda}(\mu) = - \left (\frac{\alpha_s(\mu)}{4\, \pi} \right )^k \,
\gamma_{\Lambda}^{(k)}\,,
\label{scale dependence of the baryonic current}
\end{eqnarray}
with $\gamma_{\Lambda}^{(1)}=4/3$ \cite{Chernyak:1987nu,Braun:2000kw}.
It is then a straightforward task to write down the hadronic representations for the correlation functions
defined with various weak currents
\begin{eqnarray}
\Pi_{\mu, V}(p, q) &=& \frac{f_{\Lambda}(\mu) \, (n \cdot p^{\prime}) }
{m_{\Lambda}^2/n \cdot p^{\prime} - \bar n \cdot p^{\prime} - i 0} \,\,
\frac{\!\!\! \not \bar n}{2} \,\, \bigg [  f_{\Lambda_b \to \Lambda}^{T}(q^2) \, \gamma_{\perp \, \mu}
+ \frac{f_{\Lambda_b \to \Lambda}^{0}(q^2)-f_{\Lambda_b \to \Lambda}^{+}(q^2) }
{2 \,(1 - n \cdot p^{\prime}/m_{\Lambda_b})} \,\, n_{\mu}  \nonumber  \\
&& + \frac{f_{\Lambda_b \to \Lambda}^{0}(q^2)+f_{\Lambda_b \to \Lambda}^{+}(q^2)}{2} \,\,
\bar n_{\mu} \bigg ] \Lambda_b(p) +
\int_{\omega_s}^{+\infty} \, d \omega^{\prime} \, \frac{1}{\omega^{\prime} - \bar n \cdot p^{\prime} - i 0} \,\, \nonumber \\
&& \times \,\, \frac{\!\!\! \not \bar n}{2} \,\left [  \rho_{V, \perp}^{h}(\omega^{\prime}, n \cdot p^{\prime}) \, \gamma_{\perp \, \mu}
+ \rho_{V, n}^{h}(\omega^{\prime}, n \cdot p^{\prime}) \, n_{\mu}
+  \rho_{V, \bar n}^{h}(\omega^{\prime}, n \cdot p^{\prime}) \, \bar n_{\mu}   \right ] \Lambda_b(p)  \,,
\\
\Pi_{\mu, A}(p, q) &=& \frac{f_{\Lambda}(\mu) \, (n \cdot p^{\prime}) }
{m_{\Lambda}^2/n \cdot p^{\prime} - \bar n \cdot p^{\prime} - i 0} \,\,
\gamma_5 \,\, \frac{\!\!\! \not \bar n}{2} \,\, \bigg [  g_{\Lambda_b \to \Lambda}^{T}(q^2) \, \gamma_{\perp \, \mu}
+ \frac{g_{\Lambda_b \to \Lambda}^{0}(q^2)-g_{\Lambda_b \to \Lambda}^{+}(q^2) }
{2 \,(1 - n \cdot p^{\prime}/m_{\Lambda_b})} \,\, n_{\mu}  \nonumber  \\
&& + \frac{g_{\Lambda_b \to \Lambda}^{0}(q^2)+g_{\Lambda_b \to \Lambda}^{+}(q^2)}{2} \,\,
\bar n_{\mu} \bigg ] \Lambda_b(p) +
\int_{\omega_s}^{+\infty} \, d \omega^{\prime} \, \frac{1}{\omega^{\prime} - \bar n \cdot p^{\prime} - i 0} \,\, \nonumber \\
&& \times \gamma_5 \,\, \frac{\!\!\! \not \bar n}{2} \,
\left [  \rho_{A, \perp}^{h}(\omega^{\prime}, n \cdot p^{\prime}) \, \gamma_{\perp \, \mu}
+ \rho_{A, n}^{h}(\omega^{\prime}, n \cdot p^{\prime}) \, n_{\mu}
+  \rho_{A, \bar n}^{h}(\omega^{\prime}, n \cdot p^{\prime}) \, \bar n_{\mu}   \right ] \Lambda_b(p)  \,, \hspace{0.5 cm}
\\
\Pi_{\mu, T}(p, q) &=& - \frac{m_{\Lambda_b} \, f_{\Lambda}(\mu) \, (n \cdot p^{\prime})}
{m_{\Lambda}^2/n \cdot p^{\prime} - \bar n \cdot p^{\prime} - i 0} \,\,
\frac{\!\!\! \not \bar n}{2} \,\, \bigg [  h_{\Lambda_b \to \Lambda}^{T}(q^2) \, \gamma_{\perp \, \mu} \nonumber \\
&& + \frac{h_{\Lambda_b \to \Lambda}^{+}(q^2) }{2} \,\,
\left ( \left (1- \frac{n \cdot p^{\prime}}{m_{\Lambda_b}} \right ) \bar n_{\mu} - n_{\mu} \right  ) \bigg ] \Lambda_b(p)
+ \int_{\omega_s}^{+\infty} \, d \omega^{\prime} \, \frac{1}{\omega^{\prime} - \bar n \cdot p^{\prime} - i 0} \,\, \nonumber \\
&& \times \,\, \frac{\!\!\! \not \bar n}{2} \,\left [  \rho_{T, \perp}^{h}(\omega^{\prime}, n \cdot p^{\prime}) \, \gamma_{\perp \, \mu}
+ \rho_{T, +}^{h}(\omega^{\prime}, n \cdot p^{\prime}) \,
\left ( \left (1- \frac{n \cdot p^{\prime}}{m_{\Lambda_b}} \right ) \bar n_{\mu} - n_{\mu} \right  )   \right ] \Lambda_b(p)  \,,
\hspace{0.8  cm}
\\
\Pi_{\mu, \tilde{T}}(p, q) &=&  \frac{m_{\Lambda_b} \, f_{\Lambda}(\mu) \, (n \cdot p^{\prime})}
{m_{\Lambda}^2/n \cdot p^{\prime} - \bar n \cdot p^{\prime} - i 0} \,\,
\gamma_5 \, \frac{\!\!\! \not \bar n}{2} \,\, \bigg [  \tilde{h}_{\Lambda_b \to \Lambda}^{T}(q^2) \, \gamma_{\perp \, \mu} \nonumber \\
&& + \frac{\tilde{h}_{\Lambda_b \to \Lambda}^{+}(q^2) }{2} \,\,
\left ( \left (1- \frac{n \cdot p^{\prime}}{m_{\Lambda_b}} \right ) \bar n_{\mu} - n_{\mu} \right  ) \bigg ] \Lambda_b(p)
+ \int_{\omega_s}^{+\infty} \, d \omega^{\prime} \, \frac{1}{\omega^{\prime} - \bar n \cdot p^{\prime} - i 0} \,\, \nonumber \\
&& \times \gamma_5 \,\, \frac{\!\!\! \not \bar n}{2} \,
\left [  \rho_{\tilde{T}, \perp}^{h}(\omega^{\prime}, n \cdot p^{\prime}) \, \gamma_{\perp \, \mu}
+ \rho_{\tilde{T}, +}^{h}(\omega^{\prime}, n \cdot p^{\prime}) \,
\left ( \left (1- \frac{n \cdot p^{\prime}}{m_{\Lambda_b}} \right ) \bar n_{\mu} - n_{\mu} \right  )   \right ] \Lambda_b(p)  \,,
\hspace{0.8  cm}
\end{eqnarray}
where we have defined
\begin{eqnarray}
p^{\prime}= p-q \,, \qquad
\gamma_{\perp \, \mu}=  \gamma_{\mu} - \, \frac{\! \not \bar n}{2} \, n_{\mu}
 - \, \frac{\! \not n}{2} \, \bar n_{\mu} \,.
\end{eqnarray}
Note also that we have naively assumed that effects from the negative-party baryons with $J^P=1/2^{-}$
can be absorbed into the dispersion integrals in the above expressions and we refer to \cite{Khodjamirian:2011jp}
for a detailed discussion of eliminating the ``contamination" from such background contributions in the context
of the LCSR with the nucleon DA.

\subsection{Tree-level LCSR}

Now we turn to compute the correlation function $\Pi_{\mu, a}(p, q)$
at space-like interpolating momentum with $|\bar n \cdot p^{\prime}| \sim {\cal O} (\Lambda)$
and $n \cdot p^{\prime}$ fixed by Eq. (\ref{momentum npprime}), where light-cone operator-product-expansion (OPE)
is applicable. Perturbative factorization of the partonic correlation function $\Pi^{\rm par}_{\mu, a}(p, q)$
(defined as replacing  the hadronic state $|\Lambda_b(p) \rangle$ by the on-shell partonic state$| b(p_b) u(k_1) d(k_2) \rangle$
in Eq. (\ref{definition: correlator})) at tree level takes the following form
\begin{eqnarray}
\Pi^{\rm par}_{\mu, a}(p, q)=\int d \omega_1^{\prime} \int d \omega_2^{\prime} \,
T^{(0)}_{\alpha \beta \gamma \delta}(n \cdot p^{\prime}, \bar n \cdot p^{\prime}, \omega_1^{\prime}, \omega_2^{\prime})\,
\Phi_{b u d}^{(0) \, \alpha \beta \delta}(\omega_1^{\prime}, \omega_2^{\prime}) \,,
\label{tree-level factorization at partonic level}
\end{eqnarray}
where the superscript $(0)$ indicates the tree-level approximation and the Lorenz index ``$\mu$" is suppressed on the right-hand side
in order not to overload the notation.

%%%%%%%%%%%
\begin{figure}
\begin{center}
\includegraphics[width=0.4 \columnwidth]{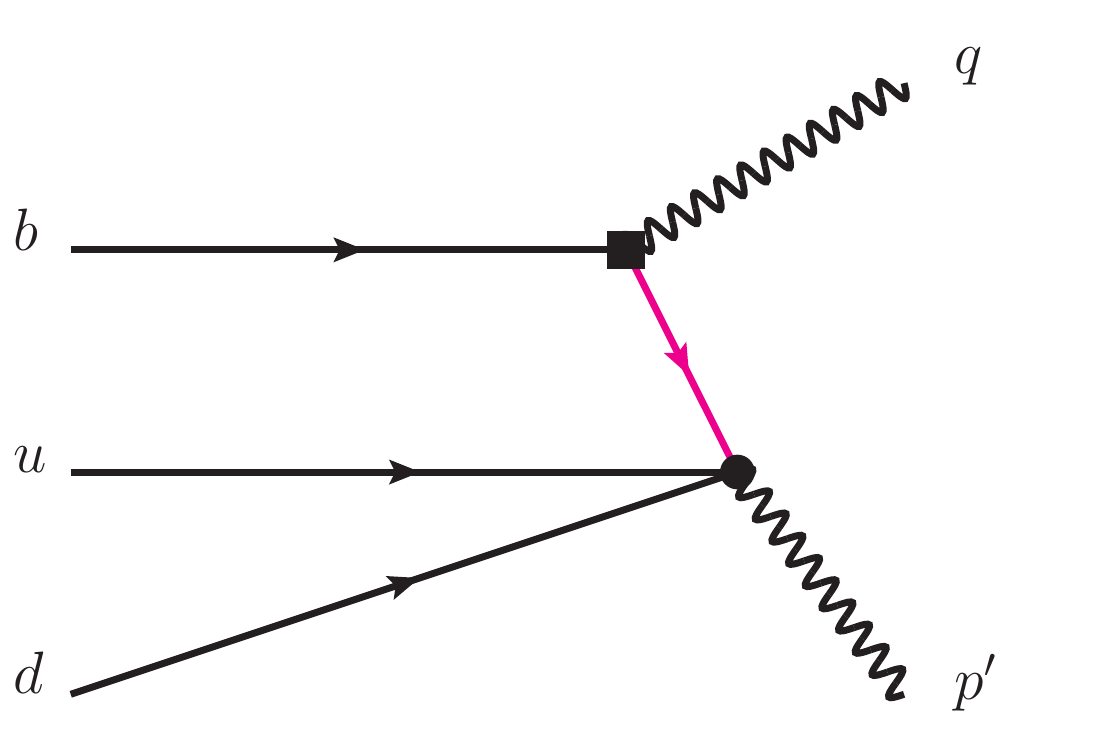}
\vspace*{0.1cm}
\caption{Diagrammatical representation of the correlation function
$\Pi_{\mu, a}(n \cdot p^{\prime},\bar n \cdot p^{\prime})$ at tree level,
where the black square denotes the weak transition vertex, the black blob represents
the Dirac structure of the $\Lambda$-baryon current and the pink internal line indicates the hard-collinear
propagator of the strange quark. }
\label{fig:tree_correlator}
\end{center}
\end{figure}
%%%%%%%%%%%

Evaluating the diagram in Fig. \ref{fig:tree_correlator} leads to the leading-order hard kernel
\begin{eqnarray}
T^{(0)}_{\alpha \beta \gamma \delta}(n \cdot p^{\prime}, \bar n \cdot p^{\prime}, \omega_1^{\prime}, \omega_2^{\prime})
= - \frac{1}{\bar n \cdot p^{\prime} - \omega_1^{\prime} - \omega_2^{\prime} + i 0} \,
\left( C \, \gamma_5\, \not \! n \right)_{\alpha \beta}  \, \,
\left ( \frac{\not \! \bar n}{2} \, \Gamma_{\mu, a} \right )_{\gamma \delta} \, ,
\label{hard kernel at tree level}
\end{eqnarray}
and the partonic DA of the $\Lambda_b$-baryon is defined as
\begin{eqnarray}
\Phi_{b u d}^{\alpha \beta \delta}(\omega_1^{\prime}, \omega_2^{\prime})
&=& \int \frac{d t_1}{ 2 \pi} \int \frac{d t_2}{ 2 \pi} \,
e^{i \left (\omega_1^{\prime} t_1 + \omega_2^{\prime} t_2 \right )} \, \nonumber \\
&& \times \, \epsilon_{i j k} \, \langle 0 | \left [u^{\rm T}_{i} (t_1 \bar n) \right ]_{\alpha} \,
[0, t_1 \bar n] \, \left [d_{j} (t_2 \bar n) \right ]_{\beta} \, [0, t_2 \bar n] \,
\left [ b_{k}(0)\right ]_{\delta} | b(v) u(k_1) d(k_2) \rangle \,, \hspace{0.5 cm}
\label{def: partonic DA}
\end{eqnarray}
where the $b$-quark field needs to be understood as an effective heavy quark field
in HQET and the light-cone Wilson line
\begin{eqnarray}
[0, t \, \bar n] = {\rm P} \, \left \{ {\rm  Exp} \left [  - i \, g_s  \, t \,
\int_0^1 \, du \, \bar n  \cdot A(u \, t \,  \bar n) \right ]  \right \}
\end{eqnarray}
is introduced with the convention of the covariant derivative in QCD as
$D_{\mu}=\partial_{\mu} - i g_s \, T^{a}\,  A^{a}_{\mu}$.
The tree-level partonic DA entering the factorized expression (\ref{tree-level factorization at partonic level})
can  be  readily found to be
\begin{eqnarray}
\Phi_{b u d}^{(0) \, \alpha \beta \delta}(\omega_1^{\prime}, \omega_2^{\prime})
= \delta(\bar n \cdot k_1 - \omega_1^{\prime}) \,  \delta(\bar n \cdot k_2 - \omega_2^{\prime}) \,
\epsilon_{i j k} \, \left [u^{\rm T}_{i} (k_1) \right ]_{\alpha}  \,
\left [d_{j} (k_2) \right ]_{\beta} \, \left [ b_{k}(v)\right ]_{\delta} \,.
\end{eqnarray}

Starting with the definition of the most-general light-cone hadronic matrix element
in coordinate space \cite{Bell:2013tfa}
\begin{eqnarray}
\Phi_{\Lambda_b}^{\alpha \beta \delta}(t_1, t_2)
&\equiv&  \epsilon_{i j k} \, \langle 0 | \left [u^{\rm T}_{i} (t_1 \bar n) \right ]_{\alpha} \,
[0, t_1 \bar n] \, \left [d_{j} (t_2 \bar n) \right ]_{\beta} \, [0, t_2 \bar n] \,
\left [ b_{k}(0)\right ]_{\delta} | \Lambda_b(v) \rangle \nonumber \\
&=&  \frac{1}{4 } \, \left \{ f_{\Lambda_b}^{(1)}(\mu) \,
\left [ \tilde{M}_1(v, t_1, t_2) \, \gamma_5 \, C^{T} \right ]_{\beta \alpha}
+  f_{\Lambda_b}^{(2)}(\mu) \,
\left [ \tilde{M}_2(v, t_1, t_2) \, \gamma_5 \, C^{T} \right ]_{\beta \alpha} \right \} \,
\left [ \Lambda_b(v) \right ]_{\delta}   \,, \nonumber \\
\end{eqnarray}
performing the Fourier transformation and including the NLO terms off the light-cone leads to
the momentum space light-cone projector in $D$ dimensions
\begin{eqnarray}
M_2(\omega_1^{\prime},\omega_2^{\prime}) &=& \frac {\! \not n}{2} \, \psi_2(\omega_1^{\prime},\omega_2^{\prime})
+ \frac {\! \not \bar  n}{2} \, \psi_4(\omega_1^{\prime},\omega_2^{\prime}) \nonumber
\\
&& -\frac{1}{D-2} \, \gamma_{\perp}^{\mu} \, \left  [ \psi_{\perp, 1}^{+-}(\omega_1^{\prime},\omega_2^{\prime}) \,
\frac{\! \not n  \, \! \not \bar  n}{4}  \, \frac{\partial}{\partial k_{1 \perp}^{\mu}}
+ \psi_{\perp, 1}^{-+}(\omega_1^{\prime},\omega_2^{\prime}) \,
\frac{\! \not \bar n  \, \! \not  n}{4}  \, \frac{\partial}{\partial k_{1 \perp}^{\mu}} \right  ] \nonumber
\\
&& -\frac{1}{D-2} \, \gamma_{\perp}^{\mu} \, \left  [ \psi_{\perp, 2}^{+-}(\omega_2^{\prime},\omega_2^{\prime}) \,
\frac{\! \not n  \, \! \not \bar  n}{4}  \, \frac{\partial}{\partial k_{2 \perp}^{\mu}}
+ \psi_{\perp, 2}^{-+}(\omega_1^{\prime},\omega_2^{\prime}) \,
\frac{\! \not \bar n  \, \! \not  n}{4}  \, \frac{\partial}{\partial k_{2 \perp}^{\mu}} \right  ]  \,,
\label{chiral-odd projector}\\
M_1(\omega_1^{\prime},\omega_2^{\prime}) &=& \frac{\! \not \bar  n \, \! \not  n }{8} \,
\psi_{3}^{+-}(\omega_1^{\prime},\omega_2^{\prime})
+ \frac{\! \not  n  \, \! \not \bar  n}{8} \,
\psi_{3}^{-+}(\omega_1^{\prime},\omega_2^{\prime}) \nonumber
\\
&& - \frac{1}{D-2} \left [ \psi_{\perp, 3}^{(1)}(\omega_1^{\prime},\omega_2^{\prime}) \! \not  v
\, \gamma_{\perp}^{\mu}  \, \frac{\partial}{\partial k_{1 \perp}^{\mu}}
+ \psi_{\perp, 3}^{(2)}(\omega_1^{\prime},\omega_2^{\prime}) \, \gamma_{\perp}^{\mu}  \,
\! \not  v \, \frac{\partial}{\partial k_{2 \perp}^{\mu}}  \right ] \nonumber
\\
&& - \frac{1}{D-2} \left [ \psi_{\perp, Y}^{(1)}(\omega_1^{\prime},\omega_2^{\prime}) \! \not \bar n
\, \gamma_{\perp}^{\mu}  \, \frac{\partial}{\partial k_{1 \perp}^{\mu}}
+ \psi_{\perp, Y}^{(2)}(\omega_1^{\prime},\omega_2^{\prime}) \,  \gamma_{\perp}^{\mu}  \,
\! \not  \bar n \, \frac{\partial}{\partial k_{2 \perp}^{\mu}}  \right ] \,,
\label{chiral-even projector}
\end{eqnarray}
where we have adjusted the notation of the $\Lambda_b$-baryon DA defined in \cite{Bell:2013tfa}.
Applying the equations of motion in the Wandzura-Wilczek approximation \cite{Wandzura:1977qf} yields
\begin{eqnarray}
\psi_{\perp, 1}^{-+}(\omega_1^{\prime},\omega_2^{\prime})=\omega_1^{\prime} \,
\psi_4(\omega_1^{\prime},\omega_2^{\prime}) \,, \qquad
\psi_{\perp, 2}^{+-}(\omega_1^{\prime},\omega_2^{\prime})=\omega_2^{\prime} \,
\psi_4(\omega_1^{\prime},\omega_2^{\prime})  \,.
\label{EOM for Lambdab DA}
\end{eqnarray}

It is now straightforward to derive the tree-level factorization formulae
\begin{eqnarray}
\Pi^{(0)}_{\mu, V(A)}(p, q) &=& f_{\Lambda_b}^{(2)}(\mu) \, \int_0^{+\infty} \, d \omega_1^{\prime} \,
\int_0^{+\infty} \, d \omega_2^{\prime} \,\,
\frac{\psi_4(\omega_1^{\prime},\omega_2^{\prime})}{\omega_1^{\prime}+\omega_2^{\prime}-\bar n \cdot p^{\prime} - i 0} \, \nonumber
\\
&& \times \left (1, \gamma_5 \right ) \, \frac{\not \! \bar n}{2} \, (\gamma_{\perp \mu} + \bar n_{\mu}) \,\, \Lambda_b(v) \,, \nonumber
\\
\Pi^{(0)}_{\mu, T(\tilde{T})}(p, q) &=& m_{\Lambda_b} \, f_{\Lambda_b}^{(2)}(\mu) \, \int_0^{+\infty} \, d \omega_1^{\prime} \,
\int_0^{+\infty} \, d \omega_2^{\prime} \,\,
\frac{\psi_4(\omega_1^{\prime},\omega_2^{\prime})}{\omega_1^{\prime}+\omega_2^{\prime}-\bar n \cdot p^{\prime} - i 0} \, \nonumber
\\
&& \times \left ( -1, \gamma_5 \right ) \, \frac{\not \! \bar n}{2} \,
\left [ \gamma_{\perp \mu} + \frac{1}{2}
\left ( \left (1- \frac{n \cdot p^{\prime}}{m_{\Lambda_b}} \right )  \, \bar n_{\mu} - n_{\mu} \right ) \right ] \,\, \Lambda_b(v) \,,
\end{eqnarray}
at leading power in $\Lambda/m_b$.
Employing the parton-hadronic duality approximation for the dispersion integrals in the hadronic representations
and performing the continuum subtraction as well as  the Borel transformation we obtain the tree-level LCSR
\begin{eqnarray}
F_{\Lambda_b \to \Lambda}^{i}(q^2)=\frac{f_{\Lambda_b}^{(2)}(\mu)}{f_{\Lambda}(\mu) \, n \cdot p^{\prime}} \,
{\rm exp} \left [ \frac{m_{\Lambda}^2}{n \cdot p^{\prime} \, \omega_M} \right ] \,
\int_0^{\omega_s} \, d \omega^{\prime} \, e^{-\omega^{\prime}/\omega_M} \, \tilde{\psi}_4(\omega^{\prime})
+ {\cal O}(\alpha_s) \,,
\end{eqnarray}
where $F_{\Lambda_b \to \Lambda}^{i}(q^2)$ represents any of the 10 $\Lambda_b \to \Lambda$  form factors
defined in section \ref{subsection: form factor definition} and
\begin{eqnarray}
\tilde{\psi}_4(\omega^{\prime}) = \omega^{\prime} \, \int_0^1 \, d u \,
\psi_4 \left (u \,\omega^{\prime},  (1-u) \,\omega^{\prime} \right )\,.
\label{definition of psi4tidle}
\end{eqnarray}
Applying the power counting scheme
\begin{eqnarray}
\omega_s \sim \omega_M \sim \frac{\Lambda^2}{n \cdot p^{\prime}} \,, \qquad
\tilde{\psi}_4(\omega^{\prime}) \sim  \omega^{\prime} \sim \omega_s \,,
\end{eqnarray}
the tree-level contribution (Feynman mechanism) to the $\Lambda_b \to \Lambda$  form factors scales as
$1/ (n \cdot p^{\prime})^3$ in the large energy  limit of the $\Lambda$-baryon, in agreement  with the obervations
 of \cite{Mannel:2011xg,Feldmann:2011xf}. Since the large-recoil symmetry relations for the form factors are preserved at tree level,
the symmetry violation effect, if it emerges at one loop, must be infrared finite
due to the vanishing soft subtraction at ${\cal O}(\alpha_s)$ in order not to invalidate QCD factorization of the correlation
functions.

\section {Factorization of the correlation function at ${\cal O}(\alpha_s)$}
\label{section: NLO factorization}

The purpose of this section is to compute the short-distance functions
entering the  factorization formulae of $\Pi_{\mu, a}(p, q)$ at one loop
\begin{eqnarray}
\Pi_{\mu, a}(p, q) =  T \otimes \Phi_{bud}  = C \cdot J \otimes \Phi_{bud} \,,
\end{eqnarray}
where $\otimes$ denotes a convolution in the light-cone variables $\omega_1^{\prime}$ and $\omega_2^{\prime}$.
We will closely follow the strategies to prove the one-loop factorization of the vacuum-to-$B$-meson
correlation function detailed in \cite{Wang:2015vgv} and employ the method of regions to evaluate the hard coefficients
and the jet functions simultaneously. We further verify  cancellation of the factorization-scale dependence of the
correlation functions by computing convolution integrals of the NLO partonic DA and the tree-level hard kernel
in (\ref{hard kernel at tree level}) explicitly. Resummation of large logarithms involved in the perturbative functions
is carried out  at NLL using the momentum-space RG approach.

\subsection{Hard and jet functions at NLO}

\begin{figure}
\begin{center}
\includegraphics[width=0.9 \columnwidth]{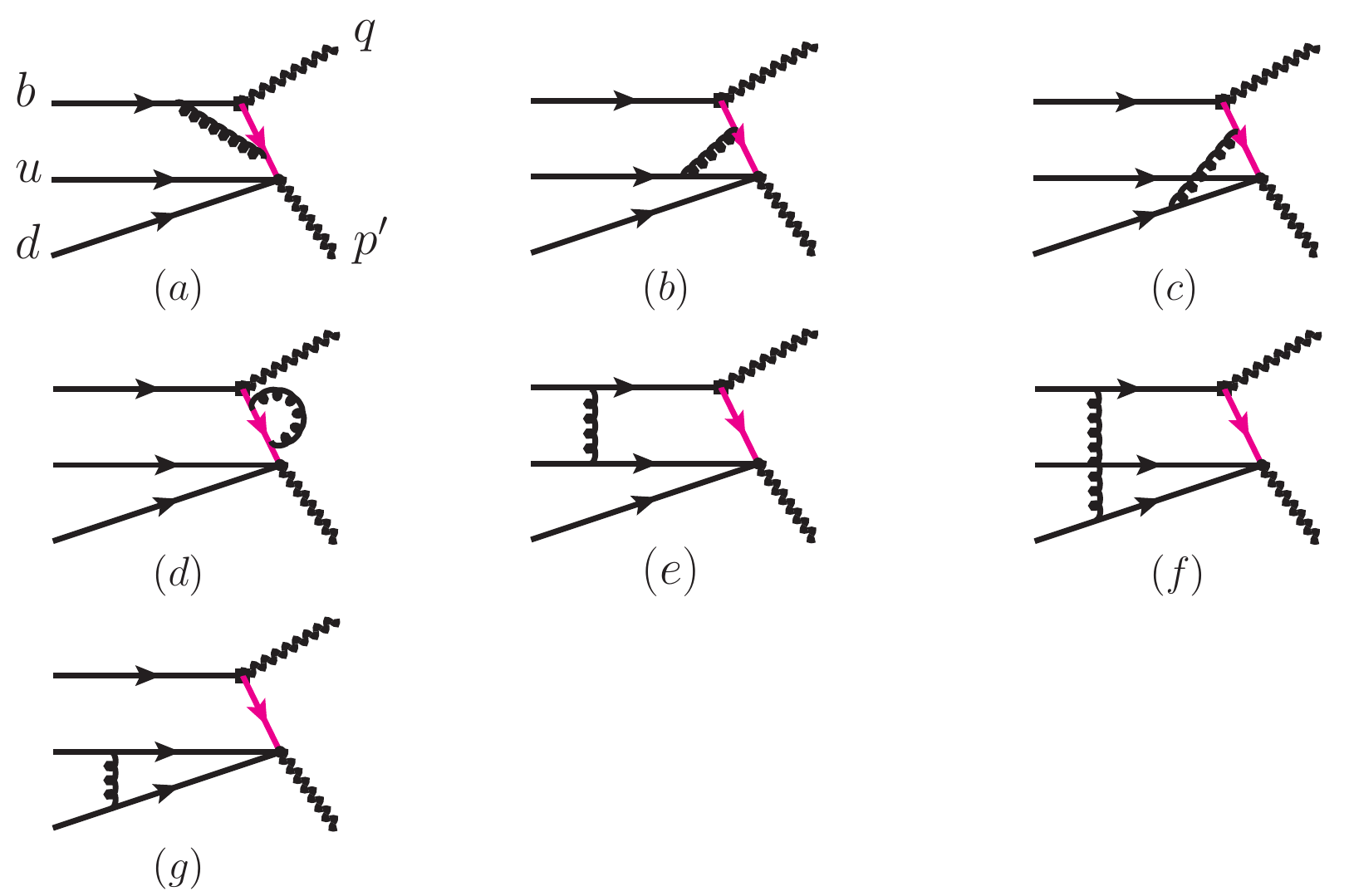}
\vspace*{0.1cm}
\caption{Diagrammatical representation of the correlation function
$\Pi_{\mu, a}(n \cdot p^{\prime},\bar n \cdot p^{\prime})$ at one loop.
Same conventions as in Fig. \ref{fig:tree_correlator}. }
\label{fig: one-loop_correlator}
\end{center}
\end{figure}

We are now ready to compute the one-loop QCD diagrams displayed in Fig. \ref{fig: one-loop_correlator}
for determinations of the perturbative matching coefficients. Since the loop integral entering the amplitude of the diagram (g)
with one-gluon exchange  between the two soft quarks does not contain any external hard and/or hard-collinear
momentum modes, no contribution to the perturbative functions can arise from this diagram and we will compute the
remaining diagrams one by one in the following. To facilitate the discussion of the one-loop calculation we will first
focus on the (axial)-vector correlation functions $\Pi_{\mu, V(A)}(p, q)$ and generalize  the computation to the
(pseudo)-tensor correlation functions $\Pi_{\mu, T(\tilde{T})}(p, q)$ in the end of this section.

%%%%%%%%%%%
%%%%%%%%%%%

\subsubsection{Weak vertex diagram}

Now we turn to compute the one-loop QCD correction to the weak vertex diagram
displayed in Fig. \ref{fig: one-loop_correlator}(a)
\begin{eqnarray}
\Pi^{{\rm par}, a}_{\mu, V(A)}(p, q) &=&
\frac{i \, g_s^2\, C_F}{\bar n \cdot p^{\prime} - (\omega_1+\omega_2) + i 0} \,
\int \frac{d^D \, l}{(2 \pi)^D} \,   \frac{1}{ [(p^{\prime}-k+l)^2+i 0] [(m_b v +l)^2-m_b^2+ i0] [l^2+i0]} \nonumber \\
&& \epsilon_{i j k} \, \left [u^{\rm T}_{i} (k_1) \, C \, \gamma_5 \,  \not \!  n \, d_{j} (k_2) \right ]
\, \frac{\!  \not \bar n} {2} \, \gamma_{\rho} \, ( \! \not  p^{\prime}- \!  \not k + \!  \not l ) \,
\gamma_{\mu} \, (1, \gamma_5) \, (m_b \!  \not v +  \! \not l + m_b) \, \gamma^{\rho} \, b_{k}(v) \,,  \hspace{1.0 cm}
\label{expression of fig. 2a}
\end{eqnarray}
where $k=k_1+k_2$, $\omega_i=\bar n \cdot k_i \, (i=1, 2)$ and $D=4 - 2 \, \epsilon$.
We have approximated the $b$-quark momentum as $p_b=m_b \, v$
by dropping out the residual momentum, since we are
only interested in extracting the leading power contributions to the correlation functions.
The standard strategies to evaluate the perturbative matching coefficients would be: (i) first computing
the loop integrals with the method of regions to determine the ``bare" perturbative kernels without the ultraviolet (UV)
and infrared (IR) subtractions, (ii) decomposing the resulting contributions in terms of independent operator bases
(including the so-called ``evanescent operators" \cite{Bonneau:1980zp,Bonneau:1979jx} whenever necessary) with the aid of
the equations of motion, (iii) implementing the UV renormalization programs and infrared subtractions (determined by matrix elements of
the effective operators), (iv) applying the momentum-space  light-cone projector of the $\Lambda_b$-baryon
to formulate factorized expressions of the correlation functions in the end.
The above-mentioned procedures can be reduced in the absence of the ``evanescent operators" as in our case, since
no Fierz rearrangement is required in the  perturbative  matching:
\begin{eqnarray}
i \int d^4 x  \, e^{i q \cdot x} \, T \{j_{\Lambda}(x), j_{\mu, a}(0) \}
& \rightarrow & \sum_{i, j} \int d t_1 \int d t_2 \, \tilde{T}_{ij}(t_1,t_2, \bar n \cdot q, v \cdot q, m_b, \mu) \,\,
[\Gamma_i]_{\alpha \beta}  \,\, [\Gamma^{\prime}_j]_{\gamma \delta} \,\, \nonumber
\\
&& \times  \epsilon_{i j k} \,  \left [u^{\rm T}_{i} (t_1 \bar n) \right ]_{\alpha} \,
[0, t_1 \bar n] \, \left [d_{j} (t_2 \bar n) \right ]_{\beta} \, [0, t_2 \bar n] \,
\left [ b_{k}(0)\right ]_{\delta} \,\,.
\end{eqnarray}
The hard function contributed from Fig. \ref{fig: one-loop_correlator}(a) can be determined  by expanding
Eq. (\ref{expression of fig. 2a}) in the hard region and by applying the light-cone projector subsequently
and this leads to
\begin{eqnarray}
&& \Pi^{a, h}_{\mu, V(A)}(p, q)  \nonumber \\
&& = i \, g_s^2\, C_F \, \frac{f_{\Lambda_b}^{(2)}(\mu) \, \psi_4(\omega_1, \omega_2)}
{\bar n \cdot p^{\prime} - (\omega_1+\omega_2) + i 0} \,
\int \frac{d^D \, l}{(2 \pi)^D} \,   \frac{1}{ [(p^{\prime}-k+l)^2+i 0] [(m_b v +l)^2-m_b^2+ i0] [l^2+i0]} \nonumber \\
&&  \hspace{0.3 cm} \times \, (1, \gamma_5) \,  \, \frac{\not \! \bar n} {2} \,
\bigg \{ \gamma_{\perp \mu} \, \bigg [ n \cdot l \left ( (D-2) \, \bar n \cdot l + 2\, m_b \right )
+ 2\, n \cdot p^{\prime} \, (\bar n \cdot l + m_b) + (D-4) \, l_{\perp}^2 \bigg ] \nonumber \\
&& \hspace{0.3 cm}  + n_{\mu} \,  \bigg [ (2-D) \, (\bar n \cdot l)^2  \bigg ]
+ \, \bar n_{\mu} \,\,  \bigg [2\, m_b \, (n \cdot p^{\prime} + n \cdot l) + (D-2) \, l_{\perp}^2 \bigg ] \bigg \} \, \Lambda_b(v) \,,
\end{eqnarray}
where the superscript ``par" of the partonic correlation functions $\Pi^{a, h}_{\mu, V(A)}$
is suppressed from now on  and we have introduced
\begin{eqnarray}
l_{\perp}^2 \equiv g_{\perp}^{\mu \nu} \, l_{\mu}  \, l_{\nu}\,,  \qquad
g_{\perp}^{\mu \nu} \equiv g^{\mu \nu}-\frac{n^{\mu} \bar n^{\nu}}{2} -\frac{n^{\nu} \bar n^{\mu}}{2} \,.
\end{eqnarray}
Evaluating the loop integrals with the formulae collected in Appendix A of \cite{Wang:2015vgv} yields
\begin{eqnarray}
\Pi^{a, h}_{\mu, V(A)}(p, q)
&=& \frac{\alpha_s \, C_F}{4 \, \pi} \,
\frac{f_{\Lambda_b}^{(2)}(\mu) \, \psi_4(\omega_1, \omega_2)} {\bar n \cdot p^{\prime} - (\omega_1+\omega_2) + i 0}  \,
\, (1, \gamma_5) \,  \, \frac{\not \! \bar n} {2} \,\,  \nonumber \\
&& \cdot \left [ \gamma_{\perp \mu}  \,\, C_{h \,, \perp}^{(a)}(n \cdot p^{\prime})
 + n_{\mu}  \,\, C_{h \,, n}^{(a)}(n \cdot p^{\prime})
+ \bar n_{\mu}  \,\, C_{h \,, \bar n}^{(a)}(n \cdot p^{\prime})  \right ] \,,
\label{result of full diagram 2a: hard}
\end{eqnarray}
where the coefficient functions read
\begin{eqnarray}
C_{h \,, \perp}^{(a)}(n \cdot p^{\prime}) &=& {1 \over \epsilon^2} +
{1 \over \epsilon} \, \left ( 2 \, \ln {\mu \over  n \cdot p^{\prime}} + 1  \right )
+ 2 \, \ln^2 {\mu \over  n \cdot p^{\prime}} + 2\, \ln {\mu \over m_b}
-2 \,  {\rm Li_2} \left (1- {1 \over r} \right ) \nonumber \\
&& - \ln^2 r + \frac{3 r-2}{1-r} \, \ln r
+\frac{\pi^2}{12} + 4 \,, \\
C_{h \,, n}^{(a)}(n \cdot p^{\prime}) &=& \frac{1}{r-1}
\left ( 1 + {r \over 1-r} \, \ln r \right ) \,, \\
C_{h \,, \bar n}^{(a)}(n \cdot p^{\prime}) &=& {1 \over \epsilon^2} +
{1 \over \epsilon} \, \left ( 2 \, \ln {\mu \over  n \cdot p^{\prime}} + 1  \right )
+ 2 \, \ln^2 {\mu \over  n \cdot p^{\prime}} + 2\, \ln {\mu \over m_b}
-2 \,  {\rm Li_2} \left (1- {1 \over r} \right ) \nonumber \\
&& - \ln^2 r + \frac{2-r}{r-1} \, \ln r
+\frac{\pi^2}{12} + 3 \,,
\end{eqnarray}
with $r=n \cdot p^{\prime}/m_b$.

By proceeding in a similar way, we can extract the hard-collinear contribution
from Fig.  \ref{fig: one-loop_correlator}(a) as follows
\begin{eqnarray}
\Pi^{a, hc}_{\mu, V(A)}(p, q)  &=& i \, g_s^2\, C_F \, \frac{f_{\Lambda_b}^{(2)}(\mu) \, \psi_4(\omega_1, \omega_2)}
{\bar n \cdot p^{\prime} - (\omega_1+\omega_2) + i 0} \, \, (1, \gamma_5) \,  \, \frac{\not \! \bar n} {2} \,
\left [ \gamma_{\perp \mu}  +  \bar n_{\mu} \right ] \, \Lambda_b(v) \,\, \nonumber
\\
&& \int \frac{d^D \, l}{(2 \pi)^D} \,   \frac{2 \, m_b \, n \cdot (p^{\prime}+l)}
{[n \cdot (p^{\prime}+l) \, \bar n \cdot (p^{\prime}-k+l) + l_{\perp}^2  + i 0]
[ m_b \, n \cdot l+ i 0] [l^2+i0]} \,,
\end{eqnarray}
where the loop integrals are identical to the corresponding case in the vacuum-to-$B$-meson correlation function \cite{Wang:2015vgv}.
We then write
\begin{eqnarray}
\Pi^{a, hc}_{\mu, V(A)}(p, q)  &=& - \frac{\alpha_s \, C_F}{4 \, \pi} \,
\frac{f_{\Lambda_b}^{(2)}(\mu) \, \psi_4(\omega_1, \omega_2)} {\bar n \cdot p^{\prime} - (\omega_1+\omega_2) + i 0}  \,
\, (1, \gamma_5) \,  \, \frac{\not \! \bar n} {2} \,\,
\left [ \gamma_{\perp \mu}  +  \bar n_{\mu} \right ] \, \Lambda_b(v) \,\,  \nonumber \\
&&  \times \bigg [ {2 \over \epsilon^2} + {2 \over \epsilon} \,
\left (\ln {\mu^2 \over  n \cdot p^{\prime} \, (\omega - \bar n \cdot p^{\prime})} + 1  \right )
+ \ln^2 {\mu^2 \over  n \cdot p^{\prime} \, (\omega - \bar n \cdot p^{\prime})} \nonumber \\
&& \hspace{0.5 cm}  + 2 \, \ln {\mu^2 \over  n \cdot p^{\prime} \, (\omega - \bar n \cdot p^{\prime})}  %\nonumber \\
-{\pi^2 \over 6} + 4 \bigg ] \,,
\label{result of full diagram 2a: hard-collinear}
\end{eqnarray}
with $\omega=\omega_1+\omega_2$.

To facilitate the determination of the jet function for the (pseudo)-tensor correlation functions
$\Pi_{\mu, T(\tilde{T})}(p, q)$,  we can just expand Eq. (\ref{expression of fig. 2a}) in the hard-collinear region
without employing the light-cone projector in momentum space
\begin{eqnarray}
\Pi^{a, hc}_{\mu, V(A)}(p, q) &=&
\frac{i \, g_s^2\, C_F}{\bar n \cdot p^{\prime} - (\omega_1+\omega_2) + i 0} \,
\int \frac{d^D \, l}{(2 \pi)^D} \,  \nonumber \\
&&  \frac{2 \, m_b \, n \cdot (p^{\prime}+l)}
{ [n \cdot (p^{\prime}+l) \, \bar n \cdot (p^{\prime}-k+l) + l_{\perp}^2  + i 0]
[ m_b \, n \cdot l+ i 0] [l^2+i0]} \nonumber \\
&& \epsilon_{i j k} \, \left [u^{\rm T}_{i} (k_1) \, C \, \gamma_5 \, \not \! n \, d_{j} (k_2) \right ]
\, \frac{\not \! \bar n} {2} \, \gamma_{\mu} \, (1, \gamma_5)  \, b_{k}(v) \,,
\label{eikonal expression of fig. 2a}
\end{eqnarray}
where no information of the weak vertex is used for  reduction of the Dirac algebra.
It is then evident that the hard-collinear contribution from Fig.  \ref{fig: one-loop_correlator}(a)
is {\it independent} of Lorenz structure of the weak vertex, at leading power in $\Lambda/m_b$.

\subsubsection{$\Lambda$-baryon vertex diagrams}

The one-loop contributions to $\Pi_{\mu, V(A)}(p, q)$ from  the $\Lambda$-baryon vertex diagrams shown
in Fig. \ref{fig: one-loop_correlator}(b) and \ref{fig: one-loop_correlator}(c) are given by
\begin{eqnarray}
\Pi^{b}_{\mu, V(A)}(p, q) &=&
- \frac{i}{2} \, g_s^2 \, \left( 1 + {1 \over N_c} \right) \,
\frac{1}{n \cdot p^{\prime} \, [\bar n \cdot p^{\prime} - (\omega_1+\omega_2) + i 0]} \, \nonumber
\\
&& \times \int \frac{d^D \, l}{(2 \pi)^D} \,   \frac{1}{ [(p^{\prime}-k_2-l)^2+i 0] [(l-k_1)^2+ i0] [l^2+i0]}  \nonumber
\\
&& \times  \epsilon_{i j k} \, \left [u^{\rm T}_{i} (k_1) \, C \, \gamma_{\rho} \, \!
\not l \,  \gamma_5 \, \not \! n \, d_{j} (k_2) \right ]
\,  (\not \!  p^{\prime}-\not \! k_2 - \! \not  l ) \, \gamma^{\rho}  \, (\not \!  p^{\prime}-\not \! k_1 - \not \! k_2 ) \,
 \gamma_{\mu} \, (1, \gamma_5) \, b_{k}(v) \,,   \nonumber
 \\
 \label{expression of fig. 2b}
 \\
\Pi^{c}_{\mu, V(A)}(p, q) &=& \Pi^{b}_{\mu, V(A)}(p, q) \left [ k_1 \leftrightarrow k_2 \right ] \,,
\label{expression of fig. 2c}
\end{eqnarray}
where the isospin symmetry has been employed to derive the second equation.
As already discussed in \cite{Wang:2015vgv} it is more transparent to compute the loop integrals
in Eq. (\ref{expression of fig. 2b}) exactly  instead of applying the method of regions, then keeping only
the leading power terms in the resulting partonic amplitude and inserting the light-cone projector of the $\Lambda_b$-baryon.
The three-point integral
\begin{eqnarray}
\frac{(4 \, \pi)^2}{i} \, \, \int \frac{d^D \, l}{(2 \pi)^D} \,  \,   \frac{l_{\alpha} \, (p^{\prime}-k_2-l)_{\beta} }
{ [(p^{\prime}-k_2-l)^2+i 0] [(l-k_1)^2+ i0] [l^2+i0]}
\end{eqnarray}
can be deduced from Eq. (120) of \cite{Wang:2015vgv} with the following replacement rules
\begin{eqnarray}
p \to p^{\prime} - k_2, \qquad k \to k_1 \,.
\end{eqnarray}
Based upon the argument from the power counting analysis, the leading power contribution to $\Pi^{b}_{\mu, V(A)}(p, q)$
can only arise from  the hard-collinear region and the resulting contribution to the jet function is found to be
\begin{eqnarray}
\Pi^{b, hc}_{\mu, V(A)}(p, q)  &=& - \frac{\alpha_s}{4 \, \pi} \,  \left( 1 + {1 \over N_c} \right) \,
\frac{f_{\Lambda_b}^{(2)}(\mu) \, \psi_4(\omega_1, \omega_2)} {\bar n \cdot p^{\prime} - (\omega_1+\omega_2) + i 0}  \,
\, (1, \gamma_5) \,  \, \frac{\not \! \bar n} {2} \,\,
\left [ \gamma_{\perp \mu}  +  \bar n_{\mu} \right ] \, \Lambda_b(v) \,\,  \nonumber \\
&&  \times \bigg \{  \left [ { 1 + \eta_2 \over \eta_1} \, \ln {1+\eta_{12} \over 1+ \eta_2} - {3 \over 4} \right ]
\bigg [ {1 \over \epsilon}  + \ln {\mu^2 \over  n \cdot p^{\prime} (\omega_2 -\bar n \cdot p^{\prime})}
-{1 \over 2} \,  \ln {1+\eta_{12} \over 1+ \eta_2} \,  \nonumber \\
&& \hspace{0.8 cm} + {5 \over 8} \, {\eta_1 \over 1+ \eta_2 }  + 2 \bigg ]
+ {15 \over 32} {\eta_1 \over 1 + \eta_2} - {1 \over 4} \bigg \}  \,,
\label{result of full diagram 2b}
\end{eqnarray}
where we have defined
\begin{eqnarray}
\eta_i=-\omega_i/\bar n \cdot p^{\prime} \,\, (i=1,2) \,,
\qquad \eta_{12}=\eta_1+\eta_2 \,,
\end{eqnarray}
and the first relation in Eq. (\ref{EOM for Lambdab DA})  due to the equations of motion have been implemented.

\subsubsection{Wave function renormalization}

The hard-collinear contribution from the self-energy correction to the intermediate quark propagator
in Fig. \ref{fig: one-loop_correlator}(d) is independent of the Dirac structures of
the weak transition current and the baryonic interpolating current.
It is straightforward to write
\begin{eqnarray}
\Pi^{d, hc}_{\mu, V(A)}(p, q)  &=& \frac{\alpha_s \, C_F}{4 \, \pi}  \,
\frac{f_{\Lambda_b}^{(2)}(\mu) \, \psi_4(\omega_1, \omega_2)} {\bar n \cdot p^{\prime} - (\omega_1+\omega_2) + i 0}  \,
\, (1, \gamma_5) \,  \, \frac{\not \! \bar n} {2} \,\,
\left [ \gamma_{\perp \mu}  +  \bar n_{\mu} \right ] \, \Lambda_b(v) \,\,  \nonumber \\
&&  \times \left [  {1 \over \epsilon}
+ \ln {\mu^2 \over  n \cdot p^{\prime} (\omega -\bar n \cdot p^{\prime})} + 1  \right ]  \,.
\end{eqnarray}
The contributions of the wave function renormalization to the external quark fields
can be taken from  \cite{Wang:2015vgv}
\begin{eqnarray}
\Pi^{bwf, (1)}_{\mu, V(A)} -  \Phi_{bud, bwf}^{(1)} \otimes T^{(0)}
 &=& \frac{\alpha_s \, C_F}{8 \, \pi}  \,
\frac{f_{\Lambda_b}^{(2)}(\mu) \, \psi_4(\omega_1, \omega_2)} {\bar n \cdot p^{\prime} - (\omega_1+\omega_2) + i 0}  \,
\, (1, \gamma_5) \,  \, \frac{\not \! \bar n} {2} \,\,
\left [ \gamma_{\perp \mu}  +  \bar n_{\mu} \right ] \, \Lambda_b(v) \,\,  \nonumber \\
&&  \times \left [  {3 \over \epsilon}
+ 3 \, \ln {\mu^2 \over  m_b^2} + 4  \right ]  \,,  \\
\Pi^{uwf, (1)}_{\mu, V(A)} -  \Phi_{bud, uwf}^{(1)} \otimes T^{(0)}
 &=& \Pi^{dwf, (1)}_{\mu, V(A)} -  \Phi_{bud, dwf}^{(1)} \otimes T^{(0)} =0 \,,
\end{eqnarray}
where $\Pi^{qwf, (1)}_{\mu, V(A)}$ ($q=b \,, u \,, d$) stands for the contribution to  $\Pi_{\mu, V(A)}$
from the wave function renormalization of the $q$-quark field at one loop, and $\Phi_{bud, qwf}^{(1)}$
denotes the one-loop contribution to $\Phi_{bud}$ defined in Eq. (\ref{def: partonic DA})
from field renormalization of the $q$-quark.

\subsubsection{Box diagrams}

We proceed to compute the one-loop contributions from the two box diagrams displayed in
Fig. \ref{fig: one-loop_correlator}(e) and \ref{fig: one-loop_correlator}(f).
We can readily write
\begin{eqnarray}
&& \Pi^{e}_{\mu, V(A)}(p, q) \nonumber \\
&& = - \frac{i}{2} \, \, g_s^2 \,  \left ( 1+ {1 \over N_c} \right )
\int \frac{d^D \, l}{(2 \pi)^D} \,
\frac{1}{ [(p^{\prime}-k+l)^2+i 0] [(m_b v +l)^2-m_b^2+ i0] [(l-k_1)^2 + i 0] [l^2+i0]}  \nonumber \\
&& \hspace{0.8 cm}  \epsilon_{i j k} \, \left [u^{\rm T}_{i} (k_1) \, C \, \gamma_{\rho} \,
(\! \not  k_1 -  \! \not l )\, \gamma_5 \, \not \! n \, d_{j} (k_2) \right ]
\, (\! \not   p^{\prime}- \!  \not k + \! \not  l ) \,
\gamma_{\mu} \, (1, \gamma_5) \, (m_b \not \! v +  \! \not l + m_b) \, \gamma^{\rho} \, b_{k}(v) \,.
\label{expression of fig. 2e}
\end{eqnarray}
With the isospin symmetry of exchanging the up and down quark fields  we can again find
\begin{eqnarray}
\Pi^{f}_{\mu, V(A)}(p, q) = \Pi^{e}_{\mu, V(A)}(p, q) \left [ k_1 \leftrightarrow k_2 \right ] \,.
\label{expression of fig. 2f}
\end{eqnarray}
It is evident that no hard contribution can arise from the box diagrams and the contribution
to the jet function from Fig. \ref{fig: one-loop_correlator}(e) can be determined by expanding Eq. (\ref{expression of fig. 2e})
in the hard collinear region systematically
\begin{eqnarray}
\Pi^{e, hc}_{\mu, V(A)}(p, q) &=&  i \, \, g_s^2 \,  \left ( 1+ {1 \over N_c} \right )
\int \frac{d^D \, l}{(2 \pi)^D} \, \nonumber \\
&& \frac{n \cdot ( p^{\prime} + l)}{ [n \cdot (p^{\prime}+l) \, \bar n \cdot (p^{\prime}-k+l) + l_{\perp}^2  + i 0]
[n \cdot l \, \bar n \cdot (l-k_1) + l_{\perp}^2 + i 0] [l^2+i0]} \nonumber \\
&& \epsilon_{i j k} \, \left [u^{\rm T}_{i} (k_1) \, C \, \gamma_5 \, \not \! n \, d_{j} (k_2) \right ]
\, \frac{\not \! \bar n} {2} \, \gamma_{\mu} \, (1, \gamma_5)  \, b_{k}(v)  \,.
\label{expanded expression of fig. 2e}
\end{eqnarray}
We therefore conclude that the hard-collinear contribution induced by Fig. \ref{fig: one-loop_correlator}(e)
is {\it independent} of the spin structure of the weak current, given the fact that only the Taylor expansion
of the integrand in Eq. (\ref{expression of fig. 2e}) at leading power in $\Lambda/m_b$ and the equation of motion for the effective $b$-quark
are needed in obtaining Eq. (\ref{expanded expression of fig. 2e}).

The loop integral entering the hard collinear contribution of Fig. \ref{fig: one-loop_correlator}(e)
can be deduced from Eq. (128) of \cite{Wang:2015vgv} with the substitution rules
\begin{eqnarray}
n \cdot p \to n \cdot p^{\prime} \,, \qquad
\bar n \cdot p \to  \bar n \cdot \left ( p^{\prime} - k_2  \right ) \,, \qquad
\bar n \cdot k \to \bar n \cdot k_1.
\end{eqnarray}
Applying the momentum-space projector of the $\Lambda_b$-baryon we find
\begin{eqnarray}
\Pi^{e, hc}_{\mu, V(A)}(p, q)  &=& \frac{\alpha_s}{4 \, \pi} \,  \left( 1 + {1 \over N_c} \right) \,
\frac{f_{\Lambda_b}^{(2)}(\mu) \, \psi_4(\omega_1, \omega_2)} {\bar n \cdot p^{\prime} - (\omega_1+\omega_2) + i 0}  \,
\, (1, \gamma_5) \,  \, \frac{\not \! \bar n} {2} \,\,
\left [ \gamma_{\perp \mu}  +  \bar n_{\mu} \right ] \, \Lambda_b(v) \,\,  \nonumber \\
&&  \times \left [ { 1 + \eta_{12} \over \eta_1} \, \ln {1+\eta_{12} \over 1+ \eta_2} \right ]
\bigg [ {1 \over \epsilon}  + \ln {\mu^2 \over  n \cdot p^{\prime} (\omega -\bar n \cdot p^{\prime})}
+ {1 \over 2} \,  \ln {1+\eta_{12} \over 1+ \eta_2} + 1  \bigg ]  \,.
\label{result of full diagram 2e}
\end{eqnarray}

\subsubsection{The NLO hard-scattering kernels}

Now we are ready to determine the one-loop hard kernels entering  QCD factorization formulae
of the correlation functions $\Pi^{\rm par}_{\mu, V(A)}(p, q) $ by collecting different pieces together
\begin{eqnarray}
\Phi_{bud}^{(0)} \otimes T^{(1)}_{V(A)} &=&
\bigg [ \Pi^{a, h}_{\mu, V(A)} +  \left ( \Pi^{bwf, (1)}_{\mu, V(A)}
-  \Phi_{bud, bwf}^{(1)} \otimes T^{(0)} \right ) \bigg ] \nonumber \\
&&  + \bigg [ \Pi^{a, hc}_{\mu, V(A)} + \Pi^{b, hc}_{\mu, V(A)}
+ \Pi^{c, hc}_{\mu, V(A)}+  \Pi^{d, hc}_{\mu, V(A)} + \Pi^{e, hc}_{\mu, V(A)}+  \Pi^{f, hc}_{\mu, V(A)} \bigg ] \,,
\end{eqnarray}
where the terms in the first and second square brackets correspond to the hard and jet functions
at ${\cal O}(\alpha_s)$, respectively. Introducing the definition
\begin{eqnarray}
\Pi_{\mu, V(A)} &=& \left (1, \gamma_5 \right ) \, \frac{\not \! \bar n}{2} \,
\left [ \Pi_{\perp, V(A)} \, \gamma_{\perp \mu} + \Pi_{\bar n, V(A)} \, \bar n_{\mu}
 + \Pi_{n, V(A)} \, n_{\mu}  \right ] \,\, \Lambda_b(v)  \,,
\end{eqnarray}
we can readily obtain the following factorization formulae for the vacuum-to-$\Lambda_b$-baryon
correlation functions at NLO
\begin{eqnarray}
\Pi_{\perp, V(A)} &=& f_{\Lambda_b}^{(2)}(\mu) \, C_{\perp, V(A)}(n \cdot p^{\prime}, \mu) \,
\int_0^{\infty} d \omega_1 \,  \int_0^{\infty} d \omega_2 \,
\frac{1}{\omega_1+\omega_2-\bar n \cdot p^{\prime} - i 0} \, \nonumber \\
&&  J\left ( {\mu^2  \over \bar n \cdot p^{\prime} \, \omega_i},
{\omega_i \over \bar n \cdot p^{\prime}} \right ) \, \psi_4(\omega_1,\omega_2,\mu) \,,
\\
\Pi_{\bar n, V(A)} &=& f_{\Lambda_b}^{(2)}(\mu) \, C_{\bar n, V(A)}(n \cdot p^{\prime}, \mu) \,
\int_0^{\infty} d \omega_1 \,  \int_0^{\infty} d \omega_2 \,
\frac{1}{\omega_1+\omega_2-\bar n \cdot p^{\prime} - i 0} \, \nonumber \\
&&  J\left ( {\mu^2  \over \bar n \cdot p^{\prime} \, \omega_i},
{\omega_i \over \bar n \cdot p^{\prime}} \right ) \, \psi_4(\omega_1,\omega_2,\mu) \,,
\\
\Pi_{n, V(A)} &=& f_{\Lambda_b}^{(2)}(\mu) \, C_{n, V(A)}(n \cdot p^{\prime}, \mu) \,
\int_0^{\infty} d \omega_1 \,  \int_0^{\infty} d \omega_2 \,
\frac{\psi_4(\omega_1,\omega_2,\mu)}{\omega_1+\omega_2-\bar n \cdot p^{\prime} - i 0} \,,
\label{NLO Corr n V-A}
\end{eqnarray}
where the renormalized hard coefficients are given by
\begin{eqnarray}
C_{\perp, V(A)}(n \cdot p^{\prime}, \mu) &=& 1 - \frac{\alpha_s(\mu) \, C_F}{4 \, \pi}
\bigg [ 2 \, \ln^2 {\mu \over  n \cdot p^{\prime}} + 5 \, \ln {\mu \over m_b}
-2 \,  {\rm Li_2} \left (1- {1 \over r} \right ) \nonumber \\
&& - \ln^2 r + \frac{3 r-2}{1-r} \, \ln r
+\frac{\pi^2}{12} + 6 \bigg ] \,,
\\
C_{\bar n, V(A)}(n \cdot p^{\prime}, \mu) &=& 1 - \frac{\alpha_s(\mu) \, C_F}{4 \, \pi}
\bigg [  2 \, \ln^2 {\mu \over  n \cdot p^{\prime}} + 5 \, \ln {\mu \over m_b}
-2 \,  {\rm Li_2} \left (1- {1 \over r} \right ) \nonumber \\
&& - \ln^2 r + \frac{2-r}{r-1} \, \ln r +\frac{\pi^2}{12} + 5  \bigg ] \,,
\\
C_{n, V(A)}(n \cdot p^{\prime}, \mu) &=& - \frac{\alpha_s(\mu) \, C_F}{4 \, \pi}
\bigg [ \frac{1}{r-1} \left ( 1 + {r \over 1-r} \, \ln r \right ) \bigg ]  \,,
\end{eqnarray}
and the renormalized jet function reads
\begin{eqnarray}
&& J\left ( {\mu^2  \over \bar n \cdot p^{\prime} \, \omega_i},
{\omega_i \over \bar n \cdot p^{\prime}} \right ) \nonumber \\
&& =  1 + {\alpha_s(\mu) \over 4 \pi }\, {4 \over 3} \,
\bigg \{\ln^2 {\mu^2 \over n \cdot p^{\prime} \, (\omega- \bar n \cdot p^{\prime})}
-2 \, \ln {\omega -\bar n \cdot p^{\prime} \over \omega_2 -\bar n \cdot p^{\prime}} \,
\ln {\mu^2 \over n \cdot p^{\prime} \, (\omega- \bar n \cdot p^{\prime})}  \nonumber \\
&& \hspace{0.5 cm} -{1 \over 2} \, \ln {\mu^2 \over n \cdot p^{\prime} \, (\omega- \bar n \cdot p^{\prime})}
-  \ln^2 {\omega -\bar n \cdot p^{\prime} \over \omega_2 -\bar n \cdot p^{\prime}} \,
+ 2 \, \ln  {\omega -\bar n \cdot p^{\prime} \over \omega_2 -\bar n \cdot p^{\prime}} \,
\left [ {\omega_2 - \bar n \cdot p^{\prime} \over \omega_1}-{3 \over 4}  \right ] \, \nonumber \\
&&  \hspace{0.5 cm}
-{\pi^2 \over 6} - {1 \over 2} \bigg \}  \,.
\label{origianl form of one-loop jet function}
\end{eqnarray}

Several comments on QCD factorization  of the correlation functions  $\Pi_{\mu, V(A)}$ at NLO
are in order.
\begin{itemize}
\item {Since one universal jet function enters  the factorization formulae of the correlation functions
at ${\cal O}(\alpha_s)$ and at leading power in $\Lambda/m_b$,  the symmetry breaking effects of the
form factor relations in Eq. (\ref{symmetry relation of FFs at tree level}) can only arise from the perturbative
fluctuations at $m_b$ scale, as reflected by the distinct hard functions for different weak currents.
To determine the hard collinear contribution to the large-energy symmetry violations, we need to evaluate
a {\it specific} sub-leading power contribution to the correlation functions induced by the  $\Lambda$-baryon current.
Technically,  this can be achieved by introducing the vacuum-to-$\Lambda_b$-baryon correlation functions
with the ``wrong" light-cone projector acting on the $\Lambda$-baryon current as proposed in \cite{Feldmann:2011xf}.
The hard-collinear symmetry breaking effects are shown to be of the same power in $\Lambda/m_b$ as the soft overlap
contributions, despite the fact that they are computed with the sum rules constructed from the power-suppressed  correlation
functions. This is by all means not surprising, because hadronic dispersion relations of the sub-leading correlation functions
also involve  an additional power-suppressed factor $m_{\Lambda}/ n \cdot p^{\prime}$. However, the numerical impacts of such
hard-collinear symmetry violations defined by a hadronic matrix element of the ``B-type" SCET current
turn out to be insignificant from the  same LCSR approach \cite{Feldmann:2011xf},
we will therefore not include it in the following analysis. Also,  evaluating hadronic matrix elements
from the power-suppressed correlation functions are less favored from the standard  philosophy of QCD sum rules,
since the systematic  uncertainty generated by the parton-hadron duality approximation is difficult to be  under control. }
\item {In naive dimension regularization the hard matching coefficients  satisfy the relations $C_{\perp, V}=C_{\perp, A}$,
$C_{\bar n, V}=C_{\bar n, A}$ and $C_{n, V}=C_{n, A}$ to all orders in perturbation theory due to the ${\rm U}(1)$
helicity symmetry for  both massless QCD and SCET Lagrangian functions \cite{Bauer:2000yr}.
It is then evident that the axial-vector $\Lambda_b \to \Lambda$
form factors at large hadronic recoil will be identical to the corresponding vector form factors within our approximations. }
\item {Only the weak vertex diagram and  the two box diagrams could in principle yield hard-collinear contributions sensitive to
the Dirac structure of the weak current, however, such sensitivity is shown to disappear at leading power
in $\Lambda/m_b$ after expanding  the involved loop integrals in the hard-collinear region, as indicated by
Eqs. (\ref{eikonal expression of fig. 2a})  and (\ref{expanded expression of fig. 2e}).
This leads us to  conclude that the hard-collinear contributions to the correlation functions
$\Pi_{\mu, a}(p, q)$ are {\it independent} of the spin structure of the weak transition current,
at leading power in $\Lambda/m_b$.  }
\end{itemize}

We now turn to consider factorization of the (pseudo)-tensor correlation functions
$\Pi_{\mu, T(\tilde{T})}$ at one loop. The hard coefficient functions can be extracted
from the matching calculation of the weak (pseudo)-tensor currents from QCD onto SCET \cite{Beneke:2004rc}
\begin{eqnarray}
&& \left [ \bar q(0) \, (1, \gamma_5) \, i \, \sigma_{\mu \nu} \, b(0) \right ]_{\rm QCD} \nonumber \\
&& \rightarrow  \int d \hat{s} \left [\bar \xi W_{hc}  \right ](s \, n) \, (1, \gamma_5)
\left \{ \, \tilde{C}_{T(\tilde{T})}^{A}(\hat s)\,  \left [ i \, \sigma_{\mu \nu}  \right ]
+ \, \tilde{C}_{T(\tilde{T})}^{B}(\hat s) \, \left [ \bar n_{\mu} \gamma_{\nu} - \bar n_{\nu} \gamma_{\mu} \right ] \right \}
\,[S^{\dag} h](0) + ... \,,  \hspace{0.8  cm}
\end{eqnarray}
where the ellipses stand for the terms absent at ${\cal O}(\alpha_s)$ as well as the
sub-leading power currents, and we have defined the dimensionless convolution variable $\hat s = s \, m_b$.
We have introduced the hard-collinear and the soft Wilson lines
\begin{eqnarray}
W_{hc}(x) &=&  {\rm P} \, \left \{ {\rm  Exp} \left [   i \, g_s \,
\int_{- \infty}^{0} \, dt \,  n  \cdot A_{hc}(x +  t \, n) \right ]  \right \} \,, \nonumber \\
S(x) &=& {\rm P} \, \left \{ {\rm  Exp} \left [   i \, g_s \,
\int_{- \infty}^{0} \, dt \,  \bar n  \cdot A_{s}(x +  t \, \bar n) \right ]  \right \} \,
\end{eqnarray}
to  construct the building blocks invariant under both soft and hard-collinear gauge transformations.
Performing the Fourier transformation from the momentum space to the position space yields
\cite{Bauer:2000yr,Beneke:2004rc}
\begin{eqnarray}
C_{T(\tilde{T})}^{A}(n \cdot p^{\prime}, \mu) &=& 1 - \frac{\alpha_s(\mu) \, C_F}{4 \, \pi}
\bigg [ 2 \, \ln^2 {\mu \over  n \cdot p^{\prime}} + 7  \, \ln {\mu \over m_b}
-\, 2 \,  {\rm Li_2} \left (1- {1 \over r} \right ) -  \ln^2 r  \nonumber \\
&& + \, \frac{4 r-2}{1-r} \, \ln r
+\frac{\pi^2}{12} + 6 \bigg ] \,,
\label{original form of the one-loop  tensor hard function A}
\\
C_{T(\tilde{T})}^{B}(n \cdot p^{\prime}, \mu) &=&  \frac{\alpha_s(\mu) \, C_F}{4 \, \pi} \,
\left [ {2 \, r \over 1-r} \, \ln r \right ] \,.
\end{eqnarray}

Decomposing the correlation functions $\Pi_{\mu, T(\tilde{T})}$ in terms of  Lorenz invariant amplitudes
\begin{eqnarray}
\Pi_{\mu, T(\tilde{T})}= \left ( -1, \gamma_5 \right ) \, \frac{\not \! \bar n}{2} \,
\left [ \Pi_{\perp, T(\tilde{T})} \,\, \gamma_{\perp \mu} +   \frac{\Pi_{+, T(\tilde{T})}}{2} \,\,
\left ( \left (1- \frac{n \cdot p^{\prime}}{m_{\Lambda_b}} \right )  \, \bar n_{\mu} - n_{\mu} \right ) \right ]
 \,\, \Lambda_b(v) \,\,,
\end{eqnarray}
it is straightforward to derive the factorization formulae
\begin{eqnarray}
\Pi_{\perp, T(\tilde{T})} &=& m_{\Lambda_b} \,\, f_{\Lambda_b}^{(2)}(\mu) \,\,
 C_{T(\tilde{T})}^{A}(n \cdot p^{\prime}, \mu) \,
\int_0^{\infty} d \omega_1 \,  \int_0^{\infty} d \omega_2 \,
\frac{1}{\omega_1+\omega_2-\bar n \cdot p^{\prime} - i 0} \, \nonumber \\
&&  J\left ( {\mu^2  \over \bar n \cdot p^{\prime} \, \omega_i},
{\omega_i \over \bar n \cdot p^{\prime}} \right ) \, \psi_4(\omega_1,\omega_2,\mu) \,\, \nonumber \\
&& + \,\, m_{\Lambda_b} \,\, f_{\Lambda_b}^{(2)}(\mu) \,\,
 C_{T(\tilde{T})}^{B}(n \cdot p^{\prime}, \mu) \,
\int_0^{\infty} d \omega_1 \,  \int_0^{\infty} d \omega_2 \,
\frac{\psi_4(\omega_1,\omega_2,\mu) }{\omega_1+\omega_2-\bar n \cdot p^{\prime} - i 0} \,\,,
\\
\Pi_{+, T(\tilde{T})} &=& m_{\Lambda_b} \,\, f_{\Lambda_b}^{(2)}(\mu) \,\,
 C_{T(\tilde{T})}^{A}(n \cdot p^{\prime}, \mu) \,
\int_0^{\infty} d \omega_1 \,  \int_0^{\infty} d \omega_2 \,
\frac{1}{\omega_1+\omega_2-\bar n \cdot p^{\prime} - i 0} \, \nonumber \\
&&  J\left ( {\mu^2  \over \bar n \cdot p^{\prime} \, \omega_i},
{\omega_i \over \bar n \cdot p^{\prime}} \right ) \, \psi_4(\omega_1,\omega_2,\mu) \,\,.
\label{NLO Corr plus T-Ttidle}
\end{eqnarray}

\subsection{Factorization-scale independence}

We are now in a position to verify the factorization-scale independence of the
correlation functions $\Pi_{\mu, a}(p,q)$ explicitly at one loop.
Having the one-loop factorization formulae at hand we can readily write
\begin{eqnarray}
&& \frac{d}{ d \, \ln \mu} \, \Pi_{\perp, V(A)} =  \frac{d}{ d \, \ln \mu} \, \Pi_{\bar n, V(A)}
\nonumber  \\
&& = {\alpha_s(\mu) \, \over 4 \, \pi} \, {4 \over 3} \,
\int_0^{\infty} d \omega_1 \,  \int_0^{\infty} d \omega_2 \,
\frac{1}{\omega_1+\omega_2-\bar n \cdot p^{\prime} - i 0} \, \nonumber \\
&&  \hspace{0.5 cm} \times \left [ 4 \, \ln {\mu \over \omega - \bar n \cdot p^{\prime}}
- 4 \,  \ln {\omega - \bar n \cdot p^{\prime} \over \omega_2 - \bar n \cdot p^{\prime}} - 6  \right ]
 \left [ f_{\Lambda_b}^{(2)}(\mu) \,  \psi_4(\omega_1,\omega_2,\mu)  \right ]  \nonumber \\
&& \hspace{0.5 cm} +  \int_0^{\infty} d \omega_1 \,  \int_0^{\infty} d \omega_2 \,
\frac{1}{\omega_1+\omega_2-\bar n \cdot p^{\prime} - i 0} \,\,
{d \over d \ln \mu} \, \left [ f_{\Lambda_b}^{(2)}(\mu) \,  \psi_4(\omega_1,\omega_2,\mu)  \right ]
+ {\cal O}(\alpha_s^2) \,\,,
\\
&& \frac{d}{ d \, \ln \mu} \, \Pi_{\perp, T(\tilde{T})} = \frac{d}{ d \, \ln \mu} \, \Pi_{+, T(\tilde{T})}  \nonumber  \\
&& = \frac{d}{ d \, \ln \mu} \, \Pi_{\perp, V(A)}
- {\alpha_s(\mu) \, \over 4 \, \pi} \, {8 \over 3} \,\, f_{\Lambda_b}^{(2)}(\mu) \,
\int_0^{\infty} d \omega_1 \,  \int_0^{\infty} d \omega_2 \,
\frac{  \psi_4(\omega_1,\omega_2,\mu) }{\omega_1+\omega_2-\bar n \cdot p^{\prime} - i 0}
\,\,,
\label{RGE of tensor correlation function}\\
&& \frac{d}{ d \, \ln \mu} \, \Pi_{n, V(A)} = {\cal O}(\alpha_s^2) \,\,,
\end{eqnarray}
where the second term in the evolution equation  (\ref{RGE of tensor correlation function})
is due to renormalization of the (pseudo)-tensor currents in QCD, since we do not distinguish
the factorization and the renormalization scales in dimensional regularization.

%%%%%%%%%%%
\begin{figure}
\begin{center}
\includegraphics[width=0.90 \columnwidth]{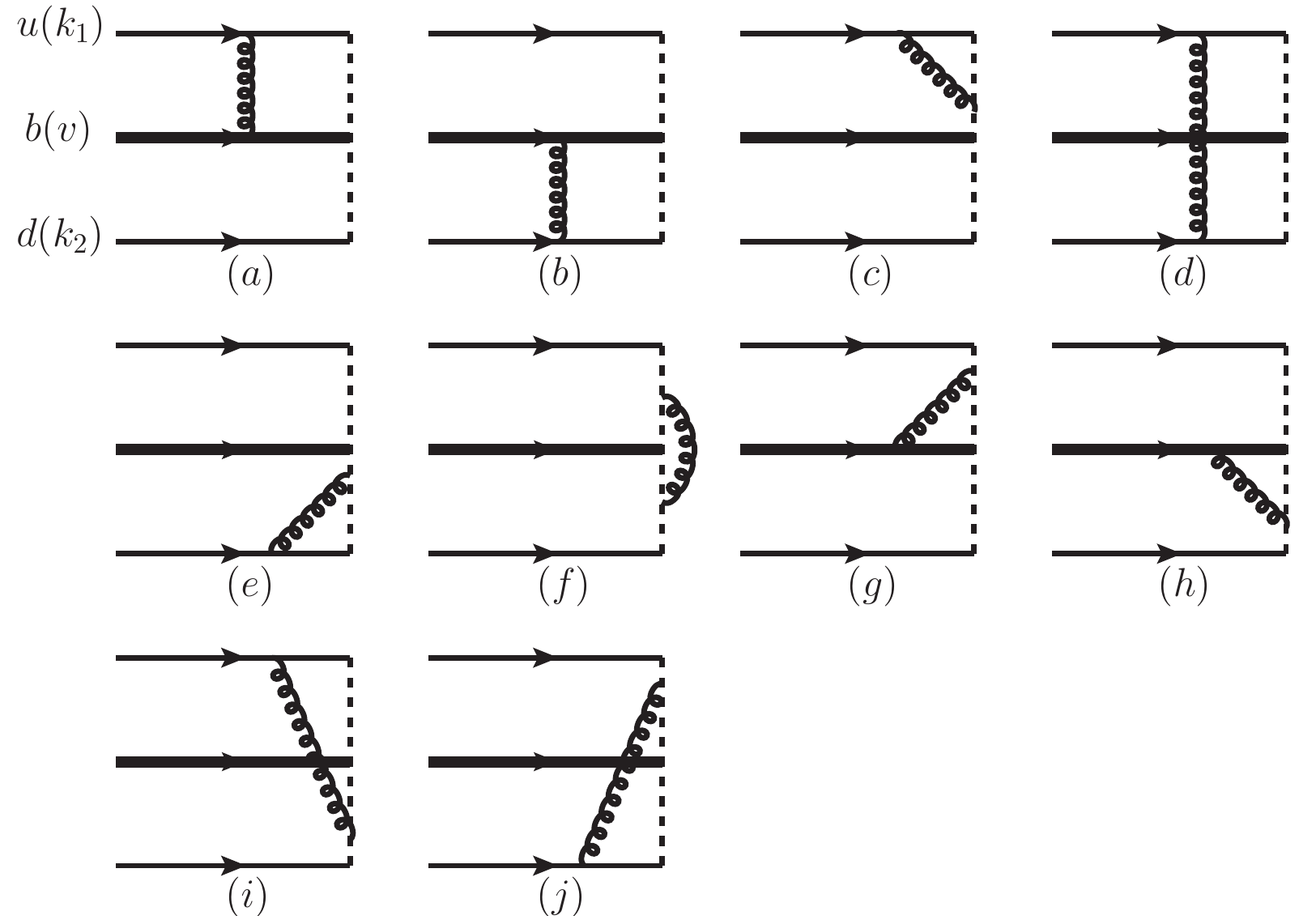}
\vspace*{0.1cm}
\caption{Radiative correction to the $\Lambda_b$-baryon DA $\psi_4(\omega_1,\omega_2,\mu)$ at one loop. }
\label{fig: one-loop_DA}
\end{center}
\end{figure}
%%%%%%%%%%%

At present the one-loop evolution equation of the $\Lambda_b$-baryon DA $\psi_4(\omega_1,\omega_2,\mu)$
is not explicitly known in the literature, we will compute the factorization-scale dependence of
the convolution integral
\begin{eqnarray}
\int_0^{\infty} d \omega_1 \,  \int_0^{\infty} d \omega_2 \,
\frac{1}{\omega_1+\omega_2-\bar n \cdot p^{\prime} - i 0} \,\,
{d \over d \ln \mu} \, \left [ f_{\Lambda_b}^{(2)}(\mu) \,  \psi_4(\omega_1,\omega_2,\mu)  \right ]\,
\end{eqnarray}
at one loop  in detail. This amounts to extract the UV divergence of the amplitude of the 10 diagrams
displayed in Fig. \ref{fig: one-loop_DA}.

Applying the Wilson-line Feynman rules we can compute the effective diagram displayed in
Fig. \ref{fig: one-loop_DA}(a) as
\begin{eqnarray}
&& \Phi_{bud,a}^{(1)} \otimes T^{(0)} \nonumber \\
&& = - {i \over 2} \, g_s^2 \, \left (1+ {1 \over N_c}  \right ) \,
\int \frac{d^D \, l}{(2 \pi)^D} \, \frac{1}{[(k_1+l)^2+i0] [\bar n \cdot (p^{\prime} - k -l) + i0]
[-v \cdot l + i0] [l^2 + i 0]} \, \nonumber \\
&& \hspace{0.5 cm} \epsilon_{i j k} \, \left [u^{\rm T}_{i} (k_1) \, C \, \! \not v  \, (\! \not k_1 + \! \not l)
\, \gamma_5 \, \not \! n \, d_{j} (k_2) \right ]
\, (1, \gamma_5)  \, \frac{\not \! \bar n} {2} \, \gamma_{\mu}
\, \left ( \gamma_{\perp \mu} + \bar n_{\mu} \right ) b_{k}(v) \,.
\end{eqnarray}
Evaluating the loop integral with the standard techniques yields
\begin{eqnarray}
\frac{d}{d \ln \mu} \left [\Phi_{bud,a}^{(1)} \otimes T^{(0)} \right ]
= - {\alpha_s(\mu) \over 2 \pi} \, \left (1+ {1 \over N_c}  \right ) \,
\frac{\bar n \cdot (p^{\prime}-k)}{\bar n \cdot k_1} \,
\ln \frac{\bar n \cdot k-\bar n \cdot p^{\prime}}{\bar n \cdot k_2-\bar n \cdot p^{\prime}} \,\,
\Phi_{bud}^{(0)} \otimes T^{(0)}\,,
\end{eqnarray}
which cancels the factorization-scale dependence of the QCD amplitude from the diagram
\ref{fig: one-loop_correlator}(e) in Eq. (\ref{result of full diagram 2e}) completely.
Based upon the isospin symmetry argument we can readily obtain
\begin{eqnarray}
\frac{d}{d \ln \mu} \left [ \Phi_{bud,b}^{(1)} \otimes T^{(0)} \right ]
= - {\alpha_s(\mu) \over 2 \pi} \, \left (1+ {1 \over N_c}  \right ) \,
\frac{\bar n \cdot (p^{\prime}-k)}{\bar n \cdot k_2} \,
\ln \frac{\bar n \cdot k-\bar n \cdot p^{\prime}}{\bar n \cdot k_1-\bar n \cdot p^{\prime}} \,\,
\Phi_{bud}^{(0)} \otimes T^{(0)}\,.
\end{eqnarray}

Along the same vein, the light-quark-Wilson-line diagram in Fig. \ref{fig: one-loop_DA}(c)
can be computed as
\begin{eqnarray}
\Phi_{bud,c}^{(1)} \otimes T^{(0)}
&=& - \frac{i \, g_s^2 \, C_F}{\bar n \cdot p^{\prime}-\bar n \cdot k + i 0}  \,
\int \frac{d^D \, l}{(2 \pi)^D} \, \frac{1}{[(k_1+l)^2+i0] [\bar n \cdot (p^{\prime} - k -l) + i0]
[l^2 + i 0]} \, \nonumber \\
&& \epsilon_{i j k} \, \left [u^{\rm T}_{i} (k_1) \, C \, \! \not \bar n  \, (\! \not k_1 + \! \not l)
\, \gamma_5 \, \not \! n \, d_{j} (k_2) \right ]
\, (1, \gamma_5)  \, \frac{\not \! \bar n} {2} \,
\, \left ( \gamma_{\perp \mu} + \bar n_{\mu} \right ) b_{k}(v) \,.
\end{eqnarray}
A few comments on evaluating $\Phi_{bud,c}^{(1)} \otimes T^{(0)}$ are in order.
\begin{itemize}
\item {The equation of motion for a soft $u$-quark field $u^{\rm T} (k_1) \, C \, \! \not k_1 =0 $
is needed to reduce the Dirac structure of the light-quark sector
\begin{eqnarray}
\left [u^{\rm T}_{i} (k_1) \, C \, \! \not \bar n  \, (\! \not k_1 + \! \not l)
\, \gamma_5 \, \not \! n \, d_{j} (k_2) \right ]
&=& \left [u^{\rm T}_{i} (k_1) \, C \, \! \not \bar n  \, (\! \not k_{1 \perp} + \! \not l_{\perp})
\, \gamma_5 \, \not \! n \, d_{j} (k_2) \right ]  \nonumber \\
& \propto & 2 \, \bar n \cdot k_1  \,
\left [ u^{\rm T}_{i} (k_1) \, C \, \gamma_5 \, \! \not  n \, d_{j} (k_2) \right ]  \,,
\nonumber
\end{eqnarray}
where the second step should be understood after performing  the integral over the loop momentum $l$.
}
\item {Since the involved loop integral develops both UV and IR singularities, a fictitious gluon mass
$m_g$ will be introduced to regularize the soft  divergence for the sake of separating
IR and UV divergences. }

\item {Employing the Georgi parametrization trick leads to
\begin{eqnarray}
\frac{d}{d \ln \mu} \left [\Phi_{bud,c}^{(1)} \otimes T^{(0)} \right ]
= {\alpha_s(\mu) \, C_F \over \pi} \,
\left [  \frac{\bar n \cdot (k_2 - p^{\prime})}{\bar n \cdot k_1} \,
\ln \frac{\bar n \cdot k_2-\bar n \cdot p^{\prime}}{\bar n \cdot k-\bar n \cdot p^{\prime}} + 1 \right ]\,\,
\Phi_{bud}^{(0)} \otimes T^{(0)}\,, \hspace{0.5 cm}
\end{eqnarray}
which further implies that under the isospin symmetry
\begin{eqnarray}
\frac{d}{d \ln \mu} \left [\Phi_{bud,e}^{(1)} \otimes T^{(0)} \right ]
= {\alpha_s(\mu) \, C_F \over \pi} \,
\left [  \frac{\bar n \cdot (k_1 - p^{\prime})}{\bar n \cdot k_2} \,
\ln \frac{\bar n \cdot k_1-\bar n \cdot p^{\prime}}{\bar n \cdot k-\bar n \cdot p^{\prime}} + 1 \right ]\,\,
\Phi_{bud}^{(0)} \otimes T^{(0)}\,. \hspace{0.5 cm}
\end{eqnarray}
}

\item {Inspecting the amplitudes of the effective diagrams in  Fig. \ref{fig: one-loop_DA}(i)
and \ref{fig: one-loop_DA}(j) yields
\begin{eqnarray}
\Phi_{bud,i}^{(1)} \otimes T^{(0)}&=& - \frac{1}{2 \, C_F} \, \left ( 1+ {1 \over N_c}  \right )
\Phi_{bud,c}^{(1)} \otimes T^{(0)} \,,  \\
\Phi_{bud,j}^{(1)} \otimes T^{(0)}&=& - \frac{1}{2 \, C_F} \, \left ( 1+ {1 \over N_c}  \right )
\Phi_{bud,e}^{(1)} \otimes T^{(0)} \,.
\end{eqnarray}
We then conclude that the single logarithmic terms in the evolution equations of
\begin{eqnarray}
\frac{d}{d \ln \mu} \left [ \left ( \Phi_{bud,c}^{(1)} +  \Phi_{bud,i}^{(1)} \right ) \otimes T^{(0)} \right ] \,,
\qquad
\frac{d}{d \ln \mu} \left [ \left ( \Phi_{bud,e}^{(1)} +  \Phi_{bud,j}^{(1)} \right ) \otimes T^{(0)} \right ]
\end{eqnarray}
cancel against the ones in the QCD amplitudes for the  diagrams \ref{fig: one-loop_correlator}(b)
and \ref{fig: one-loop_correlator}(c) as presented in (\ref{result of full diagram 2b})
and (\ref{expression of fig. 2c}), respectively.
}
\end{itemize}

We proceed to evaluate the contribution from the effective diagram displayed in Fig. \ref{fig: one-loop_DA}(d)
\begin{eqnarray}
&& \Phi_{bud,d}^{(1)} \otimes T^{(0)} \nonumber \\
&& = - {i \over 2} \, \frac{g_s^2}{\bar n \cdot p^{\prime}-\bar n \cdot k + i 0}
\, \left (1+ {1 \over N_c}  \right ) \,
\int \frac{d^D \, l}{(2 \pi)^D} \, \frac{1}{[(k_1+l)^2+i0] [(k_2-l)^2 + i0]
 [l^2 + i 0]} \, \nonumber \\
&& \hspace{0.5 cm} \epsilon_{i j k} \, \left [u^{\rm T}_{i} (k_1) \, C \, \gamma_{\alpha}  \, (\! \not k_1 + \! \not l)
\, \gamma_5 \, \not \! n \, (\! \not k_2 - \! \not l) \, \gamma^{\alpha} \, d_{j} (k_2) \right ]
\, (1, \gamma_5)  \, \frac{\not \! \bar n} {2} \, 
\, \left ( \gamma_{\perp \mu} + \bar n_{\mu} \right ) b_{k}(v) \,.
\end{eqnarray}
The factorization-scale dependence of $\Phi_{bud,d}^{(1)} \otimes T^{(0)}$ can be readily determined as
\begin{eqnarray}
\frac{d}{d \ln \mu} \left [ \Phi_{bud,d}^{(1)}  \otimes T^{(0)} \right ]
={\alpha_s(\mu) \over 4 \pi} \, \left (1+ {1 \over N_c}  \right ) \,
\Phi_{bud}^{(0)} \otimes T^{(0)}\,.
\end{eqnarray}
The self-energy correction to the light-cone Wilson lines shown in Fig. \ref{fig: one-loop_DA}(f)
vanishes in Feynman gauge due to $\bar n^2=0$.

We further turn to compute the contributions from the heavy-quark-Wilson-line diagrams shown
in Fig. \ref{fig: one-loop_DA}(g) and (h)
\begin{eqnarray}
&& \Phi_{bud,g}^{(1)} \otimes T^{(0)} = \Phi_{bud,h}^{(1)} \otimes T^{(0)}  \nonumber \\
&& = {i \over 2} \, \frac{g_s^2}{\bar n \cdot p^{\prime}-\bar n \cdot k + i 0}
\, \left (1+ {1 \over N_c}  \right ) \,
\int \frac{d^D \, l}{(2 \pi)^D} \, \frac{1}{[v \cdot l+i0] [\bar n \cdot (p^{\prime}-k+l) + i0]
 [l^2 + i 0]} \, \nonumber \\
&& \hspace{0.5 cm} \epsilon_{i j k} \, \left [u^{\rm T}_{i} (k_1) \, C \,
\, \gamma_5 \, \not \! n  \, d_{j} (k_2) \right ]
\, (1, \gamma_5)  \, \frac{\not \! \bar n} {2} \, \gamma_{\mu}
\, \left ( \gamma_{\perp \mu} + \bar n_{\mu} \right ) b_{k}(v) \,.
\end{eqnarray}
Evaluating the UV divergent terms of  $\Phi_{bud,g(h)}^{(1)} \otimes T^{(0)}$ explicitly leads to
\begin{eqnarray}
&& \frac{d}{d \ln \mu} \left [ \Phi_{bud,g}^{(1)}  \otimes T^{(0)} \right ]
= \frac{d}{d \ln \mu} \left [ \Phi_{bud,h}^{(1)}  \otimes T^{(0)} \right ] \nonumber \\
&& = - {\alpha_s(\mu) \over 2 \pi} \, \left (1+ {1 \over N_c}  \right ) \,
\ln {\mu \over \bar n \cdot k - \bar n \cdot p^{\prime}} \,
\Phi_{bud}^{(0)} \otimes T^{(0)}\,,
\end{eqnarray}
which gives the desired cusp anomalous dimension to compensate the corresponding
terms in the QCD amplitude of the diagram in Fig. \ref{fig: one-loop_correlator}(a)
as presented in Eqs. (\ref{result of full diagram 2a: hard}) and
(\ref{result of full diagram 2a: hard-collinear}).

Finally, we need to consider the LSZ term due to renormalization of the external light quark fields
in QCD and of the heavy quark in HQET
\begin{eqnarray}
Z_{q} =1 -{\alpha_s(\mu) \, C_F \over 4 \pi } \, {1 \over \epsilon} \,, \qquad
Z_{Q} =1 + {\alpha_s(\mu) \, C_F \over 2 \pi } \, {1 \over \epsilon}  \,,
\end{eqnarray}
which gives rise to
\begin{eqnarray}
 \frac{d}{d \ln \mu} \left [ Z_{q} \, Z_{Q}^{1/2} \, \Phi_{bud}^{(0)} \otimes T^{(0)} \right ]
 ={\cal O}(\alpha_s^2) \,.
\end{eqnarray}

Putting all the pieces together we obtain
\begin{eqnarray}
&& \int_0^{\infty} d \omega_1 \,  \int_0^{\infty} d \omega_2 \,
\frac{1}{\omega_1+\omega_2-\bar n \cdot p^{\prime} - i 0} \,\,
{d \over d \ln \mu} \, \left [ f_{\Lambda_b}^{(2)}(\mu) \,  \psi_4(\omega_1,\omega_2,\mu)  \right ] \nonumber \\
&& = - {\alpha_s(\mu) \, \over 4 \, \pi} \, {4 \over 3} \,
\int_0^{\infty} d \omega_1 \,  \int_0^{\infty} d \omega_2 \,
\frac{1}{\omega_1+\omega_2-\bar n \cdot p^{\prime} - i 0} \, \nonumber \\
&&  \hspace{0.5 cm} \times \left [ 4 \, \ln {\mu \over \omega - \bar n \cdot p^{\prime}}
- 4 \,  \ln {\omega - \bar n \cdot p^{\prime} \over \omega_2 - \bar n \cdot p^{\prime}} - 5  \right ]
 \left [ f_{\Lambda_b}^{(2)}(\mu) \,  \psi_4(\omega_1,\omega_2,\mu)  \right ]  \,,
\end{eqnarray}
from which we can readily deduce
\begin{eqnarray}
&& \frac{d}{ d \, \ln \mu} \, \Pi_{\perp, V(A)} =  \frac{d}{ d \, \ln \mu} \, \Pi_{\bar n, V(A)}
\nonumber  \\
&& = - {\alpha_s(\mu) \, \over 4 \, \pi} \, {4 \over 3} \, f_{\Lambda_b}^{(2)}(\mu) \,
\int_0^{\infty} d \omega_1 \,  \int_0^{\infty} d \omega_2 \,
\frac{ \psi_4(\omega_1,\omega_2,\mu) }{\omega_1+\omega_2-\bar n \cdot p^{\prime} - i 0} \,\,.
\label{scale dependence of the vector and axial correlator}
\end{eqnarray}
The residual $\mu$-dependence of $\Pi_{\perp, V(A)}$  in Eq. (\ref{scale dependence of the vector and axial correlator})
stems from the UV renormalization of the baryonic current as displayed in (\ref{scale dependence of the baryonic current}).
Differentiating the renormalization scales for the interpolating current of the $\Lambda$-baryon and
for the weak transition current in QCD from the factorization scale (see the next section for details),
we reach the desired conclusion that  the factorization-scale dependence cancels out completely in the
factorized expressions of the  correlation functions $\Pi_{\mu, a}(p,q)$  at one loop.

\subsection{Resummation of large logarithms}

The objective of this section is to sum the parametrically large logarithms to all orders
at NLL in perturbative matching coefficients by solving RG evolution equations in momentum space.
Following the argument of \cite{Wang:2015vgv} the characterized scale of the jet function $\mu_{hc}$
is comparable to the hadronic  scale $\mu_0$ entering the initial condition of the $\Lambda_b$-baryon DA
in practice, we will not resum logarithms of $\mu_{hc}/\mu_0$ from the RG running of the hadronic wave function
when the factorization scale is chosen as  a hard-collinear scale of order $\sqrt{n \cdot p^{\prime} \, \Lambda}$.
Also,  the normalization parameter $f_{\Lambda_b}^{(2)}(\mu)$ will be taken from the HQET sum rule calculation
directly instead of converting it to the corresponding QCD coupling, thus in contrast to \cite{Wang:2015vgv}
no RG evolution of $f_{\Lambda_b}^{(2)}(\mu)$ at the two-loop order is in demand.

Prior to presenting the RG evolution equations of the hard functions
we need to distinguish the renormalization and the factorization scales
which are set to be equal in dimensional regularization.
In doing so we introduce $\nu$ and $\nu^{\prime}$ to denote the renormalization scales
for the baryonic current and the  weak current in QCD, respectively.
It is evident that the  dependence  of $\ln \nu$ needs to be separated from the jet function,
while the $\ln \nu^{\prime}$ dependence requires to be factorized from the hard functions
$C_{T(\tilde{T})}$. Following \cite{Bell:2010mg} the distinction between the renormalization and
the factorization scales can be accounted by writing
\begin{eqnarray}
J \left ( {\mu^2  \over \bar n \cdot p^{\prime} \, \omega_i},
{\omega_i \over \bar n \cdot p^{\prime}}, \nu \right ) &=& J\left ( {\mu^2  \over \bar n \cdot p^{\prime} \, \omega_i},
{\omega_i \over \bar n \cdot p^{\prime}} \right ) + \delta J \left ( {\mu^2  \over \bar n \cdot p^{\prime} \, \omega_i},
{\omega_i \over \bar n \cdot p^{\prime}}, \nu \right ) \,,  \\
C_{T(\tilde{T})}^{A}(n \cdot p^{\prime}, \mu, \nu^{\prime})  &=& C_{T(\tilde{T})}^{A}(n \cdot p^{\prime}, \mu)
+ \delta C_{T(\tilde{T})}^{A}(n \cdot p^{\prime}, \mu, \nu^{\prime}) \,,
\end{eqnarray}
where $J\left ( {\mu^2  \over \bar n \cdot p^{\prime} \, \omega_i},
{\omega_i \over \bar n \cdot p^{\prime}} \right )$  and
$C_{T(\tilde{T})}^{A}(n \cdot p^{\prime}, \mu)$ on the right-hand sides refer to the matching coefficients given by Eqs.
(\ref{origianl form of one-loop jet function}) and (\ref{original form of the one-loop  tensor hard function A}).
Exploiting the RG evolution equations
\begin{eqnarray}
\frac{d}{d \ln \nu} \, \ln \delta J\left ( {\mu^2  \over \bar n \cdot p^{\prime} \, \omega_i},
{\omega_i \over \bar n \cdot p^{\prime}}, \nu \right ) &=&
- \sum_k \left (\frac{\alpha_s(\mu)}{4\, \pi} \right )^k \,
\gamma_{\Lambda}^{(k)} \,, \\
\frac{d}{d \ln \nu^{\prime}} \, \ln \delta C_{T(\tilde{T})}^{A}(n \cdot p^{\prime}, \mu, \nu^{\prime})
&=& - \sum_k \left (\frac{\alpha_s(\mu)}{4\, \pi} \right )^k \,
\gamma_{T(\tilde{T})}^{(k)} \,,
\end{eqnarray}
and implementing the renormalization conditions
\begin{eqnarray}
\delta J \left ( {\mu^2  \over \bar n \cdot p^{\prime} \, \omega_i},
{\omega_i \over \bar n \cdot p^{\prime}}, \mu \right )=0 \,,
\qquad \delta C_{T(\tilde{T})}^{A}(n \cdot p^{\prime}, \mu, \mu)=0\,,
\end{eqnarray}
we find
\begin{eqnarray}
\delta J\left ( {\mu^2  \over \bar n \cdot p^{\prime} \, \omega_i},
{\omega_i \over \bar n \cdot p^{\prime}}, \nu \right ) &=&
-\left (\frac{\alpha_s(\mu)}{4\, \pi} \right ) \,
\gamma_{\Lambda}^{(1)} \, \ln {\nu \over \mu}
+ {\cal O}\left (\alpha_s^2 \right ) \,\,,  \\
\delta C_{T(\tilde{T})}^{A}(n \cdot p^{\prime}, \mu, \nu^{\prime})
&=& -\left (\frac{\alpha_s(\mu)}{4\, \pi} \right ) \,
\gamma_{T(\tilde{T})}^{(1)} \, \ln {\nu^{\prime} \over \mu}
+ {\cal O}\left (\alpha_s^2 \right ) \,\,,
\end{eqnarray}
%where $\alpha_s^{(5)}(\mu)$ refers to the $\rm {\overline {MS}}$ renormalized
%strong coupling constant in five-flavor QCD,
%in constrast with $\alpha_s$ defined in the four-flavor scheme as used in the remainder of this paper.
%%%%%%%%%%%%%%%%%%%%%%%%%%%%%%%%%%%%%%%%%%%%%%%%%%%%%%%%%%%%%%%%%%%%%%%%%%%%%%%%%%%
% Comment: we do not distinguish alpha_s between 5 flavor and 4 flavor schemes, since the
%difference only arises at the two-loop order.
%%%%%%%%%%%%%%%%%%%%%%%%%%%%%%%%%%%%%%%%%%%%%%%%%%%%%%%%%%%%%%%%%%%%%%%%%%%%%%%%%%%
 The anomalous dimensions $\gamma_{\Lambda}^{(k)}$ are already defined in
Eq. (\ref{scale dependence of the baryonic current}), and the renormalization
constants  $\gamma_{T(\tilde{T})}^{(k)}$ at two loops are given by \cite{Bell:2010mg}
\begin{eqnarray}
\gamma_{T(\tilde{T})}^{(1)}=2 \, C_F\,, \qquad
\gamma_{T(\tilde{T})}^{(2)}= - C_F \left [ 19 \, C_F - {257 \over 9} \, C_A
+ {52 \over 9} \, n_f \, T_F  \right ] \,,
\end{eqnarray}
where $n_f=5$ denotes the number of active quark flavours.

Now we are ready to present the jet function and the hard function for the weak tensor current
with the renormalization scales distinct from the factorization scale
\begin{eqnarray}
&& J \left ( {\mu^2  \over \bar n \cdot p^{\prime} \, \omega_i},
{\omega_i \over \bar n \cdot p^{\prime}}, \nu \right ) \nonumber \\
&& =  1 + {\alpha_s(\mu) \over 4 \pi }\, {4 \over 3} \,
\bigg \{\ln^2 {\mu^2 \over n \cdot p^{\prime} \, (\omega- \bar n \cdot p^{\prime})}
-2 \, \ln {\omega -\bar n \cdot p^{\prime} \over \omega_2 -\bar n \cdot p^{\prime}} \,
\ln {\mu^2 \over n \cdot p^{\prime} \, (\omega- \bar n \cdot p^{\prime})}  \nonumber \\
&& \hspace{0.5 cm} -  {1 \over 2} \,
\ln  {\nu^2 \over n \cdot p^{\prime} \, (\omega- \bar n \cdot p^{\prime})}
-  \ln^2 {\omega -\bar n \cdot p^{\prime} \over \omega_2 -\bar n \cdot p^{\prime}} \,
+ 2 \, \ln  {\omega -\bar n \cdot p^{\prime} \over \omega_2 -\bar n \cdot p^{\prime}} \,
\left [ {\omega_2 - \bar n \cdot p^{\prime} \over \omega_1}-{3 \over 4}  \right ] \, \nonumber \\
&&  \hspace{0.5 cm}  -{\pi^2 \over 6} - {1 \over 2} \bigg \}  \,,
\label{revised form of one-loop jet function}
\\
&& C_{T(\tilde{T})}^{A}(n \cdot p^{\prime}, \mu, \nu^{\prime}) \nonumber \\
&& = 1 - \frac{\alpha_s(\mu) \, C_F}{4 \, \pi}
\bigg [2 \, \ln^2 {\mu \over  n \cdot p^{\prime}} +5  \, \ln {\mu \over m_b} + 2\, \ln {\nu^{\prime} \over m_b}
-\, 2 \,  {\rm Li_2} \left (1- {1 \over r} \right ) -  \ln^2 r  \nonumber \\
&& \hspace{0.5 cm} + \, \frac{4 r-2}{1-r} \, \ln r
+ \frac{\pi^2}{12} + 6 \bigg ] \,.
\label{revised form of the one-loop  tensor hard function A}
\end{eqnarray}
Resummation of large logarithms in the hard functions at NLL can be achieved
by solving the RG equations
\begin{eqnarray}
\frac{d}{d \, \ln \mu} C_i(n \cdot p^{\prime}, \mu, \nu^{\prime})
&=& \left[ - \Gamma_{\rm cusp}(\alpha_s) \, \ln {\mu \over n \cdot p^{\prime}}
+ \gamma(\alpha_s) \right ] \, C_i(n \cdot p^{\prime}, \mu, \nu^{\prime}) \,,
\label{RGE of hard functions: mu}\\
\frac{d}{d \, \ln \nu^{\prime}}C_{T(\tilde{T})}^{A}(n \cdot p^{\prime}, \mu, \nu^{\prime})
&=& \left [  -\sum_k \left (\frac{\alpha_s(\mu)}{4\, \pi} \right )^k \,
\gamma_{T(\tilde{T})}^{(k)} \right ] \,\, C_{T(\tilde{T})}^{A}(n \cdot p^{\prime}, \mu, \nu^{\prime}) \,\,,
\label{RGE of hard functions: nuprime}
\end{eqnarray}
where $C_i$ stands for $C_{\perp, V(A)}$, $C_{\bar n, V(A)}$  and $C_{T(\tilde{T})}^{A}$,
the cusp anomalous dimension $ \Gamma_{\rm cusp}(\alpha_s)$
at the three-loop order and the remaining anomalous dimensions $\gamma(\alpha_s)$
and $\gamma_{T(\tilde{T})}^{(k)}$ at two loops are needed
(see \cite{Beneke:2011nf} for the detailed expressions).
Solving Eqs. (\ref{RGE of hard functions: mu}) and (\ref{RGE of hard functions: nuprime}) yields
\begin{eqnarray}
C_{\perp(\bar n), V(A)}(n \cdot p^{\prime}, \mu) &=& U_1(\bar n \cdot p^{\prime}/2, \mu_{h}, \mu) \,
C_{\perp(\bar n), V(A)}(n \cdot p^{\prime}, \mu_{h}) \,,
\\
C_{T(\tilde{T})}^{A}(n \cdot p^{\prime}, \mu, \nu^{\prime}) &=& U_1(\bar n \cdot p^{\prime}/2, \mu_{h}, \mu) \,
U_2(\nu_{h}^{\prime}, \nu^{\prime}) \,C_{T(\tilde{T})}^{A}(n \cdot p^{\prime}, \mu_{h}, \nu_h^{\prime}) \,,
\end{eqnarray}
where $U_1(\bar n \cdot p^{\prime}/2, \mu_{h}, \mu)$ can be deduced from
$U_1(E_{\gamma}, \mu_{h}, \mu)$ in \cite{Beneke:2011nf} with
$E_{\gamma} \to \bar n \cdot p^{\prime}/2$, and $U_2(\nu_{h}^{\prime}, \nu^{\prime})$
can be read from $U_2(\mu_{h2}, \mu)$ in \cite{Wang:2015vgv}
with the following substituent rules
\begin{eqnarray}
\mu_{h2} \to \nu_{h}^{\prime} \,, \qquad \mu \to \nu^{\prime} \,, \qquad
\tilde{\gamma}^{(k)} \to  - \gamma_{T(\tilde{T})}^{(k-1)} \,.
\end{eqnarray}

Finally we present NLL resummmation improved factorized formulae for the invariant amplitudes
entering the Lorenz decomposition of the correlation functions  $\Pi_{\mu, a}(p, q)$
%at leading power in $\Lambda/m_b$
\begin{eqnarray}
\Pi_{\perp, V(A)} &=& f_{\Lambda_b}^{(2)}(\mu) \, \left [ U_1(\bar n \cdot p^{\prime}/2, \mu_{h}, \mu) \,
C_{\perp, V(A)}(n \cdot p^{\prime}, \mu_h)  \right ] \, \nonumber \\
&& \int_0^{\infty} d \omega_1 \,  \int_0^{\infty} d \omega_2 \,
\frac{1}{\omega_1+\omega_2-\bar n \cdot p^{\prime} - i 0} \,
 J\left ( {\mu^2  \over \bar n \cdot p^{\prime} \, \omega_i},
{\omega_i \over \bar n \cdot p^{\prime}}, \nu \right ) \, \psi_4(\omega_1,\omega_2,\mu) \,, \hspace{0.8 cm}
\label{NLL resummed Coree perp V-A}
\\
\Pi_{\bar n, V(A)} &=& f_{\Lambda_b}^{(2)}(\mu) \, \left [ U_1(\bar n \cdot p^{\prime}/2, \mu_{h}, \mu) \,
C_{\bar n, V(A)}(n \cdot p^{\prime}, \mu_h) \right ] \, \nonumber \\
&& \int_0^{\infty} d \omega_1 \,  \int_0^{\infty} d \omega_2 \,
\frac{1}{\omega_1+\omega_2-\bar n \cdot p^{\prime} - i 0} \,
J\left ( {\mu^2  \over \bar n \cdot p^{\prime} \, \omega_i},
{\omega_i \over \bar n \cdot p^{\prime}} , \nu \right ) \, \psi_4(\omega_1,\omega_2,\mu) \,, \hspace{0.8 cm}
\label{NLL resummed Coree nabr V-A}
\\
\Pi_{\perp, T(\tilde{T})} &=& m_{\Lambda_b} \,\, f_{\Lambda_b}^{(2)}(\mu) \,\,
 \left [ U_1(\bar n \cdot p^{\prime}/2, \mu_{h}, \mu) \,
U_2(\nu_{h}^{\prime}, \nu^{\prime}) \,C_{T(\tilde{T})}^{A}(n \cdot p^{\prime}, \mu_{h}, \nu_h^{\prime})  \right ] \, \nonumber  \\
&& \int_0^{\infty} d \omega_1 \,  \int_0^{\infty} d \omega_2 \,
\frac{1}{\omega_1+\omega_2-\bar n \cdot p^{\prime} - i 0} \,
J\left ( {\mu^2  \over \bar n \cdot p^{\prime} \, \omega_i},
{\omega_i \over \bar n \cdot p^{\prime}} , \nu \right ) \, \psi_4(\omega_1,\omega_2,\mu) \,\, \nonumber \\
&& + \,\, m_{\Lambda_b} \,\, f_{\Lambda_b}^{(2)}(\mu) \,\,
 C_{T(\tilde{T})}^{B}(n \cdot p^{\prime}, \mu) \,
\int_0^{\infty} d \omega_1 \,  \int_0^{\infty} d \omega_2 \,
\frac{\psi_4(\omega_1,\omega_2,\mu) }{\omega_1+\omega_2-\bar n \cdot p^{\prime} - i 0} \,\,.
\label{NLL resummed Coree perp T-Ttidle}
\end{eqnarray}
where $\mu$ needs to be taken as a hard-collinear scale of order $\sqrt{n \cdot p^{\prime} \, \Lambda}$
and $\mu_{h}$ should be set to a hard scale of order $n \cdot p^{\prime} \sim m_b$.
Choosing $\nu_h^{\prime}=m_b$ to eliminate the single logarithmic term $\ln \left ( \nu_h^{\prime} / m_b \right )$
in  $C_{T(\tilde{T})}^{A}(n \cdot p^{\prime}, \mu_{h}, \nu_h^{\prime})$, the evolution function
$U_2(\nu_{h}^{\prime}, \nu^{\prime})$  can be further reduced to one provided that $\nu^{\prime}=m_b$.

\section {The LCSR of $\Lambda_b \to \Lambda$ form factors at ${\cal O}(\alpha_s)$}
\label{section: resummation improved LCSR}

It is now a straightforward task to derive the NLL resummmation improved sum rules for the
$\Lambda_b \to \Lambda$ form factors. Working out dispersion forms of the factorized
correlation functions with the aid of the relations in Appendix \ref{appendix: spectral representations}
and applying the standard strategies to construct QCD sum rules, we find
\begin{eqnarray}
&& f_{\Lambda}(\nu) \, (n \cdot p^{\prime}) \, e^{- m_{\Lambda}^2 / \left ( n \cdot p^{\prime} \, \omega_M \right )} \,
\left \{ f_{\Lambda_b \to \Lambda}^{T}(q^2),  g_{\Lambda_b \to \Lambda}^{T}(q^2) \right \}\, \nonumber \\
&& = f_{\Lambda_b}^{(2)}(\mu) \, \left [ U_1(\bar n \cdot p^{\prime}/2, \mu_{h}, \mu) \,
C_{\perp, V(A)}(n \cdot p^{\prime}, \mu_h)  \right ] \,  \int_0^{\omega_s} d \omega^{\prime} \,
e^{-\omega^{\prime}/\omega_M} \, \, \psi_{4, \rm eff}(\omega^{\prime},\mu, \nu) \,,
\label{final NLL sum rules of fperp}
\\
&& f_{\Lambda}(\nu) \, (n \cdot p^{\prime}) \, e^{- m_{\Lambda}^2 / \left ( n \cdot p^{\prime} \, \omega_M \right )} \,
\left \{ f_{\Lambda_b \to \Lambda}^{0}(q^2),  g_{\Lambda_b \to \Lambda}^{0}(q^2) \right \}\, \nonumber \\
&& = f_{\Lambda_b}^{(2)}(\mu) \, \left [ U_1(\bar n \cdot p^{\prime}/2, \mu_{h}, \mu) \,
C_{\bar n, V(A)}(n \cdot p^{\prime}, \mu_h)  \right ] \,  \int_0^{\omega_s} d \omega^{\prime} \,
e^{-\omega^{\prime}/\omega_M} \, \, \psi_{4, \rm eff}(\omega^{\prime},\mu, \nu) \, \nonumber \\
&& \hspace{0.5 cm} +  \, f_{\Lambda_b}^{(2)}(\mu) \, \left ( 1- {n \cdot p^{\prime} \over m_{\Lambda_b}} \right ) \,
C_{n, V(A)}(n \cdot p^{\prime}, \mu_h) \,  \int_0^{\omega_s} d \omega^{\prime} \,
e^{-\omega^{\prime}/\omega_M} \, \, \tilde{\psi}_{4}(\omega^{\prime},\mu) \,,
\\
&& f_{\Lambda}(\nu) \, (n \cdot p^{\prime}) \, e^{- m_{\Lambda}^2 / \left ( n \cdot p^{\prime} \, \omega_M \right )} \,
\left \{ f_{\Lambda_b \to \Lambda}^{+}(q^2),  g_{\Lambda_b \to \Lambda}^{+}(q^2) \right \}\, \nonumber \\
&& = f_{\Lambda_b}^{(2)}(\mu) \, \left [ U_1(\bar n \cdot p^{\prime}/2, \mu_{h}, \mu) \,
C_{\bar n, V(A)}(n \cdot p^{\prime}, \mu_h)  \right ] \,  \int_0^{\omega_s} d \omega^{\prime} \,
e^{-\omega^{\prime}/\omega_M} \, \, \psi_{4, \rm eff}(\omega^{\prime},\mu, \nu) \, \nonumber \\
&& \hspace{0.5 cm} -  \, f_{\Lambda_b}^{(2)}(\mu) \, \left ( 1- {n \cdot p^{\prime} \over m_{\Lambda_b}} \right ) \,
C_{n, V(A)}(n \cdot p^{\prime}, \mu_h) \,  \int_0^{\omega_s} d \omega^{\prime} \,
e^{-\omega^{\prime}/\omega_M} \, \, \tilde{\psi}_{4}(\omega^{\prime},\mu) \,,
\\
&& f_{\Lambda}(\nu) \, (n \cdot p^{\prime}) \, e^{- m_{\Lambda}^2 / \left ( n \cdot p^{\prime} \, \omega_M \right )} \,
\left \{ h_{\Lambda_b \to \Lambda}^{T}(q^2),  \tilde{h}_{\Lambda_b \to \Lambda}^{T}(q^2) \right \}\, \nonumber \\
&& = f_{\Lambda_b}^{(2)}(\mu) \, \left \{  \left [ U_1(\bar n \cdot p^{\prime}/2, \mu_{h}, \mu) \,
U_2(\nu_{h}^{\prime}, \nu^{\prime}) \,C_{T(\tilde{T})}^{A}(n \cdot p^{\prime}, \mu_{h}, \nu_h^{\prime})  \right ]
+ C_{T(\tilde{T})}^{B}(n \cdot p^{\prime}, \mu)   \right \} \nonumber \\
&&   \hspace{0.5 cm} \, \times \,  \int_0^{\omega_s} d \omega^{\prime} \, e^{-\omega^{\prime}/\omega_M} \, \,
 \psi_{4, \rm eff}(\omega^{\prime},\mu, \nu) \,\,,
\\
&& f_{\Lambda}(\nu) \, (n \cdot p^{\prime}) \, e^{- m_{\Lambda}^2 / \left ( n \cdot p^{\prime} \, \omega_M \right )} \,
\left \{ h_{\Lambda_b \to \Lambda}^{+}(q^2),  \tilde{h}_{\Lambda_b \to \Lambda}^{+}(q^2) \right \}\, \nonumber \\
&& = f_{\Lambda_b}^{(2)}(\mu) \,\left [ U_1(\bar n \cdot p^{\prime}/2, \mu_{h}, \mu) \,
U_2(\nu_{h}^{\prime}, \nu^{\prime}) \,C_{T(\tilde{T})}^{A}(n \cdot p^{\prime}, \mu_{h}, \nu_h^{\prime})  \right ] \, \, \nonumber \\
&& \hspace{0.5 cm} \, \times \,\,  \int_0^{\omega_s} d \omega^{\prime} \, e^{-\omega^{\prime}/\omega_M}
 \psi_{4, \rm eff}(\omega^{\prime},\mu, \nu) \,\,,
\end{eqnarray}
where we need to multiply out all $\left [1+{\cal O}(\alpha_s) \right]$ factors involved in the NLO perturbative matching coefficients
and the RG evolution functions, and drop out ${\cal O}(\alpha_s^2)$ terms beyond the NLL approximation \cite{Beneke:2011nf}.
The effective ``distribution amplitude" $\psi_{4, \rm eff}(\omega^{\prime},\mu, \nu)$ is given by
\begin{eqnarray}
\psi_{4, \rm eff}(\omega^{\prime},\mu, \nu) &=& \tilde{\psi}_{4}(\omega^{\prime},\mu)
+ \frac{\alpha_s(\mu)}{4 \, \pi} \, {4 \over 3} \,
\bigg \{ \int_0^{\omega^{\prime}} \, d \omega \,
\left [ {2  \over \omega^{\prime} - \omega}  \,
\ln {\mu^2  \over n \cdot p^{\prime} \, (\omega^{\prime} - \omega) }  \right ]_{\oplus}  \,
\tilde{\psi}_{4}(\omega,\mu) \nonumber
 \\
&& - \, 2 \,\omega^{\prime} \, \int_0^{\omega^{\prime}} \, d \omega \,
\left [ {1 \over \omega^{\prime} - \omega}  \,
\ln { \omega^{\prime} - \omega  \over \omega^{\prime}}  \right ]_{\oplus} \, \phi_4(\omega,\mu) \nonumber
\\
&& - \, \omega^{\prime} \, \int_{\omega^{\prime}}^{\infty} \, d \omega \,
\bigg [ {\omega \over \omega^{\prime}} \, \ln^2 {\mu^2  \over n \cdot p^{\prime} \, \omega^{\prime}}
- 2\, \ln {\mu^2  \over n \cdot p^{\prime} \, \omega^{\prime}} \,
\ln {\omega- \omega^{\prime}  \over \omega^{\prime} }
- {11 \over 2} \,\ln {\omega- \omega^{\prime}  \over \omega^{\prime} } \nonumber
\\
&& \hspace{2.6 cm} - \, {\pi^2 + 1 \over 2} \,  {\omega \over \omega^{\prime}} \,
+ \left ( {2 \, \pi^2 \over 3}  - {11 \over 2} \right ) \bigg ] \, {d \phi_4(\omega,\mu) \over d \omega} \nonumber
\\
&& - \, \int_{\omega^{\prime}}^{\infty} \, \bigg [ \, \ln^2 {\mu^2  \over n \cdot p^{\prime} \, \omega^{\prime}}
+ 2\, \ln {\omega- \omega^{\prime}  \over \omega } - \, {\pi^2 + 1 \over 2}  \bigg ]\, \, \phi_4(\omega,\mu) \nonumber
\\
&& - \int_{0}^{\omega^{\prime}} \, d \omega \,
\bigg [ 2\, \ln^2 {\mu^2  \over n \cdot p^{\prime} \, (\omega^{\prime} -\omega)}
+ {1 \over 2} \, \ln^2 {\nu^2  \over n \cdot p^{\prime} \, (\omega^{\prime} -\omega)} \bigg ] \,
\, {d \tilde{\psi}_4(\omega,\mu) \over d \omega} \bigg \} \,, \hspace{0.5 cm}
\end{eqnarray}
where $\psi_4(\omega_1, \omega_2, \mu)=\psi_4(u \, \omega, (1-u) \,\omega,\mu)$ is supposed to be independent
of the momentum fraction $u$ as motivated by \cite{Feldmann:2011xf,Bell:2013tfa,Ball:2008fw}
and will be set to $\phi_4(\omega,\mu)$ for brevity,
and $\tilde{\psi}_4(\omega,\mu)$  defined in Eq. (\ref{definition of psi4tidle}) can be identified as
$\tilde{\psi}_4(\omega,\mu)=\omega \, \phi_4(\omega,\mu)$ within this approximation.
The $\oplus$ function is defined as
\begin{eqnarray}
\int_0^{\infty} \, d \omega \, \left [ f(\omega,\omega^{\prime}) \right ]_{\oplus} \, g(\omega)
= \int_0^{\infty} \, d \omega \, f(\omega,\omega^{\prime})  \,
\left [ g(\omega) - g(\omega^{\prime}) \right ] \,.
\end{eqnarray}

The following observations on the structures of the NLL sum rules can be made.
\begin{itemize}
\item {Due to the integration bounds of $\omega^{\prime}$ after the continuum subtraction,
the scaling behaviour $\omega^{\prime} \sim \omega_s \sim \Lambda^2/(n \cdot p^{\prime})$
implies that the natural choice for the factorization scale $\mu$ of
$\ln^k \left [ {\mu^2 /  \left (n \cdot p^{\prime} \, \omega^{\prime} \right )} \right ]$ ($k=1\,, 2$)
in  $\psi_{4, \rm eff}(\omega^{\prime},\mu, \nu)$ should be
$\mu_{s} \sim s_0 = n \cdot p^{\prime} \, \omega^{\prime} \sim \Lambda^2 $ in contrast to the
favored choice $\mu_{hc} \sim n \cdot p^{\prime} \, \Lambda $ in the factorization formulae of the
correlation functions presented in Eqs. (\ref{NLL resummed Coree perp V-A}), (\ref{NLL resummed Coree nabr V-A})
and (\ref{NLL resummed Coree perp T-Ttidle}). }
\item {Due to the power counting $\omega \sim  \Lambda$ determined by
the canonical behavious of the $\Lambda_b$-baryon DA $\phi_4(\omega,\mu)$,
the logarithmic term $\ln \left [{(\omega- \omega^{\prime})  /  \omega^{\prime} } \right ]$ appeared in
$\psi_{4, \rm eff}(\omega^{\prime},\mu, \nu)$ is counted as $\ln (n \cdot p^{\prime} / \Lambda)$
in the heavy-quark limit.  Such enhanced logarithm arises from the contributions of the $\Lambda$-baryon vertex diagrams
and the two box diagrams displayed in Fig. \ref{fig: one-loop_correlator} and it shares the same origin as the
rapidity singularities preventing a complete factorization  of heavy-to-light form factors in $\rm SCET_{II}$
(see also \cite{DeFazio:2007hw}). It is evident that the standard momentum-space resummation technique cannot be
applied to cope with this term which is independent of the factorization scale.
Investigating resummation of such logarithm with the rapidity RG evolution equations
\cite{Chiu:2012ir,Li:2012md,Li:2013xna,Shen:2014wga} is
apparently of conceptual interest and we will pursue this endeavour  in a future work. }
\end{itemize}

\section {Numerical results}
\label{section: numerical analysis}

Having at hand the NLL resummation improved sum rules for the $\Lambda_b \to \Lambda$ form factors
we are ready to explore their phenomenological implications.
We will begin the numerical analysis with  specifying the non-perturbative models for the $\Lambda_b$-baryon DA,
determining the ``internal" sum rule parameters and evaluating the normalization parameters
$f_{\Lambda}(\nu)$ and $f_{\Lambda_b}^{(2)}(\mu)$.
Theory predictions for the $\Lambda_b \to \Lambda$ form factors at large hadronic recoil will be further presented
and extrapolation of the form factors toward large momentum transfer will be performed by applying the $z$-series expansion
and matching the calculated form factors from the LCSR approach at low $q^2$.

\subsection{Theory input parameters}

\begin{figure}
\begin{center}
\includegraphics[width=0.50 \columnwidth]{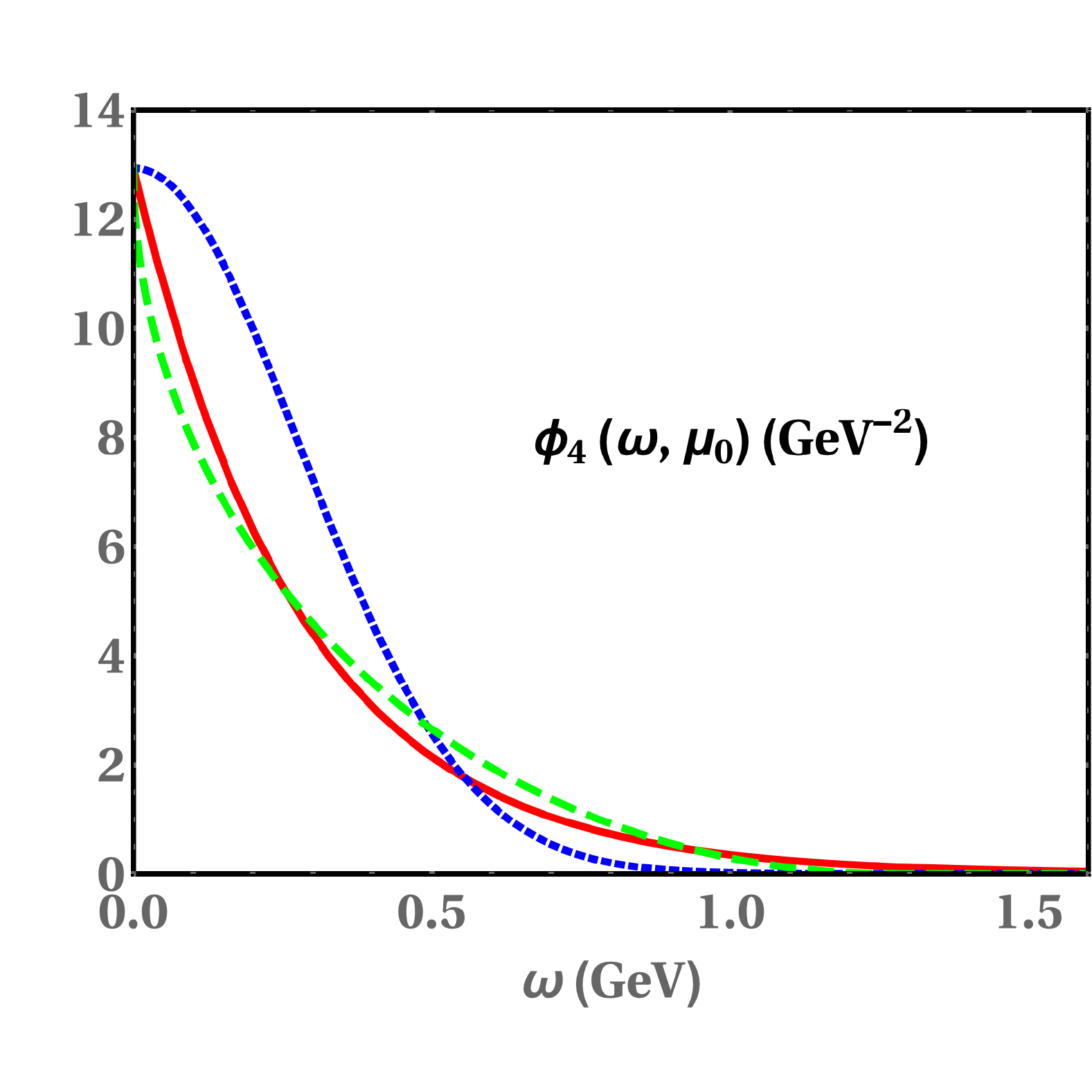}
\vspace*{0.1cm}
\caption{The small $\omega$ behaviors  for three different models of  $ \phi_4(\omega, \mu_0)$.
Solid (red), dotted (blue) and dashed (green) curves refer to $\phi_4^{\rm I}$, $\phi_4^{\rm II}$
and $\phi_4^{\rm III}$, respectively. }
\label{fig: shape of phi4}
\end{center}
\end{figure}

Light-cone wave functions of the $\Lambda_b$-baryon at small transverse separations
have attracted renewed attention \cite{Ball:2008fw,Bell:2013tfa,Braun:2014npa} due to the available measurements
of the baryonic $\Lambda_b \to \Lambda \, \ell^{+} \ell^{-}$ decays at the LHC and the Tevatron
 \cite{Aaij:2013mna,Aaij:2015xza,Aaltonen:2011qs}.
Improved models of the twist-2 $\Lambda_b$-baryon DA in compatible with the RG evolution equation at one loop
have been discussed in  \cite{Bell:2013tfa,Braun:2014npa}, however, no dedicated study of the twist-4 DA
$\psi_4(\omega_1,\omega_2, \mu)$ (or $\phi_4(\omega,\mu)$), taking into account the QCD constraints,
exists in the literature to the best of our knowledge. Motivated by the ``on-shell-wave-function" analysis
of  \cite{Bell:2013tfa} we consider three different parameterizations of the $\Lambda_b$-baryon DA
$\phi_4(\omega,\mu_0)$ at a soft scale
\begin{eqnarray}
\phi_4^{\rm I}(\omega,\mu_0) &=& {1 \over \omega_0^2 } \, e^{-\omega/\omega_0} \,, \nonumber \\
\phi_4^{\rm II}(\omega,\mu_0) &=& {1 \over \omega_0^2 } \, e^{- \left ( \omega/\omega_1 \right )^2}  \,,
\qquad \omega_1 = \sqrt{2} \, \omega_0 \,, \nonumber  \\
\phi_4^{\rm III}(\omega,\mu_0) &=& {1 \over \omega_0^2 } \,
 \left [ \, 1- \sqrt{\left (2- {\omega \over \omega_2}\right ) \, {\omega \over \omega_2}}  \, \right ] \,
\theta(\omega_2-\omega)  \,\,,
\qquad  \omega_2 = \sqrt{{12 \over 10 - 3 \, \pi} } \, \omega_0 \,,
\end{eqnarray}
where $\phi_4^{\rm II}(\omega,\mu_0)$  and $\phi_4^{\rm III}(\omega,\mu_0)$
are analogies to the mesonic counterparts proposed in \cite{DeFazio:2007hw}
for the sake of maximizing the model dependence of $\phi_4(\omega,\mu_0)$
and the normalization constants are determined by
\begin{eqnarray}
\int_0^{\infty} \,  d \omega \,\, \omega \, \phi_4(\omega, \mu) = 1 \,.
\end{eqnarray}
Applying the equations of motion with the  Wandzura-Wilczek approximation yields
\begin{eqnarray}
\psi_2(\omega_1,\omega_2,\mu_0)= \omega_1 \, \omega_2 \,
 {d \psi_4(\omega_1,\omega_2,\mu_0) \over d \omega_1 \, d \omega_2} \,,
\end{eqnarray}
in analogy to the Wandzura-Wilczek relation for the $B$-meson DA \cite{Beneke:2000wa}.
We will take $\phi_4^{\rm I}(\omega,\mu_0)$ as our default model in
computing the $\Lambda_b \to \Lambda$ form factors from the LCSR approach and
take into account the numerical impact of the alternative parameterizations
$\phi_4^{\rm II, III}(\omega,\mu_0)$ in the uncertainty analysis.
To illustrate the main features of the above-mentioned three models we present the small $\omega$
behaviors of $ \phi_4(\omega, \mu_0)$ in Fig. \ref{fig: shape of phi4} with a reference value $\omega_0 = 280 \, {\rm MeV}$.
We remark that these models do not develop the radiative tail at large $\omega$ due to perturbative corrections,
and they should  be merely treated as an effective description of $\phi_4(\omega,\mu_0)$
at small $\omega$ where QCD factorization of the correlation
functions is established.

%%%%%%%%%%%
%%%%%%%%%%%

Regarding the determination of the internal sum rule parameters we follow closely the strategies
proposed to explore the sum rules for the $B \to \pi$ form factors \cite{Wang:2015vgv}.
\begin{itemize}
\item {To reduce the sysmematic uncertainty induced by the parton-hadron duality approximation,
the continuum contributions to the dispersion forms of the correlation functions, displayed
in Eqs. (\ref{NLL resummed Coree perp V-A}), (\ref{NLL resummed Coree nabr V-A}),
(\ref{NLL resummed Coree perp T-Ttidle}), (\ref{NLO Corr n V-A}) and (\ref{NLO Corr plus T-Ttidle})
need to be under  reasonable control, i.e., less than 40 \%. }
\item {The sum rule predictions should be stable with respect to the variation of the Borel mass parameter
$\omega_M$. More concretely, we impose the following condition on the logarithmic derivative to a  given
form factor
\begin{eqnarray}
\frac{\partial \, \ln F_{\Lambda_b \to \Lambda}^{i}}{\partial \ln \omega_M}
\leq 40 \% \,,
\end{eqnarray}
where $F_{\Lambda_b \to \Lambda}^{i}$ stands for a general $\Lambda_b \to \Lambda$ form factor.}
\end{itemize}
The allowed regions of the Borel parameter and the effective duality threshold are found to be
\begin{eqnarray}
M^2 \equiv n \cdot p^{\prime} \, \omega_M = (2.6 \pm 0.4) \, {\rm GeV^2} \,,
\qquad
s_0 \equiv n \cdot p^{\prime} \, \omega_s = (2.56 \pm 0.10) \, {\rm GeV^2} \,,
\end{eqnarray}
where the obtained interval of $s_0$ is in agreement with that adopted in \cite{Feldmann:2011xf,Liu:2014uha}.

The coupling $f_{\Lambda_b}^{(2)}(\mu_0)$ will be taken from the NLO HQET sum rule calculation \cite{Groote:1997yr}
\begin{eqnarray}
 f_{\Lambda_b}^{(2)}(1 \, {\rm GeV})= (3.0 \pm 0.5) \times 10^{-2} \, {\rm GeV^3} \,.
\end{eqnarray}
In order to reduce the theory uncertainty induced by the Borel mass parameter $\omega_M$ we will employ the
two-point QCD sum rules of the normalization parameter $f_{\Lambda}(\nu)$ \cite{Liu:2008yg}
\begin{eqnarray}
f_{\Lambda}^2 \, e^{-m_{\Lambda}^2/M^2} &=& {1 \over 640 \, \pi^4} \,
\int_{m_s^2}^{s_0} \, d s \,\, e^{-s/M^2} \, s \, \left (1-{m_s^2 \over s} \right )^5 \, \nonumber \\
&& - {1 \over 192 \, \pi^2} \, \left \langle  {\alpha_s \over \pi} GG  \right \rangle \,
\int_{m_s^2}^{s_0} \, d s \, e^{-s/M^2} \,  {m_s^2 \over s^2} \, \left ( 1- {m_s^2 \over s} \right ) \,
\left (1-{2 \, m_s^2 \over s} \right ) \,
\end{eqnarray}
at tree level,  where the gluon condensate density
$\left \langle \alpha_s / \pi GG \right \rangle = \left (1.2^{+0.6}_{-1.2} \right) \times 10^{-2}  \,\, {\rm GeV^4}$
will be used in the numerical analysis.

We now turn to discuss the choices of the renormalization and the factorization scales
entering the NLL sum rules. The renormalization scale $\nu$ of the baryonic current
and the factorization scale $\mu$ will be varied in the interval
$1 \,{\rm GeV} \leq \mu, \nu \leq 2 \, {\rm GeV}$  around the default value $\mu=\nu=1.5 \, {\rm GeV}$.
The renormalization scale $\nu^{\prime}$ of the weak (pseudo)-tensor currents and the hard scale $\mu_h$ in the
hard matching coefficients will be taken as $\mu_h=\nu^{\prime}=m_b$ with the variation in the range
$[m_b/2 \,\,, 2 \, m_b]$.  In addition, we adopt the $\overline{{\rm MS}}$ bottom-quark mass
$\overline{m_b}(\overline{m_b}) = 4.193^{+0.022}_{-0.035} \,\, {\rm GeV}$ determined from non-relativistic
sum rules for the inclusive $e^{+} \, e^{-} \to b \, \bar b$ production cross section
at next-to-next-to-next-to-leading order \cite{Beneke:2014pta}.

\subsection{Predictions for the $\Lambda_b \to \Lambda$ form factors}

\begin{figure}
\begin{center}
\includegraphics[width=0.75 \columnwidth]{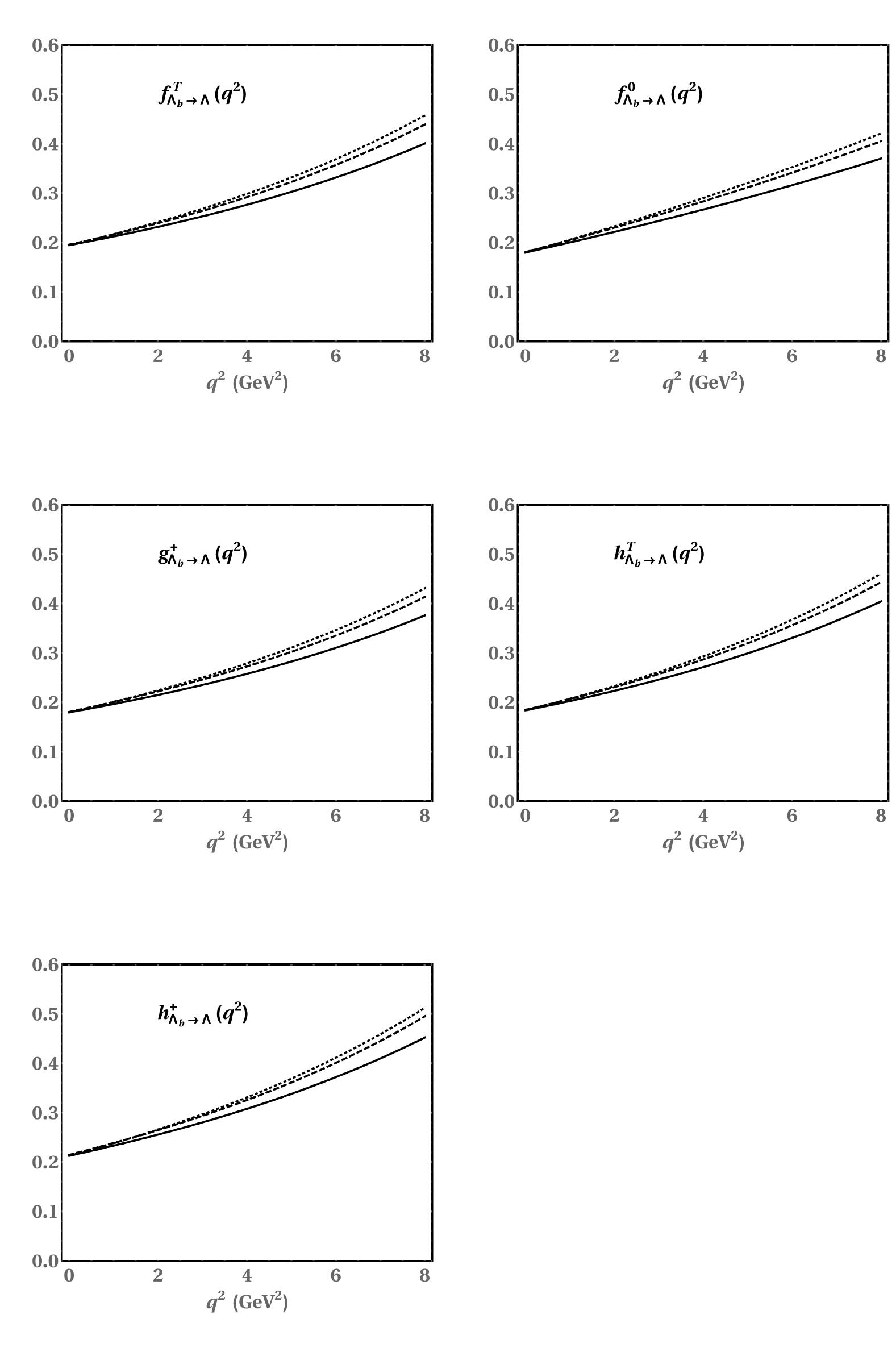}
\vspace*{0.1cm}
\caption{The momentum-transfer dependence of the $\Lambda_b \to \Lambda$ form factors computed from LCSR
with the fitted values of $\omega_0$ parameter presented  in (\ref{fitted values of omega0}) for three
different models of $\phi_4(\omega,\mu_0)$.     Solid, dotted and dashed curves correspond to the sum rule
predictions with the $\Lambda_b$-baryon DA $\phi_4^{\rm I}(\omega,\mu_0) $, $\phi_4^{\rm II}(\omega,\mu_0) $
and $\phi_4^{\rm III}(\omega,\mu_0) $, respectively. }
\label{fig: model dependence of FFs}
\end{center}
\end{figure}

After specifying all the necessary input parameters we will first turn to determine
the shape parameter $\omega_0$ of the $\Lambda_b$-baryon DA $\phi_4(\omega,\mu_0)$.
Given the sizeable uncertainty of $\omega_0$ estimated from the sum rule analysis in \cite{Ball:2008fw},
we prefer to, following \cite{Wang:2015vgv},  extract this parameter by matching the LCSR prediction of the form factor
$f_{\Lambda_b \to \Lambda}^{+}(q^2)$ at zero momentum transfer to that determined from an alternative method.
In doing so, we apply the ${\rm SU}(3)$ flavour symmetry relation between the $\Lambda_b \to \Lambda$ and the $\Lambda_b \to p$
form factors
\begin{eqnarray}
\frac{f_{\Lambda_b \to \Lambda}^{+}(0)}{f_{\Lambda_b \to p}^{+}(0)}
\simeq {f_{\Lambda} \over f_N} \,,
\end{eqnarray}
motivated by an analogous relation for the $B$-meson decay form factors
\begin{eqnarray}
\frac{f_{B \to K}^{+}(0)}{f_{B \to \pi}^{+}(0)}
\simeq {f_{K} \over f_{\pi} } \,,
\end{eqnarray}
which turns out to be a rather satisfactory approximation when confronted with the predictions
from both the LCSR \cite{Khodjamirian:2011ub,Khodjamirian:2010vf} and the
TMD factorization \cite{Li:2012nk,Wang:2012ab} approaches.
Employing the prediction of $f_{\Lambda_b \to p}^{+}(0)$ from the LCSR with the nucleon DA \cite{Khodjamirian:2011jp}
and the result of $f_{\Lambda} / f_N$ computed from QCD sum rules \cite{Chernyak:1987nu} yields
$f_{\Lambda_b \to \Lambda}^{+}(0)=0.18 \pm 0.04$. Proceeding with the above-mentioned matching procedure
we then find
\begin{eqnarray}
\omega_0 &=& 280^{+47}_{-38} \,\, {\rm MeV} \,\,,  \qquad {\rm (Model \,\,\, I)} \,  \nonumber  \\
\omega_0 &=& 386^{+45}_{-37} \,\, {\rm MeV} \,\,,  \qquad {\rm (Model \,\,\, II)} \,  \nonumber  \\
\omega_0 &=&  273^{+38}_{-29} \,\, {\rm MeV} \,\,.  \qquad {\rm (Model \,\,\, III)} \,
\label{fitted values of omega0}
\end{eqnarray}
The apparent dependence of the extracted values of $\omega_0$ on the specific parametrization of $\phi_4(\omega,\mu_0)$
implies that the $\Lambda_b \to \Lambda$ form factors cannot be  determined by the shape parameter $\omega_0$
satisfactorily to a reasonable approximation and the detailed information of the small $\omega$ behaviours of
$\phi_4(\omega,\mu_0)$ is in demand to have a better control on the form factors from the sum rule analysis.
Having this in  mind, our main purpose is  to predict the momentum-transfer dependence
of all the ten $\Lambda_b \to \Lambda$ form factors  in anticipation of the reduced model dependence of
$\phi_4(\omega,\mu_0)$ in the form factor ratios.
Anatomy of the sum rules numerically indeed indicates the expected insensitivity  of the form-factor
shapes as displayed in  Fig. \ref{fig: model dependence of FFs}.

%%%%%%%%%%%
%%%%%%%%%%%

%%%%%%%%%%%
\begin{figure}
\begin{center}
\includegraphics[width=1.0 \columnwidth]{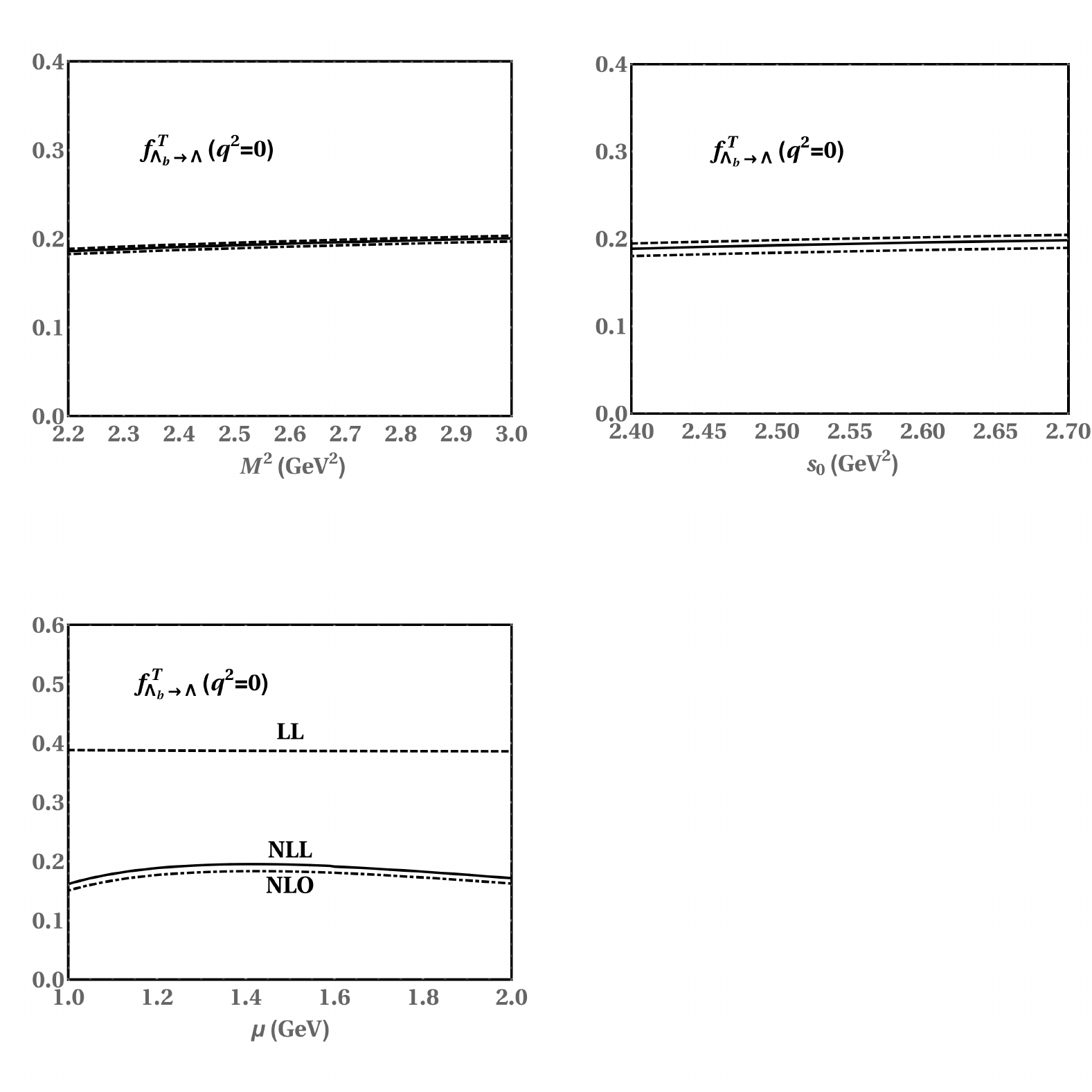}
\vspace*{0.1cm}
\caption{Dependence of $f_{\Lambda_b \to \Lambda}^{T}(0)$ on the Borel parameter (top left),
on the threshold parameter (top right) and on the factorization scale (bottom left).
Solid, dashed and dotted curves are obtained from the NLL sum rules with $s_0=2.56 \, {\rm GeV^2}$,
$2.66 \, {\rm GeV^2}$,  $2.46 \, {\rm GeV^2}$ (top left) and $M^2=2.6  \, {\rm GeV^2}$,
$3.0 \, {\rm GeV^2}$, $2.2 \, {\rm GeV^2}$ (top right) while all the other input parameters
are fixed at their central values. The curves labelled by  ``LL",  ``NLO'' and ``NLL" (bottom)
correspond to the sum rule predictions at LL, NLO and NLL accuracy. }
\label{fig: sum rule parameter and mu dependence of FFperp}
\end{center}
\end{figure}
%%%%%%%%%%%

To demonstrate some important numerical features of the LCSR predictions, we show the dependencies of
$f_{\Lambda_b \to \Lambda}^{T}(0)$ on the sum rule parameters $M^2$ and $s_0$ and on the factorization
scale $\mu$ in Fig. \ref{fig: sum rule parameter and mu dependence of FFperp}
as an illustrative example and analogous profiles are also observed for the remaining
$\Lambda_b \to \Lambda$ form factors. It is evident that the sum rules of $f_{\Lambda_b \to \Lambda}^{T}(0)$
exhibits extraordinary mild dependence on the Borel mass parameter due to a strong cancellation of the sysmematic
uncertainty between the LCSR of $f_{\Lambda_b \to \Lambda}^{T}(0)$  and the QCD
sum rules of the coupling $f_{\Lambda}$. One can further find that both the leading-logarithmic (LL) and the NLL
resummation improved  sum rules are insensitive to the factorization scale $\mu$ in the allowed interval
and resummation of parametrically large  logarithms in the hard matching coefficients  only induces a minor impact
on the sum rules for  $f_{\Lambda_b \to \Lambda}^{T}(0)$ numerically compared with  the one-loop fixed-order correction.
More importantly, the perturbative ${\cal O}(\alpha_s)$ correction  is found to reduce
the tree-level sum rule prediction by a factor of $1/2$, implying the importance of QCD radiative effect in baryonic sum rule
applications (see also \cite{Groote:1996em} for a similar observation on the perturbative spectral function of the vacuum-to-vacuum
correlation function defined with two baryonic currents in HQET).

%%%%%%%%%%%
\begin{figure}
\begin{center}
\includegraphics[width=1.0 \columnwidth]{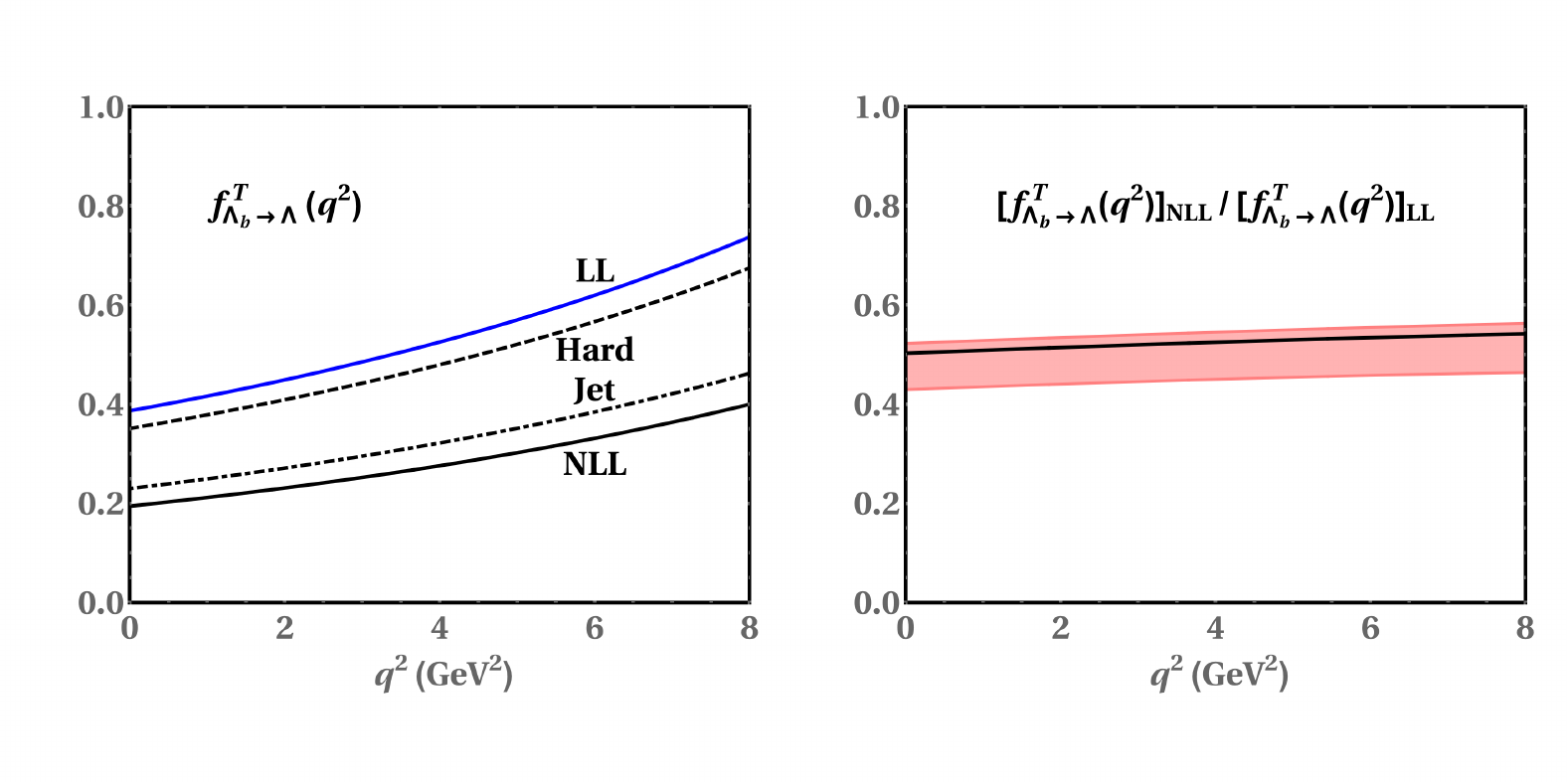}
\vspace*{0.1cm}
\caption{Breakdown of the one-loop contribution to the sum rules of $f_{\Lambda_b \to \Lambda}^{T} (q^2)$
from the NLO hard and the NLO jet functions (left panel) and the momentum transfer dependence of
the ratio $\left [f_{\Lambda_b \to \Lambda}^{T} (q^2) \right ]_{\rm NLL}/
\left [f_{\Lambda_b \to \Lambda}^{T} (q^2) \right ]_{\rm LL}$ with theory uncertainties
from varying the renormalization and the factorization scales (right panel). }
\label{fig: breakdown of NLO corrections}
\end{center}
\end{figure}
%%%%%%%%%%%

To develop a better understanding of the origin of the significant perturbative correction,
we break the complete one-loop contribution to the sum rules of $f_{\Lambda_b \to \Lambda}^{T}(0)$
down into the hard and the hard-collinear corrections, which are defined as replacing
$\psi_{4, {\rm eff}}(\omega^{\prime},\mu,\nu)$  in Eq. (\ref{final NLL sum rules of fperp}) by
$\tilde{\psi}_{4}(\omega^{\prime})$ for the former and as replacing
$\left [ U_1(\bar n \cdot p^{\prime}/2, \mu_{h}, \mu) \,
C_{\perp, V(A)}(n \cdot p^{\prime}, \mu_h)  \right ]$
by one for the latter. In Fig. \ref{fig: breakdown of NLO corrections} (left panel)
we plot the separate perturbative contributions
from  hard and hard-collinear fluctuations  as  functions of the momentum transfer squared.
We can readily find that the dominant $\alpha_s$ correction at one loop is from the
NLO jet (hard-collinear) function instead of the NLO  hard function and this highlights
the importance of the perturbative matching calculations at the hard-collinear scale performed
in this paper.  The $q^2$-dependence of the ratio
$\left [f_{\Lambda_b \to \Lambda}^{T} (q^2) \right ]_{\rm NLL}/
\left [f_{\Lambda_b \to \Lambda}^{T} (q^2) \right ]_{\rm LL}$
with the theory uncertainty estimated from varying both the renormalization and
the factorization scales in the acceptable ranges are displayed in the right panel of
 Fig. \ref{fig: breakdown of NLO corrections}.

\begin{figure}
\begin{center}
\includegraphics[width=1.0 \columnwidth]{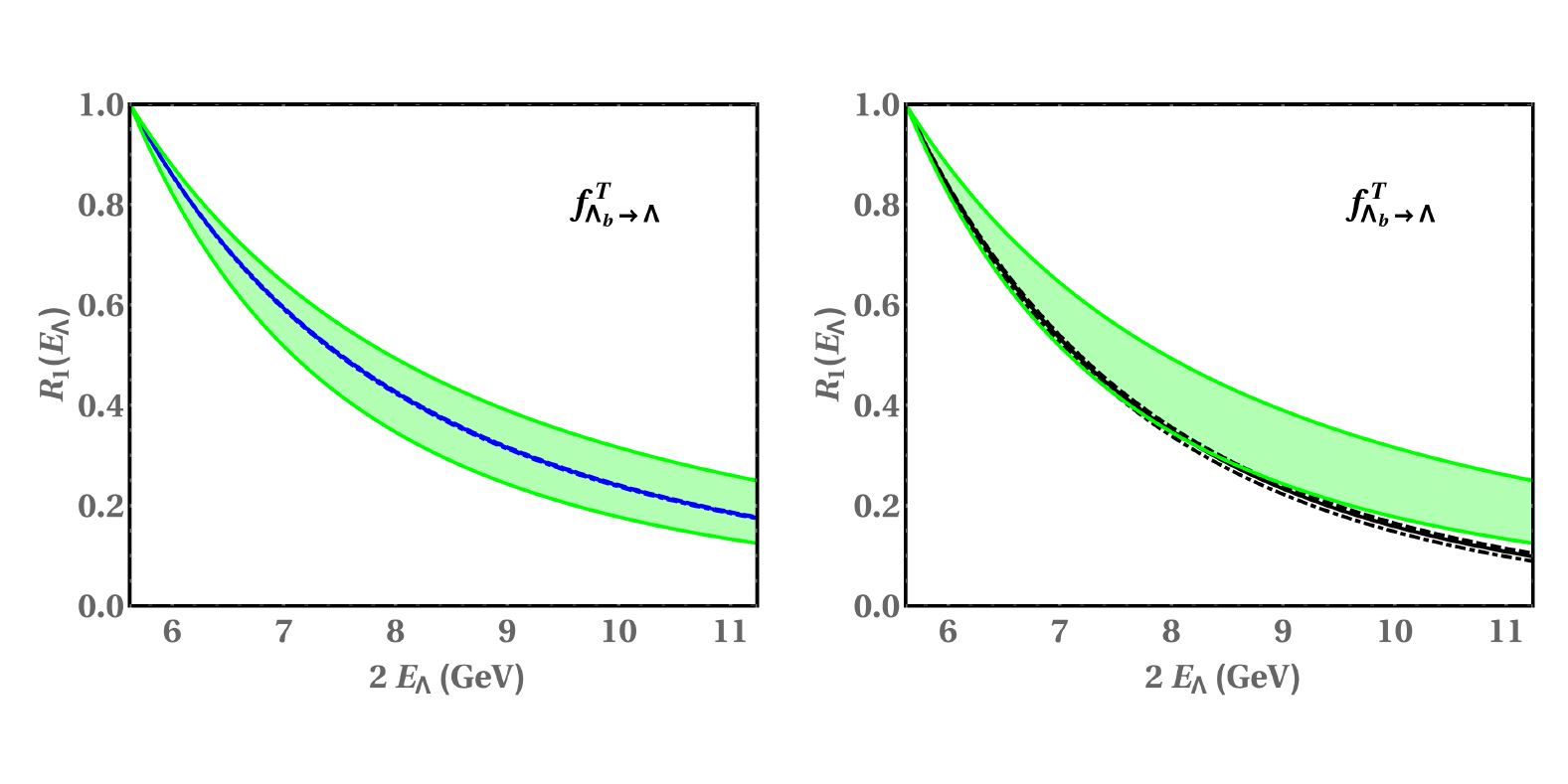}
\vspace*{0.1cm}
\caption{Dependence of the ratio $R_1(E_{\Lambda})$ on the $\Lambda$-baryon energy $E_{\Lambda}$.
The blue (left panel) and the black (right panel) curves  are obtained from the LL and NLL sum rule predictions,
respectively. The two green curves refer to a pure $1/E_{\Lambda}^2$ and a pure $1/E_{\Lambda}^3$
dependence.}
\label{fig: energy dependence of fperp}
\end{center}
\end{figure}

We now turn to investigate the $\Lambda$-baryon energy dependence of the form factor
$f_{\Lambda_b \to \Lambda}^{T} (q^2)$, from the sum rules at LL and at NLL accuracy,
which is of particular conceptual interest in that the soft overlap contributions
and the hard-spectator scattering effects in the heavy-to-light baryonic form factors
differ in the  scaling of $1/E_{\Lambda}$ at large hadronic recoil.
For this purpose, we introduce the following ratio originally proposed in \cite{DeFazio:2007hw}
\begin{eqnarray}
R_1(E_{\Lambda}) = \frac{f_{\Lambda_b \to \Lambda}^{T} (n \cdot p^{\prime})}
{f_{\Lambda_b \to \Lambda}^{T} (m_{\Lambda_b})} \,,
\end{eqnarray}
where we have switched the argument of the form factor from  $q^2$ as
used in the remainder of this paper to $n \cdot p^{\prime}\simeq 2 \, E_{\Lambda}$.
As shown in Fig. \ref{fig: energy dependence of fperp},
the predicted energy dependence of $f_{\Lambda_b \to \Lambda}^{T}$
from the LL sum rules exhibits a scaling behaviour in between $1/E_{\Lambda}^2$ and $1/E_{\Lambda}^3$ for the
default choices of theory input parameters, while the NLL sum rule prediction favors evidently a
$1/E_{\Lambda}^3$ behavior in consistent with the power counting analysis.
We have also verified that such observation can be made for the energy dependence
of all the other $\Lambda_b \to \Lambda$ form factors.

%%%%%%%%%%%
%%%%%%%%%%%

Since the light-cone operator-product expansion of the correlation functions $\Pi_{\mu, a}(p, q)$
can only  be  justified at low $q^2$, we need to extrapolate the sum rule predictions
for the $\Lambda_b \to \Lambda$  form factors at $q^2 \leq q_{\rm max}^2 = 8\, {\rm GeV^2}$
toward large momentum transfer $q^2$. To this end, we apply the simplified $z$-series parametrization
\cite{Bourrely:2008za} based upon the conformal mapping of the  cut $q^2$-plane onto
the disk $|z(q^2, t_0)| \leq 1$ in the complex $z$-plane with the standard transformation
\begin{eqnarray}
z(q^2, t_0) = \frac{\sqrt{t_{+}-q^2}-\sqrt{t_{+}-t_0}}{\sqrt{t_{+}-q^2}+\sqrt{t_{+}-t_0}}\,.
\end{eqnarray}
The parameter $t_{+}$ is determined by the threshold of the lowest continuum state which
can be excited by the weak transition current in QCD.
It is evident that all  the  channels $|B K\rangle $, $| B_s \pi \rangle$
and  $| \Lambda_b \, \bar \Lambda \rangle$ can be produced by the $\bar s \,\, \Gamma_{\mu, a} \,\, b$
current, the form factors can be analytical functions in the complex $q^2$-plane cut along
the real axis for
\begin{eqnarray}
q^2 \geq  {\rm min} \left \{ (m_{B_s}+m_{\pi})^2 \,,  (m_{B}+m_{K})^2 \,,
(m_{\Lambda_b}+m_{\Lambda})^2  \right \} = (m_{B_s}+m_{\pi})^2 \,,
\end{eqnarray}
in addition to the potential resonances below the branch cut.
We theretofore need to set $t_{+}=(m_{B_s}+m_{\pi})^2$ for all the $\Lambda_b \to \Lambda$ form factors.
The auxiliary parameter $t_0$ determines the $q^2$ point that will be mapped onto the origin of the complex $z$-plane,
and in practice we will choose $t_0=(m_{\Lambda_b}-m_{\Lambda})^2$ following \cite{Detmold:2015aaa}.
Since the helicity form factors are constructed from the hadronic matrix elements of weak transition currents
 with definite spin-parity quantum numbers by projecting on the polarization vector for a spin-one particle
 with the four-momentum $q_{\mu}$, we collect some fundamental information of the lowest resonances produced
 by the helicity-projected weak currents in Table \ref{table of resonance information}.

%%%%%%%%%%%%%%%%%%%%%%%%%%%%%%%%%%%%
\begin{table}[t!bph]
\begin{center}
\begin{tabular}{|c|c|c|c|}
  \hline
  \hline
  &&& \\
  form \,\, factor  &  \,\, {$B_s (J^{P})$} \,\,  & \,\, Mass \, (GeV) \,\, & \,\,  Ref. \,\, \\
    &&& \\
    \hline
     &&& \\
$f_{\Lambda_b \to \Lambda}^{+, T}(q^2) , \,\, h_{\Lambda_b \to \Lambda}^{+, T}(q^2) $
& $B_s^{\ast} \, (1^{-})$ & $5.42$ & \cite{Agashe:2014kda}  \\
  &&& \\
$f_{\Lambda_b \to \Lambda}^{0}(q^2)$ & $B_{s0} \, (0^{+})$ & $5.72$  & \,\, (our \, estimate) \,\,  \\
  &&& \\
\,\, $g_{\Lambda_b \to \Lambda}^{+, T}(q^2)$, \,\,
$\tilde{h}_{\Lambda_b \to \Lambda}^{+, T}(q^2)$  \,\, & $B_{s1} \, (1^{+})$  & $5.83$ & \cite{Agashe:2014kda} \\
  &&& \\
$g_{\Lambda_b \to \Lambda}^{0}(q^2)$ &  $B_s \,  (0^{-})$ & $5.37$ &  \cite{Agashe:2014kda}  \\
  &&& \\
  \hline
  \hline
\end{tabular}
\end{center}
\caption{Summary of the masses of low-lying resonances produced by
the helicity-projected weak currents \, $\bar s \, \Gamma_{\mu, a} \, b$ \, in QCD.
Since the scalar $B_{s0}$ meson has not been observed
experimentally yet,  we estimate its mass using an approximate ${\rm SU}(3)$
symmetry relation $m_{B_{s0}}-m_{B_s}=m_{B_{d0}}-m_{B_d}$, which is found to be comparable to
that predicted by the heavy quark/chiral symmetry \cite{Bardeen:2003kt}.}
\label{table of resonance information}
\end{table}
%%%%%%%%%%%%%%%%%%%%%%%%

\begin{figure}
\begin{center}
\includegraphics[width=0.75 \columnwidth]{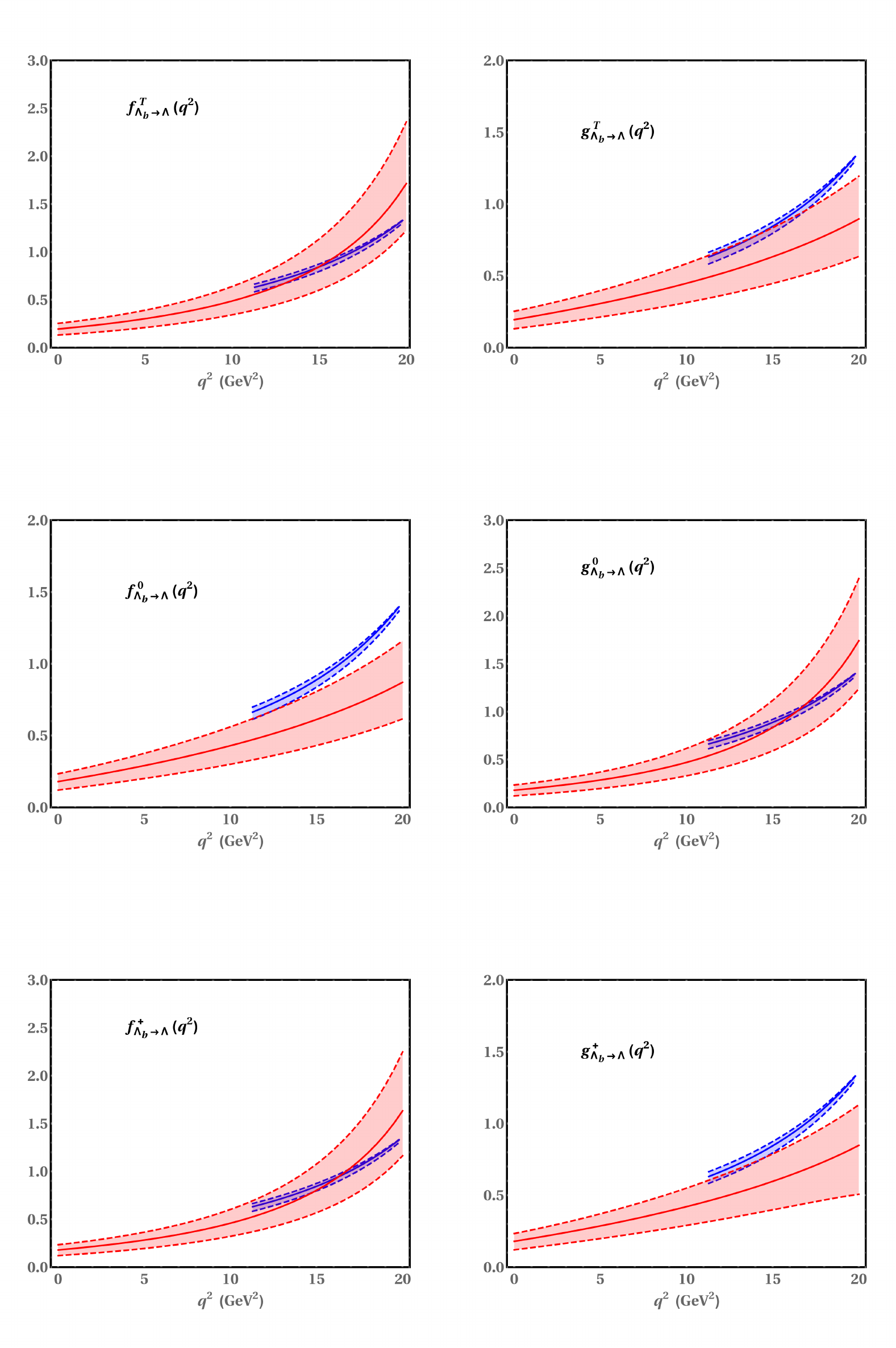}
\vspace*{0.1cm}
\caption{The $\Lambda_b \to \Lambda$ form factors induced by the (axial)-vector currents
computed from the LCSR approach at NLL accuracy and fitted to the $z$-series parameterizations.
The pink (solid) and the blue (solid) curves refer to the predictions from the LCSR with an extrapolation
and from the Lattice calculations \cite{Detmold:2012vy}, respectively, and the uncertainty bands are obtained by adding all separate
theory uncertainties in quadrature.}
\label{fig: final shape of form factors induced by V(A) currents}
\end{center}
\end{figure}

\begin{figure}
\begin{center}
\includegraphics[width=0.75 \columnwidth]{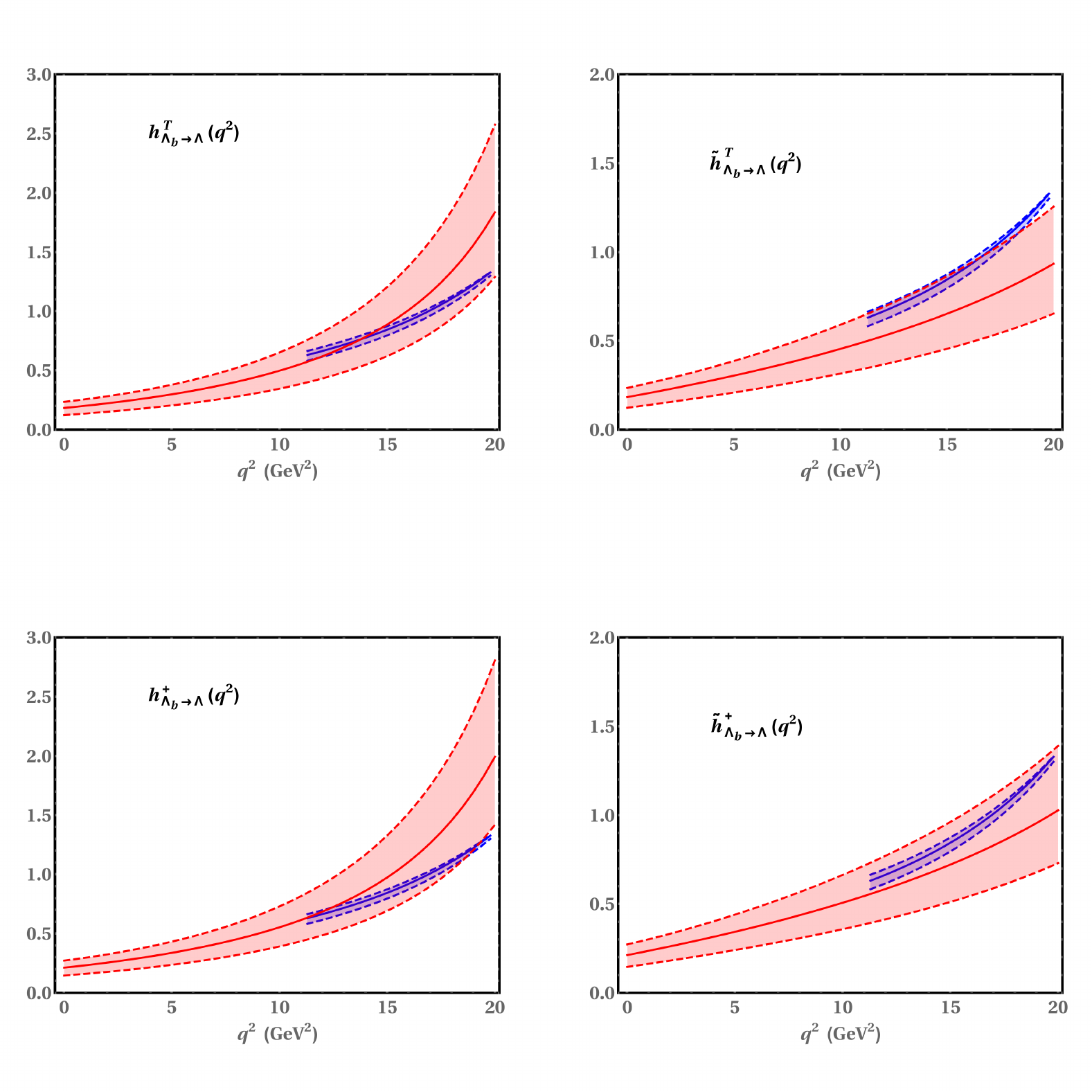}
\vspace*{0.1cm}
\caption{The $\Lambda_b \to \Lambda$ form factors induced by the (pseudo)-tensor currents
computed from the LCSR approach at NLL accuracy and fitted to the $z$-series parameterizations.
Same conventions as in Fig. \ref{fig: final shape of form factors induced by V(A) currents}.}
\label{fig: final shape of form factors induced by T(PT) currents}
\end{center}
\end{figure}

Since the lowest resonances of the  scalar and the axial-vector  channels are above
the continuum threshold $\sqrt{t_{+}}$,  it is therefore not necessary to introduce a pole factor
in the $z$-series parameterizations of the corresponding form factors.
Keeping the series expansion of the form factors to the first power of $z$-parameter
we propose the following parameterizations
\begin{eqnarray}
F_{\Lambda_b \to \Lambda}^{{\rm (I)}, \, i}(q^2)= \frac{F_{\Lambda_b \to \Lambda}^{i}(0)}{1-q^2/m_{B_s^{\ast}}^2} \,
\left \{ 1 + b_1^{i} \, \left [ z(q^2, t_0) - z(0, t_0) \right ] \right \}  \,
\end{eqnarray}
for the form factors $f_{\Lambda_b \to \Lambda}^{+, T}(q^2)$
and $ h_{\Lambda_b \to \Lambda}^{+, T}(q^2)$,
\begin{eqnarray}
F_{\Lambda_b \to \Lambda}^{{\rm (II)}, \, i}(q^2)= \frac{F_{\Lambda_b \to \Lambda}^{i}(0)}{1-q^2/m_{B_s}^2} \,
\left \{ 1 + b_1^{i} \, \left [ z(q^2, t_0) - z(0, t_0) \right ] \right \}  \,
\end{eqnarray}
for the form factor $g_{\Lambda_b \to \Lambda}^{0}(q^2)$, and
\begin{eqnarray}
F_{\Lambda_b \to \Lambda}^{{\rm (III)}, \, i}(q^2)= F_{\Lambda_b \to \Lambda}^{i}(0) \,
\left \{ 1 + b_1^{i} \, \left [ z(q^2, t_0) - z(0, t_0) \right ] \right \}  \,
\end{eqnarray}
for the form factors $f_{\Lambda_b \to \Lambda}^{0}(q^2)$, $g_{\Lambda_b \to \Lambda}^{+, T}(q^2)$,
and $ \tilde{h}_{\Lambda_b \to \Lambda}^{+, T}(q^2)$.
The shape parameters $b_1^{i}$ can be determined by matching the $z$-series parameterizations
to the NLL sum rule predictions at large hadronic recoil, i.e.,
 $0\leq q^2 \leq q_{\rm max}^2 = 8\, {\rm GeV^2}$.
The resulting form factors in the allowed kinematical region $0\leq q^2 \leq t_0$ are displayed in
Figs. \ref{fig: final shape of form factors induced by V(A) currents} and
\ref{fig: final shape of form factors induced by T(PT) currents},
 where independent calculations of these QCD form factors from Lattice determinations of
the two  HQET form factors at low hadronic recoil \cite{Detmold:2012vy}  are also presented for a comparison.

%%%%%%%%%%%
%%%%%%%%%%%

%%%%%%%%%%%
%%%%%%%%%%%

To facilitate such a comparison  we first need to perform the perturbative
matching of  the heavy-to-light currents from QCD onto HQET \cite{Manohar:2000dt}
\begin{eqnarray}
\bar s\, \gamma_{\mu} \, \left (1,  \gamma_5 \right ) \, b
&=& c_{\gamma}\,\, \bar s\, \gamma_{\mu} \, \left (1,  \gamma_5 \right ) \, h
+ c_v \,\, \bar s\, v_{\mu} \, \left (1,  - \gamma_5 \right ) \, b + ... \,, \nonumber \\
\bar s\, \sigma_{\mu \nu} \, \left (1,  \gamma_5 \right ) \, b
&=& c_{\sigma} \, \bar s\, \sigma_{\mu \nu}  \, \left (1,  \gamma_5 \right ) \, h + ... \,,
\end{eqnarray}
at leading power in $\Lambda/m_b$,
where the matching coefficients at one loop are given by
\begin{eqnarray}
c_{\gamma} &=& 1- {\alpha_s \, C_F \over 4 \, \pi} \, \left [3 \, \ln {\mu \over m_b} + 4  \right ]
+ {\cal O}(\alpha_s^2) \,, \nonumber \\
c_{v} &=&  {\alpha_s \, C_F \over 2 \, \pi}
+ {\cal O}(\alpha_s^2) \,, \nonumber \\
c_{\sigma} &=&  1- {\alpha_s \, C_F \over 4 \, \pi} \, \left [5 \, \ln {\mu \over m_b} + 4  \right ]
+ {\cal O}(\alpha_s^2) \,.
\end{eqnarray}
The HQET matrix element defined with an arbitrary Dirac structure of the leading-power  effective current
can be expressed by two Isgur-Wise functions at low hadronic recoil  \cite{Mannel:1997xy,Manohar:2000dt}
\begin{eqnarray}
\langle \Lambda(p^{\prime}, s^{\prime})|\bar s \, \Gamma \,  h| \Lambda_b (v, s) \rangle
= \bar \Lambda(p^{\prime}, s^{\prime}) \, \left [ F_1(v \cdot p^{\prime})
+ F_2(v \cdot p^{\prime}) \, \! \not  v \right ] \, \Gamma \, \Lambda_b (v, s) \,,
\end{eqnarray}
due to the heavy-quark spin symmetry. It is then straightforward to write
\begin{eqnarray}
f_{\Lambda_b \to \Lambda}^{T} &=& c_{\gamma}\, \left (F_1 - F_2  \right ) \,,  \nonumber \\
f_{\Lambda_b \to \Lambda}^{0} &=& \left ( c_{\gamma} + c_{v}  \right ) \, \left (F_1 + F_2  \right ) \,, \nonumber \\
f_{\Lambda_b \to \Lambda}^{+} &=& c_{\gamma}\,   \left (F_1 - F_2  \right ) \,,   \nonumber \\
g_{\Lambda_b \to \Lambda}^{T} &=& c_{\gamma}\, \left (F_1 + F_2  \right ) \,,  \nonumber \\
g_{\Lambda_b \to \Lambda}^{0} &=& \left ( c_{\gamma} + c_{v}  \right ) \, \left (F_1 - F_2  \right ) \,, \nonumber \\
g_{\Lambda_b \to \Lambda}^{+} &=& c_{\gamma}\,   \left (F_1 + F_2  \right ) \,,   \nonumber \\
h_{\Lambda_b \to \Lambda}^{+, T} &=& c_{\sigma} \,  \left (F_1 - F_2  \right )   \,,  \nonumber \\
\tilde{h}_{\Lambda_b \to \Lambda}^{+, T} &=& c_{\sigma} \,  \left (F_1 + F_2  \right )   \,,
\end{eqnarray}
at low recoil.
Inspection of Figs. \ref{fig: final shape of form factors induced by V(A) currents} and
\ref{fig: final shape of form factors induced by T(PT) currents} indicates that
the LCSR calculations with the aid of an  extrapolation  inspired by the $z$-series expansion
and the Lattice determinations in HQET reach a reasonable agreement in general at low hadronic recoil.
However,  the Lattice calculations \cite{Detmold:2012vy}, on the other hand,
reveal faster growing form factors of
$f_{\Lambda_b \to \Lambda}^{0}$, $g_{\Lambda_b \to \Lambda}^{+, T}$
and $\tilde{h}_{\Lambda_b \to \Lambda}^{+,T}$
but slower increasing form factors of
$f_{\Lambda_b \to \Lambda}^{+,T}$, $g_{\Lambda_b \to \Lambda}^{0}$
and  $h_{\Lambda_b \to \Lambda}^{+,T}$ at high momentum transfer squared
when confronted with the LCSR-assisted $z$-parametrization predictions.
The observed shape discrepancies might be attributed to the unaccounted power-enhanced
but $\alpha_s$-suppressed hard scattering effects, and  to the
yet unknown higher order/power corrections, to the sysmematic uncertainties
induced by the parton-hadronic quality approximation and  truncations of the $z$-series
expansion in our calculations, as well as to the power-suppressed contributions
and to the uncounted systemical uncertainties in the Lattice determinations.

We now collect the calculated form factors at zero momentum transfer $F_{\Lambda_b \to \Lambda}^{i}(0)$
and the fitted shape parameters $b_1^{i}$ in Tables \ref{tab of fitted parameters for (axial)vector FFs}
and \ref{tab of fitted parameters for (pseudo)tensor FFs},
where the numerically important uncertainties
due to  variations  of the theory input parameters are also displayed.

%%%%%%%%%%%%%%%%%%%%%%%%%%%%%%%%%%%%
\begin{table}[t!bph]
\begin{center}
\begin{tabular}{|c|c|c|c|c|c|c|c|}
  \hline
  \hline
  &  &  &  &  &  &  &  \\
 \,\,  Parameter \,\, & \,\, Central value \,\, & \,\, $\phi_4(\omega)$ \,\,
 & \,\, $\omega_0$ \,\, & \,\, $ \left \{ \mu, \nu \right \}$  \,\,
  & \,\, $ \left \{ \mu _h, \nu^{\prime} \right \}$ \,\, & \,\, $M^2$ \,\, & \,\, $s_0$ \,\, \\
   &  &  &  &  &  &  &  \\
   \hline
   &  &  &  &  &  &  &  \\
  $f_{\Lambda_b \to \Lambda}^{T}(0)$ & $0.20$ & $-$ & ${}^{-0.04}_{+0.04}$ & ${}^{-0.03}_{-0.02}$
  & ${}^{+0.02}_{-0.01}$  & ${}^{+0.01}_{-0.01}$  &  ${}^{+0.00}_{-0.00}$  \\
   &  &  &  &  &  &  &  \\
  $b_1^{f^{T}_{\Lambda_b \to \Lambda}}$ & $-6.82$ & ${}^{-2.93}_{-1.87}$ & ${}^{-1.16}_{+1.02}$
  & ${}^{+0.12}_{+2.22}$ &  ${}^{-0.30}_{+0.48}$  & ${}^{+0.22}_{-0.34}$ & ${}^{+0.25}_{-0.28}$ \\
   &  &  &  &  &  &  &  \\
  $b_1^{g^{T}_{\Lambda_b \to \Lambda}}$ & $-13.66$  & ${}^{-3.72}_{-2.36}$ & ${}^{-1.46}_{+1.30}$
  & ${}^{+0.15}_{-2.81}$ & ${}^{-0.38}_{+0.61}$ &  ${}^{+0.28}_{-0.42}$ & ${}^{+0.32}_{-0.35}$ \\
   &  &  &  &  &  &  &  \\
   \hline
   &  &  &  &  &  &  &  \\
  $f_{\Lambda_b \to \Lambda}^{0}(0)$ & $0.18$ & $-$ & ${}^{-0.04}_{+0.04}$  & ${}^{-0.03}_{+0.02}$
  & ${}^{+0.02}_{-0.01}$ & ${}^{+0.01}_{-0.01}$ & ${}^{+0.00}_{-0.00}$  \\
   &  &  &  &  &  &  &  \\
  $b_1^{f^{0}_{\Lambda_b \to \Lambda}}$ & $-14.59$ & ${}^{-3.91}_{-2.51}$  & ${}^{-1.61}_{+1.40}$
  &  ${}^{+0.11}_{-3.51}$ & ${}^{-0.36}_{+0.60}$ & ${}^{+0.33}_{-0.51}$ & ${}^{+0.35}_{-0.40}$  \\
   &  &  &  &  &  &  &  \\
  $b_1^{g^{0}_{\Lambda_b \to \Lambda}}$ & $-7.43$ & ${}^{-3.06}_{-1.97}$  & ${}^{-1.26}_{+1.11}$
   & ${}^{+0.10}_{-2.75}$ & ${}^{-0.28}_{+0.48}$ & ${}^{+0.26}_{-0.40}$  & ${}^{+0.28}_{-0.31}$   \\
   &  &  &  &  &  &  &  \\
    \hline
   &  &  &  &  &  &  &  \\
  $f_{\Lambda_b \to \Lambda}^{+}(0)$ & $0.18$ & $-$ & ${}^{-0.04}_{+0.04}$ & ${}^{-0.03}_{-0.02}$
  & ${}^{+0.02}_{-0.01}$ & ${}^{+0.01}_{-0.01}$ &  ${}^{+0.00}_{-0.00}$  \\
   &  &  &  &  &  &  &  \\
  $b_1^{f^{+}_{\Lambda_b \to \Lambda}}$ & $-7.17$ & ${}^{-3.07}_{-1.97}$ & ${}^{-1.24}_{+1.09}$
  & ${}^{+0.09}_{-2.53}$ & ${}^{-0.29}_{+0.49}$ & ${}^{+0.25}_{-0.38}$  &  ${}^{+0.28}_{-0.31}$ \\
   &  &  &  &  &  &  &  \\
  $b_1^{g^{+}_{\Lambda_b \to \Lambda}}$ & $-14.10$ & ${}^{-3.88}_{-2.48}$ & ${}^{-1.56}_{+1.38}$
  & ${}^{+0.11}_{-3.19}$ & ${}^{-0.36}_{+0.62}$  & ${}^{+0.32}_{-0.47}$  &  ${}^{+0.35}_{-0.38}$ \\
   &  &  &  &  &  &  &  \\
   \hline
   \hline
\end{tabular}
\end{center}
\caption{Summary of the calculated form factors induced
by the (axial)-vector weak transition currents  at $q^2=0$ and the fitted shape parameters $b_1^{i}$
with the uncertainties from the variations of various input parameters.}
\label{tab of fitted parameters for (axial)vector FFs}
\end{table}
%%%%%%%%%%%%%%%%%%%%%%%%

%%%%%%%%%%%%%%%%%%%%%%%%%%%%%%%%%%%%
\begin{table}[t!bph]
\begin{center}
\begin{tabular}{|c|c|c|c|c|c|c|c|}
  \hline
  \hline
  &  &  &  &  &  &  &  \\
 \,\,  Parameter \,\, & \,\, Central value \,\, & \,\, $\phi_4(\omega)$ \,\,
 & \,\, $\omega_0$ \,\, & \,\, $ \left \{ \mu, \nu \right \}$  \,\,
  & \,\, $ \left \{ \mu _h, \nu^{\prime} \right \}$ \,\, & \,\, $M^2$ \,\, & \,\, $s_0$ \,\, \\
   &  &  &  &  &  &  &  \\
   \hline
   &  &  &  &  &  &  &  \\
  $h_{\Lambda_b \to \Lambda}^{T}(0)$ & $0.18$ & $-$ & ${}^{-0.04}_{+0.04}$  & ${}^{-0.03}_{-0.01}$
  & ${}^{+0.00}_{+0.01}$ & ${}^{+0.01}_{-0.01}$  &  ${}^{+0.00}_{-0.00}$ \\
   &  &  &  &  &  &  &  \\
  $b_1^{h^{T}_{\Lambda_b \to \Lambda}}$ & $-8.26$ & ${}^{-3.08}_{-1.99}$ & ${}^{-1.31}_{+1.14}$
   & ${}^{+0.53}_{-3.52}$ & ${}^{-0.60}_{+0.76}$ & ${}^{+0.28}_{-0.44}$ & ${}^{+0.29}_{-0.33}$  \\
   &  &  &  &  &  &  &  \\
  $b_1^{\tilde{h}^{T}_{\Lambda_b \to \Lambda}}$ & $-15.49$ & ${}^{-3.90}_{-2.52}$
  & ${}^{-1.66}_{+1.45}$  & ${}^{+0.67}_{-4.45}$ & ${}^{-0.76}_{+0.97}$  & ${}^{+0.36}_{-0.56}$  &   ${}^{+0.37}_{-0.41}$   \\
   &  &  &  &  &  &  &  \\
   \hline
   &  &  &  &  &  &  &  \\
  $h_{\Lambda_b \to \Lambda}^{+}(0)$ & $0.21$ & $-$   & ${}^{-0.05}_{+0.05}$  & ${}^{-0.00}_{-0.00}$
  & ${}^{+0.00}_{+0.01}$ & ${}^{+0.01}_{-0.01}$ & ${}^{+0.00}_{-0.00}$  \\
   &  &  &  &  &  &  &  \\
  $b_1^{h^{+}_{\Lambda_b \to \Lambda}}$ & $-7.51$ & ${}^{-2.84}_{-1.81}$ & ${}^{-1.15}_{+1.03}$
   & ${}^{+0.52}_{-2.81}$  & ${}^{-0.50}_{+0.66}$ & ${}^{+0.23}_{-0.35}$ &  ${}^{+0.26}_{-0.28}$ \\
   &  &  &  &  &  &  &  \\
  $b_1^{\tilde{h}^{+}_{\Lambda_b \to \Lambda}}$ & $-14.53$ & ${}^{-3.61}_{-2.29}$  & ${}^{-1.46}_{+1.30}$
    & ${}^{+0.65}_{-3.57}$ & ${}^{-0.64}_{+0.84}$ & ${}^{+0.29}_{-0.45}$  & ${}^{+0.32}_{-0.36}$   \\
   &  &  &  &  &  &  &  \\
   \hline
   \hline
\end{tabular}
\end{center}
\caption{Summary of the calculated form factors induced
by the (pseudo)-tensor weak transition currents  at $q^2=0$ and the fitted shape parameters $b_1^{i}$
with the uncertainties from the variations of various input parameters.}
\label{tab of fitted parameters for (pseudo)tensor FFs}
\end{table}
%%%%%%%%%%%%%%%%%%%%%%%%

Several comments on the numerical results obtained above are in order.

\begin{itemize}
\item {The dominant theory uncertainty for the  form factors at $q^2=0$ computed from the NLL LCSR is due to
the variation of the $\omega_0$ parameter entering the $\Lambda_b$-baryon DA $\phi_4(\omega, \mu_0)$,
while the most significant sources of the theory errors for the shape parameters $b_1^{i}$ are from the
different parameterizations of $\phi_4(\omega, \mu_0)$ and from the variations of the renormalization scale $\mu$
and of the factorization scale $\nu$. }
\item {Large-recoil symmetry violation effects for the $\Lambda_b \to \Lambda$ form factors are found to
be relatively small, at the level of 20 \%, albeit with the observed substantial perturbative QCD corrections
to the form factors themselves. This can be readily understood from the fact that the NLO perturbative contributions to
the $\Lambda_b \to \Lambda$ form factors are dominated by the hard-collinear corrections which preserve
the large-recoil symmetry in the heavy quark limit. }
\item {Large discrepancies of the slope parameters are observed for the two form factors defined by
the matrix elements of the two weak currents  with  the same helicity projections but with the opposite space-time parities,
e.g., $f_{\Lambda_b \to \Lambda}^{T}$ and $g_{\Lambda_b \to \Lambda}^{T}$. This is in a nutshell due to the distinct
analytical structures of two types of form factors below the branch cut in the complex-$q^2$ plane which lead to
the different $z$-series parameterizations adopted in the fitting programmes.}
\end{itemize}

\section{Phenomenological applications}
\label{section: phenomenologies}

\begin{figure}[h]
\begin{center}
\includegraphics[width=0.46 \textwidth]{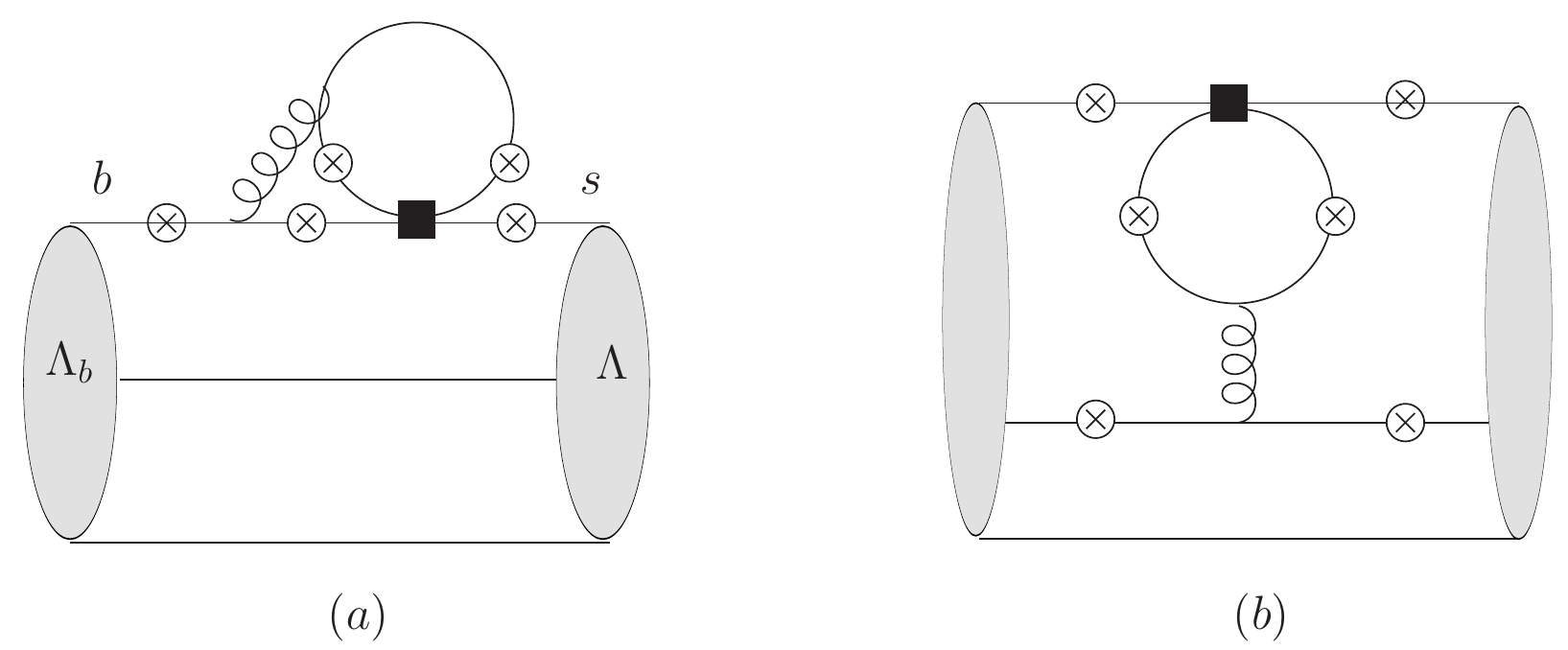}\mbox{ } \hspace{0.6 cm}
\includegraphics[width=0.46 \textwidth]{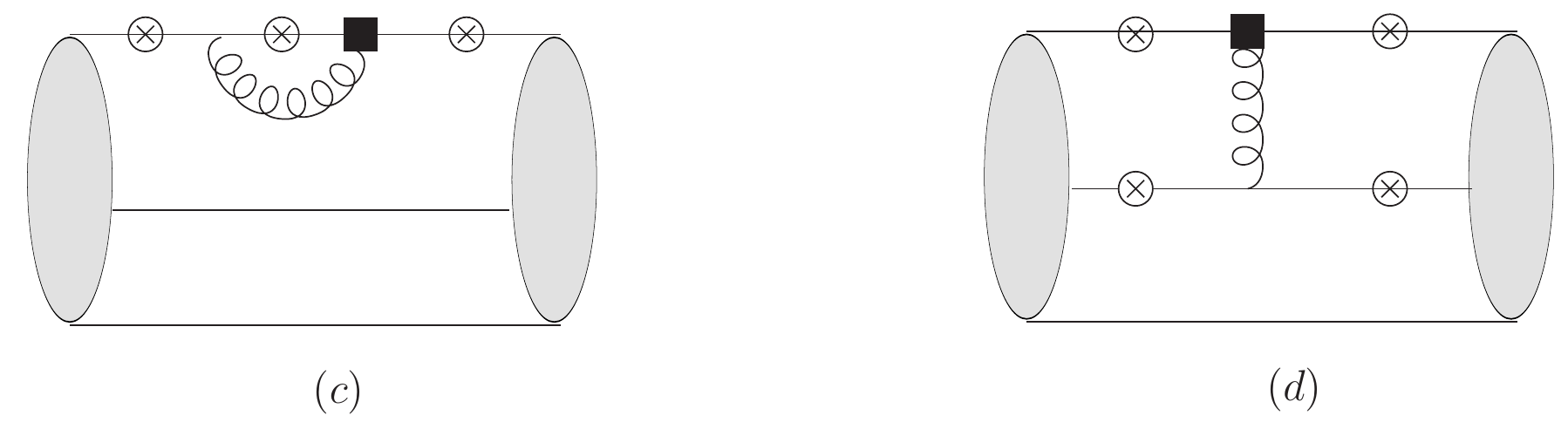}\mbox{ } \\
\vspace{0.5 cm}
\includegraphics[width=0.46 \textwidth]{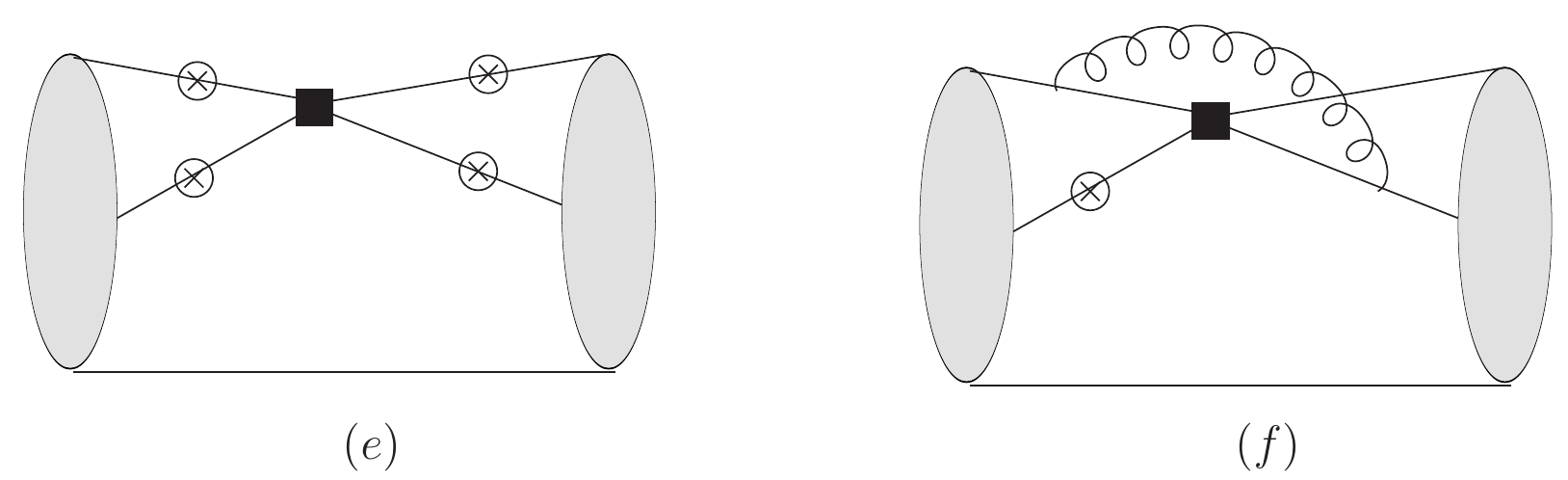}\mbox{ }
\caption{Various non-factorizable diagrams contributed to the $\Lambda_b \to \Lambda \, \ell^{+} \ell^{-}$
decays. The crossed circles indicate possible insertions of the virtual photon line
and the black squares stand for the hadronic operator vertices.
Taken from \cite{Wang:2014jya}.}
\label{NLO diagrams of Lambda_b to Lambda ll}
\end{center}
\end{figure}

In this section we aim at exploring phenomenological applications of the calculated $\Lambda_b \to \Lambda$
form factors which serve as fundamental ingredients for the theory description of the electro-weak penguin
induced $\Lambda_b \to \Lambda \, \ell^{+} \ell^{-}$ decays.
QCD dynamics of the hadronic $\Lambda_b \to \Lambda \, \ell^{+} \ell^{-}$ decay amplitude is, however,
more complicated  due to the non-factorizable strong interaction effects which arise from QED corrections
to the matrix elements of the four-quark operators and the gluonic penguin operator in the weak effective  Hamiltonian.
Some typical non-factorizable contributions to the  $\Lambda_b \to \Lambda \, \ell^{+} \ell^{-}$ matrix elements
at ${\cal O}(\alpha_s)$ are presented in Fig. \ref{NLO diagrams of Lambda_b to Lambda ll},
in analogy to the counterpart   $B \to K^{\ast} \, \ell^{+} \ell^{-}$ decays discussed in \cite{Beneke:2001at}.
It is evident that the spectator interaction effects displayed in the  diagrams  (b) and (d)
and the weak annihilation contributions shown in (e) and (f) cannot be computed with  QCD factorization
formalism  described in \cite{Beneke:2001at} and some non-perturbative QCD approaches are in demand to deal with
such non-local hadronic matrix elements. We will restrict ourselves to the factorizable contributions to
the $\Lambda_b \to \Lambda \, \ell^{+} \ell^{-}$ decay amplitude, at ${\cal O}(\alpha_s^{0})$, in this work, and leave a
systematic treatment of the non-form-factor corrections for a future work.

The double differential decay distribution of $\Lambda_b \to \Lambda \, \ell^{+} \ell^{-}$  in terms of
the momentum  transfer squared $q^2$ and the angle $\theta$ between the positively changed lepton and the
$\Lambda$-baryon in the rest frame of the lepton pair is given by \cite{Feldmann:2011xf}
\begin{eqnarray}
\frac{d \Gamma(\Lambda_b \to \Lambda \, \ell^{+} \ell^{-})}{d q^2 \, d \cos \theta}
={3 \over 8} \, \left [ H_T(q^2) \, \left ( 1 + \cos^2 \theta \right )
+ 2 \, H_A(q^2) \, \, \cos \theta + 2 \, H_L(q^2) \, \left ( 1 - \cos^2 \theta \right ) \, \right ] ,
\hspace{0.8 cm}
\end{eqnarray}
where in the factorization limit the helicity amplitudes can be computed as
\begin{eqnarray}
H_T(q^2) &=& {\cal N} \, q^2 \, { \lambda^{1/2}(m_{\Lambda_b}^2, m_{\Lambda}^2, q^2)  \over
96 \, \pi^3 \, m_{\Lambda_b}^3 }  \,  \nonumber \\
&& \bigg [ s_{-}  \, \bigg ( \bigg |  C_{9}^{\rm eff}(q^2) \, f_{\Lambda_b \to \Lambda}^{T}
+ { 2 \, m_{\Lambda_b} \, (m_{\Lambda_b}+m_{\Lambda})  \over q^2}
\, C_{7}^{\rm eff} \, h_{\Lambda_b \to \Lambda}^{T} \bigg |^2
 +  \left |C_{10} \, f_{\Lambda_b \to \Lambda}^{T} \right |^2 \, \bigg )  \nonumber \\
&& \hspace{0.3 cm} +  \, s_{+}  \, \bigg ( \bigg |  C_{9}^{\rm eff}(q^2) \, g_{\Lambda_b \to \Lambda}^{T}
+ { 2 \, m_{\Lambda_b} \, (m_{\Lambda_b}-m_{\Lambda})  \over q^2}
\, C_{7}^{\rm eff} \, \tilde{h}_{\Lambda_b \to \Lambda}^{T} \bigg |^2
 +  \left |C_{10} \, g_{\Lambda_b \to \Lambda}^{T} \right |^2 \, \bigg )   \bigg ] \,, \hspace{1.0 cm}
\\
H_A(q^2) &=& - {\cal N} \, q^2 \, { \lambda(m_{\Lambda_b}^2, m_{\Lambda}^2, q^2)  \over
48 \, \pi^3 \, m_{\Lambda_b}^3 }  \,  \nonumber \\
&& {\rm Re} \bigg [ \bigg ( C_{9}^{\rm eff}(q^2) \, f_{\Lambda_b \to \Lambda}^{T}
+ { 2 \, m_{\Lambda_b} \, (m_{\Lambda_b}+m_{\Lambda})  \over q^2}
\, C_{7}^{\rm eff} \, h_{\Lambda_b \to \Lambda}^{T}  \bigg )^{\ast}
\left (C_{10} \, g_{\Lambda_b \to \Lambda}^{T} \right ) \,  \nonumber \\
&& \hspace{0.8 cm} + \, \bigg (  C_{9}^{\rm eff}(q^2) \, g_{\Lambda_b \to \Lambda}^{T}
+ { 2 \, m_{\Lambda_b} \, (m_{\Lambda_b}-m_{\Lambda})  \over q^2}
\, C_{7}^{\rm eff} \, \tilde{h}_{\Lambda_b \to \Lambda}^{T} \bigg )^{\ast}
\left ( C_{10} \, f_{\Lambda_b \to \Lambda}^{T} \,  \right ) \, \bigg ] \,,
\\
H_L(q^2) &=& {\cal N} \,   { \lambda^{1/2}(m_{\Lambda_b}^2, m_{\Lambda}^2, q^2)  \over
192 \, \pi^3 \, m_{\Lambda_b}^3 }  \,  \nonumber \\
&& \bigg [ s_{-}  \, (m_{\Lambda_b}+m_{\Lambda})^2 \,
\bigg ( \bigg |  C_{9}^{\rm eff}(q^2) \, f_{\Lambda_b \to \Lambda}^{+}
+ { 2 \, m_{\Lambda_b}  \over m_{\Lambda_b}+m_{\Lambda}  }
\, C_{7}^{\rm eff} \, h_{\Lambda_b \to \Lambda}^{+} \bigg |^2
 +  \left |C_{10} \, f_{\Lambda_b \to \Lambda}^{+} \right |^2 \, \bigg )  \nonumber \\
&& \hspace{0.3 cm} + \, s_{+}  \,
(m_{\Lambda_b} - m_{\Lambda})^2  \,
\bigg ( \bigg |  C_{9}^{\rm eff}(q^2) \, g_{\Lambda_b \to \Lambda}^{+}
+  { 2 \, m_{\Lambda_b}  \over m_{\Lambda_b} - m_{\Lambda}  }
\, C_{7}^{\rm eff} \, \tilde{h}_{\Lambda_b \to \Lambda}^{+} \bigg |^2
+  \left |C_{10} \, g_{\Lambda_b \to \Lambda}^{+} \right |^2 \, \bigg )   \bigg ] \,, \nonumber \\
\end{eqnarray}
with
\begin{eqnarray}
 {\cal N} ={G_F^2 \, \alpha_{em}^2 \over 8 \, \pi^2 } \, \left | V_{ts} \, V_{tb} \right |^2 \,,  & \qquad &
 \lambda(a,b,c)=a^2 + b^ 2+ c^2 - 2\, a b- 2 \, a \, c -2 \, b  c \,. % \nonumber  \\
% s_{\pm} = (m_{\Lambda_b} \pm  m_{\Lambda})^2 - q^2 \,.  & \qquad &
 \end{eqnarray}
The detailed expressions for the effective Wilson coefficients $ C_{9}^{\rm eff}(q^2)$
and  $ C_{7}^{\rm eff}$ in the NDR scheme with anti-commuting $\gamma_5$ can be found in \cite{Beneke:2001at}.

\begin{figure}
\begin{center}
\includegraphics[width=1.0 \columnwidth]{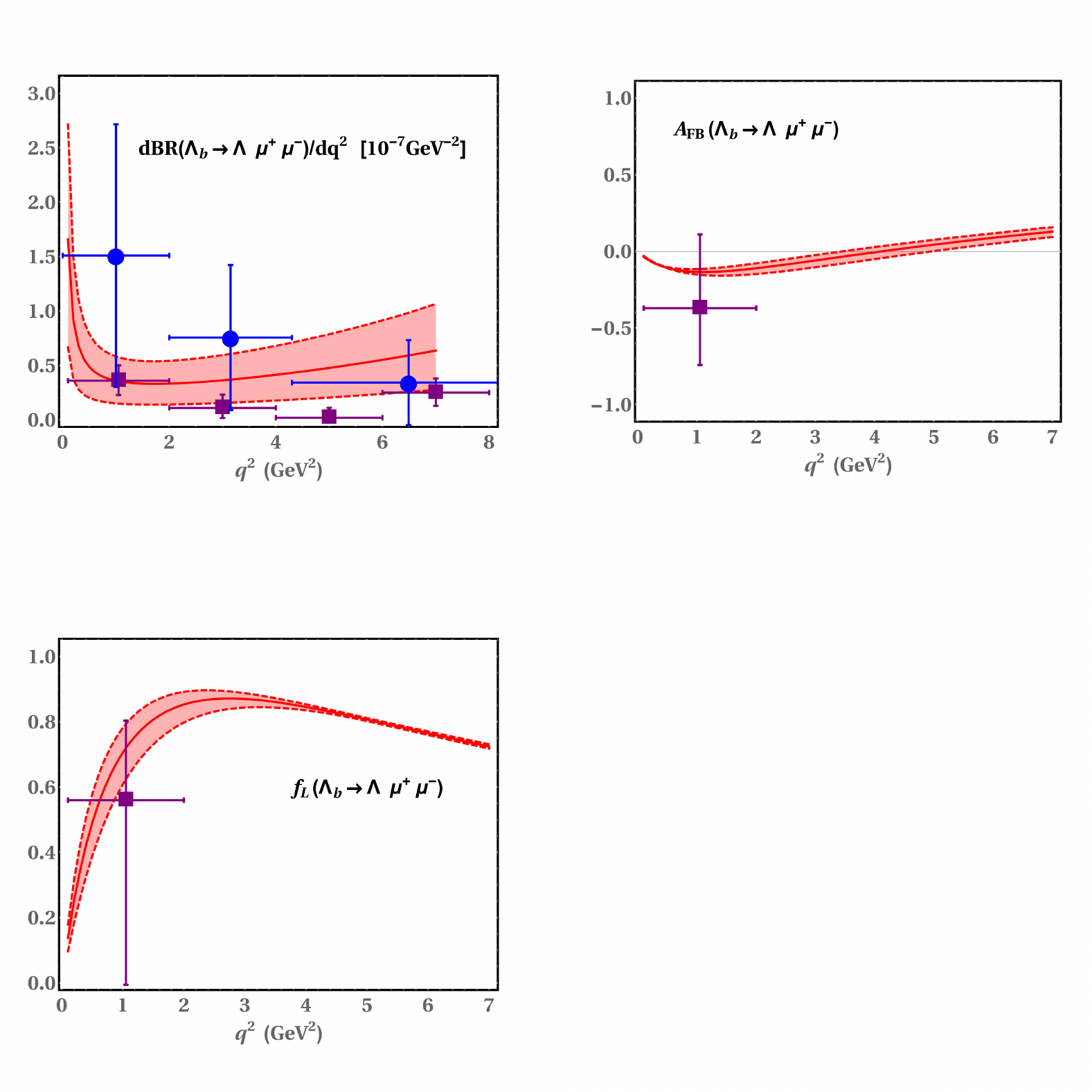}
\vspace*{0.1cm}
\caption{The differential branching fraction,
the leptonic forward-backward asymmetry and the longitudinal polarization fraction
of the di-lepton system for $\Lambda_b \to \Lambda \, \ell^{+} \ell^{-}$  as functions of $q^2$
in the factorization limit. The solid (pink) curve corresponds to the NLL sum rule predictions
with the central input and the shaded region (pink) indicates the theory uncertainties from
the calculated form factors. The experimental data bins are taken from LHCb \cite{Aaij:2015xza}
(purple squares) and CDF  \cite{CDF note 10994} (blue full circles). }
\label{fig: differential observables}
\end{center}
\end{figure}

Evaluating the helicity amplitudes with the form factors computed from the NLL LCSR
obtained in the above  yields the differential branching fraction of
$\Lambda_b \to \Lambda \, \ell^{+} \ell^{-}$  as a function of $q^2$ plotted in
Fig. \ref{fig: differential observables} and the partially
integrated decay rate over the $q^2$ intervals from \cite{Aaij:2015xza}
displayed in Table \ref{tab of binned observables}.
The theory predictions are also confronted with the experimental measurements from CDF \cite{CDF note 10994}
and LHCb \cite{Aaij:2015xza}. The LHCb data except for the first $q^2$-bin
are found to be systematically lower than the theory predictions at large hadronic recoil,
while the sizeable uncertainties of the CDF measurements prevent us from drawing a definite conclusion.

%%%%%%%%%%%
%%%%%%%%%%%

Following  \cite{Aaij:2015xza} we further consider the forward-backward asymmetry
and the longitudinal polarization fraction of the di-lepton system
\begin{eqnarray}
A_{\rm FB}(q^2) &=& \frac{ \int_{0}^{1} \, {\rm d} \cos \theta \,\,
\frac{d \Gamma(\Lambda_b \to \Lambda \, \ell^{+} \ell^{-})}{d q^2 \, d \cos \theta}
- \int_{-1}^{0} \, {\rm d}  \cos \theta \,\,
\frac{d \Gamma(\Lambda_b \to \Lambda \, \ell^{+} \ell^{-})}{d q^2 \, d \cos \theta} }
{d \Gamma(\Lambda_b \to \Lambda \, \ell^{+} \ell^{-})/d q^2} \,\,, \nonumber \\
f_L(q^2) &=& {H_L(q^2) \over H_L(q^2) + H_T(q^2)} \,,
\end{eqnarray}
where the definition of $A_{\rm FB}(q^2)$ differs  from \cite{Aaij:2015xza}
due to the distinct convention of the $\theta$  angle.
We plot the $q^2$ dependence  of the differential forward-backward asymmetry and
the longitudinal polarization fraction in Fig. \ref{fig: differential observables},
and collect the theory predictions for the binned distributions of these two observables
in Table  \ref{tab of binned observables}.

%%%%%%%%%%%%%%%%%%%%%%%%%%%%%%%%%%%%
\begin{table}[t!bph]
\begin{center}
\begin{tabular}{|c|c|c|c|}
  \hline
  \hline
  &&& \\
  $[q^2_{\rm min}, q^2_{\rm max}] \,$ & $d {\rm BR}/dq^2$ \, ($10^{-7} \, {\rm GeV}^{-2}$)
  & $ A_{\rm FB}$ & $f_L$ \\
  &&& \\
  $ \left ({\rm GeV^2} \right)$ &this work  \hspace{1 cm} LHCb  & this work  \hspace{1 cm} LHCb
  & this work  \hspace{1 cm} LHCb \\
  &&& \\
  \hline
  &&& \\
  $[0.1, 2.0]$ & $0.45^{+0.28}_{-0.26}$ \hspace{0.5 cm} $0.36^{+0.14}_{-0.13}$
  & $-0.10^{+0.01}_{-0.01}$ \hspace{0.5 cm} $-0.37^{+0.48}_{-0.37}$
  & $0.57^{+0.08}_{-0.10}$ \hspace{0.8 cm} $0.56^{+0.24}_{-0.57}$  \\
  &&& \\
  $[2.0, 4.0]$ &  $0.37^{+0.23}_{-0.21}$ \hspace{0.5 cm} $0.11^{+0.12}_{-0.09}$
  & $-0.06^{+0.04}_{-0.04}$  \hspace{1.5 cm}  $-$ \hspace{0.5 cm}
  &  $0.86^{+0.02}_{-0.03}$  \hspace{1.5 cm}  $-$ \hspace{0.5 cm} \\
  &&& \\
  $[4.0, 6.0]$ & $0.48^{+0.31}_{-0.27}$  \hspace{0.5 cm} $0.02^{+0.09}_{-0.01}$
  & $0.05^{+0.03}_{-0.04}$  \hspace{1.5 cm}  $-$ \hspace{0.2 cm}
  &  $0.80^{+0.00}_{-0.00}$  \hspace{1.5 cm}  $-$ \hspace{0.5 cm} \\
  &&& \\
  $[1.1, 6.0]$ & $0.41^{+0.26}_{-0.23}$  \hspace{0.5 cm} $0.09^{+0.06}_{-0.05}$
  & $-0.02^{+0.03}_{-0.04}$  \hspace{1.5 cm}  $-$ \hspace{0.5 cm}
  &  $0.83^{+0.02}_{-0.02}$  \hspace{1.5 cm}  $-$ \hspace{0.5 cm} \\
    &&& \\
  \hline
  \hline
\end{tabular}
\end{center}
\caption{Summary of the theory predictions for the binned distributions of the branching fraction,
the forward-backward asymmetry and the longitudinal polarization fraction.
We also present the  experimental data bins from LHCb \cite{Aaij:2015xza} for a comparison,
where various experimental uncertainties are added in quadrature.}
\label{tab of binned observables}
\end{table}
%%%%%%%%%%%%%%%%%%%%%%%%

Several comments on the numerical results computed in the above are in order.
\begin{itemize}
\item{In contrast to the $B \to K^{\ast} \, \ell^{+} \ell^{-}$ decays, the theory uncertainty
 of the leptonic forward-backward asymmetry at the zero crossing point
is not reduced compared to that at a different value of $q^2$.
This can be readily understood from the fact that $A_{\rm FB}$ is not an optimized
observable which is insensitive to the soft form factors in the former case,
while it becomes an  optimized observable in the latter case due to a single
soft form factor governing the strong interaction dynamics of the $\Lambda_b \to \Lambda$
form factors in the  SCET limit. The location of the zero-crossing point of $A_{\rm FB}$
is determined as
\begin{eqnarray}
q_0^2= 4.1^{+0.9}_{-0.7} \, {\rm GeV^2} \,. \nonumber
\end{eqnarray}
}
\item{The uncertainty band of the longitudinal polarization fraction $f_L(q^2)$ shown in
Fig. \ref{fig: differential observables} indicates rather interesting features of the different
dominant mechanisms contributing to $f_L(q^2)$ at different momentum transfer.
At very large hadronic recoil $q^2  \ll 1 \, {\rm GeV^2}$ the longitudinal helicity amplitude $H_{L}$
is strong suppressed compared to the transverse amplitude   $H_{T}$ which   receives a large contribution
from the photon pole. This indicates that $f_L(q^2)$ at very large recoil receives a suppression
factor of $q^2/m_{\Lambda_b}^2$ and the resulting theory uncertainty is also negligible.
In the vicinity of the zero-crossing point of $A_{\rm FB}$,   both helicity amplitudes $H_{L}$
and $H_{T}$ will be dominated by the contribution from the semileptonic operator $O_{10}$.
The longitudinal polarization fraction $f_L(q^2)$ is then, to a large extent, determined by a unique form-factor ratio
$f_{\Lambda_b \to \Lambda}^{+}(q^2)/f_{\Lambda_b \to \Lambda}^{T}(q^2)$ which suffers from a much smaller theory
uncertainty compared to the other two ratios $h_{\Lambda_b \to \Lambda}^{T}(q^2)/f_{\Lambda_b \to \Lambda}^{T}(q^2)$
and $h_{\Lambda_b \to \Lambda}^{+}(q^2)/f_{\Lambda_b \to \Lambda}^{T}(q^2)$ that will also play an essential role
in determining the value of  $f_L(q^2)$ for generical momentum transfer. The most significant uncertainty
of the latter two ratios is induced by the variation of the renormalization and the factorization scales.
Based upon the above discussion, we conclude that the theory prediction of $f_L(q^2)$ will involve
a sizeable uncertainty only in the region $1 \, {\rm GeV^2} < q^2 < q_0^2$ displayed
in Fig. \ref{fig: differential observables}. }
\item {The theory prediction of the differential $q^2$ distribution shown in  Fig. \ref{fig: differential observables}
involves a large  uncertainty due to the sensitivity to the $\Lambda_b \to \Lambda$ form factors.
To reduce the most important theory uncertainty from the poorly known shape parameter $\omega_0$
in the $\Lambda_b$-baryon DA $\phi_4(\omega, \mu_0)$, one can introduce  an optimized observable,
the normalized differential  $q^2$ distribution, in analogy to  that in $B \to \pi \ell \nu$ \cite{Wang:2015vgv}.
It is however not the main objective of this work to explore the rich phenomenology encoded in the angular distributions
of  $\Lambda_b \to \Lambda \, \ell^{+} \ell^{-}$ emphasizing on the implications of optimized observables
for new physics hunting.
}
\end{itemize}

\section {Concluding discussion}
\label{section: conclusion}

In this paper we have performed, for this first time,  perturbative QCD corrections to the $\Lambda_b \to \Lambda$
form factors from the LCSR with the $\Lambda_b$-baryon DA at NLL accuracy.
Applying the method of regions we have extracted both the hard coefficients and the jet functions entering the
factorization formulae for the vacuum-to-$\Lambda_b$-baryon correlation functions  at one loop.
In particular, we have verified a complete cancellation of the factorization-scale dependence
for the factorized expressions of the considered correlation functions by computing the one-loop corrections
to the $\Lambda_b$-baryon DA in QCD manifestly.
Also, we demonstrated at the diagrammatic level that QCD factorization of the vacuum-to-$\Lambda_b$-baryon
correction functions with an arbitrary weak vertex can only depend on  a universal jet function
at leading power in $\Lambda/m_b$.
Employing the RG evolution equations in momentum space and distinguishing the renormalization and the
factorization scales, we further achieved the NLL resummation improved factorization formulae
for the correlation functions defined with both the (axial)-vector and the (pseudo)-tensor weak currents.
Making use of the parton-hadron duality approximation  and implementing the continuum subtraction,
we further obtained the NLL QCD sum rules of the $\Lambda_b \to \Lambda$ form factors at large hadronic recoil.
Since we concentrate on factorization of the correlation functions at leading power in $\Lambda/m_b$,
we do not take into account the numerically insignificant contribution corresponding to
the matrix element of the ``$B$-type" SCET  current, which can be computed with  LCSR constructed
from the same correlation functions at sub-leading power
or from the correlation functions with the ``wrong" light-cone projector acting
on the interpolating current of the $\Lambda$-baryon \cite{Feldmann:2011xf}.

Proceeding with the obtained NLL sum rules on the light cone, we carried out an exploratory numerical analysis
of the $\Lambda_b \to \Lambda$ form factors, putting an emphasis on the various sources of
perturbative and systematic uncertainties. To gain a better control of the shape parameter
$\omega_0$ for the $\Lambda_b$-baryon DA $\phi_4(\omega, \mu_0)$, the  prediction
of $\Lambda_b \to p$ form factor $f_{\Lambda_b \to p}^{+}(0)$ from the LCSR with the nucleon
DA and the ${\rm SU}(3)$ flavour symmetry relation were taken as theory input in the matching determination
of $\omega_0$. In analogy to the $B \to \pi$ form factors, the sum rules of $\Lambda_b \to \Lambda$ form factors
are  not only sensitive to  the shape parameter $\omega_0$ but also to the specific behavior of $\phi_4(\omega, \mu_0)$
at small $\omega$. Of particular phenomenological interest  are that the perturbative ${\cal O}(\alpha_s)$
corrections result in a significant ($\sim 50 \, \%$) reduction  of the tree-level sum rule predictions
and the dominant one-loop correction is from the NLO jet function instead of the NLO hard functions
entering the sum rules of  the $\Lambda_b \to \Lambda$ form factors.
Such observations evidently highlight the importance of the perturbative  matching calculation
at the hard-collinear scale as accomplished in this work.
Employing the $z$-series expansion, we extrapolated the LCSR predictions of the form factors
toward large momentum transfer  where our predictions are already confronted with the
Lattice determinations of  two HQET form factors. Expressing the QCD transition form factors in terms of
the  Isgur-Wise functions at low hadronic recoil, we observed a reasonable
agreement for the predicted form factors at large momentum transfer between two independent calculations,
albeit with the perceivable discrepancies on the $q^2$ shapes  of  the $\Lambda_b \to \Lambda$ form factors.
In addition, the large-energy symmetry breaking effects for the form factors were found to be relatively small
at one loop, since the NLO QCD corrections to the sum rules are dominated  by the hard-collinear corrections
preserving  the symmetry relations.

Having at our disposal the theory predictions for the $\Lambda_b \to \Lambda$ form factors,
we investigated their phenomenological applications to the electro-weak penguin decays
$\Lambda_b \to \Lambda \, \ell^{+} \ell^{-}$ in the factorization limit.
The calculated differential $q^2$ distribution in $\Lambda_b \to \Lambda \, \ell^{+} \ell^{-}$
turned out to be systematically lower than the LHCb measurements, except for the first data bin.
We further computed the forward-backward asymmetry and the longitudinal polarization fraction
for the di-lepton system which are comparable to the LHCb data for the lowest $q^2$ bin.
The  longitudinal polarization fraction $f_L(q^2)$ was found to be of particular phenomenological interest
due to a large cancellation of the theory uncertainties for the $\Lambda_b \to \Lambda$ form factors.

The heavy-to-light baryonic form factors are apparently not sufficient to provide a complete
description of  the strong interaction dynamics involved in $\Lambda_b \to \Lambda \, \ell^{+} \ell^{-}$
due to the non-factorizable contributions induced by the QED corrections to the matrix elements of hadronic
operators in the weak Hamiltonian. The techniques developed in this work can be readily applied to
evaluate the non-form-factor effects induced by the hard spectator interaction and the weak annihilation
as displayed in Fig. \ref{NLO diagrams of Lambda_b to Lambda ll}.
Since both the factorizable and the non-factorizable contributions to the
$\Lambda_b \to \Lambda \, \ell^{+} \ell^{-}$ decay amplitude will be parameterized by the
$\Lambda_b$-baryon DA without introducing any additional non-perturbative quantities,
we are expected to have more opportunities to construct optimized observables insensitive to
the hadronic uncertainties, provided that the systematic uncertainty of the sum rule approach
is also cancelled to a large extent for these observables.
We  postpone a systematic treatment of such non-factorizable contribution as well as
a detailed discussion of the angular observables in  $\Lambda_b \to \Lambda \, \ell^{+} \ell^{-}$
for a future work.

The strategies of computing the heavy-to-light baryonic form factors at ${\cal O}(\alpha_s)$  presented here
can be further extend  to study the topical $\Lambda_b \to p \, \ell \nu$ decays \cite{Aaij:2015bfa},
which provide an alternative approach to determine the CKM matrix element $|V_{ub}|$.
To this end, a comprehensive analysis of the evolution equations for all the DA defined in
Eqs. (\ref{chiral-odd projector}) and (\ref{chiral-even projector}) at one loop are in demand,
since the spin structure of the light di-quark system in the $\Lambda_b$-baryon is distorted
in the decay product, i.e., the nucleon.
In this respect, the techniques developed in  \cite{Knodlseder:2011gc} based upon the spinor formalism
and the conformal symmetry can be applied to facilitate the construction of the renormalization kernels
in coordinate space.
To summarize, we believe that the present work serves as an essential step  towards understanding the strong
interaction dynamics in various  exclusive $\Lambda_b$-baryon decays and interesting extensions of the
present calculations into different directions are expected, especially under the encouragement of the considerable
progress on the beauty baryon decays from the experimental side.

%\newpage

\section*{Acknowledgements}

This work has been supported in part by the Gottfried Wilhelm Leibniz programme
of the Deutsche Forschungsgemeinschaft (DFG).

\appendix

\section {Spectral representations}
\label{appendix: spectral representations}

In this appendix we will collect  the dispersion representations of convolution integrals
appeared in the NLL resummation improved factorization formulae
shown in (\ref{NLL resummed Coree perp V-A}), (\ref{NLL resummed Coree nabr V-A})
and (\ref{NLL resummed Coree perp T-Ttidle}).
As already mentioned in section \ref{section: resummation improved LCSR},  the spectral representations
derived in the following are reduced with the assumption that $\psi_4(\omega_1,\omega_2,\mu_0)$ only depends on the sum of
two momentum variables $\omega=\omega_1+\omega_2$ as inspired from \cite{Feldmann:2011xf,Bell:2013tfa,Ball:2008fw}.

\begin{eqnarray}
&& {1 \over \pi} \, {\rm Im}_{\omega^{\prime}} \, \int_0^{\infty} \, d \omega_1  \,
\int_0^{\infty} \, d \omega_2 \, \frac{1}{\omega-\omega^{\prime}-i0} \,
\ln^2{\mu^2 \over n \cdot p^{\prime} \, (\omega-\omega^{\prime})} \,\, \psi_4(\omega_1,\omega_2, \mu) \nonumber \\
&& =  \int_0^{\omega^{\prime}} \, d \omega \, \left [ {2 \over \omega-\omega^{\prime}} \,
\ln {\mu^2 \over n \cdot p^{\prime} \, (\omega^{\prime}-\omega)} \right ]_{\oplus} \, \tilde{\psi}_4(\omega,\mu)
+ \left [ \ln^2 {\mu^2 \over n \cdot p^{\prime} \, \omega^{\prime}} - {\pi^2 \over 3} \right ]
\, \tilde{\psi}_4(\omega^{\prime},\mu) \,,
\\
&& {1 \over \pi} \, {\rm Im}_{\omega^{\prime}} \, \int_0^{\infty} \, d \omega_1  \,
\int_0^{\infty} \, d \omega_2 \, \frac{1}{\omega-\omega^{\prime}-i0} \,
\ln {\omega-\omega^{\prime} \over \omega_2-\omega^{\prime} }\,
\ln{\mu^2 \over n \cdot p^{\prime} \, (\omega-\omega^{\prime})} \,\, \psi_4(\omega_1,\omega_2, \mu) \nonumber \\
&& = \omega^{\prime} \, \int_0^{\omega^{\prime}}  \, d \omega \,
\left [ {1 \over \omega-\omega^{\prime}} \,
\ln {\omega^{\prime}-\omega \over \omega^{\prime} } \right ]_{\oplus} \, \phi_4(\omega, \mu) \,
+ \int_0^{\omega^{\prime}}  \, d \omega \, \ln {\mu^2 \over n \cdot p^{\prime} \, (\omega^{\prime}-\omega)} \,
{d \tilde{\psi}_4(\omega, \mu) \over d \omega} \nonumber \\
&& \hspace{0.5 cm} + \, {\omega^{\prime} \over 2} \, \int_{\omega^{\prime}}^{\infty}  \, d \omega \,
\left [ \ln^2 {\mu^2 \over n \cdot p^{\prime} \, (\omega-\omega^{\prime})}
- \ln{\mu^2 \over n \cdot p^{\prime} \, \omega^{\prime}} + {\pi^2 \over 3} \right ]
\, {d \phi_4(\omega, \mu) \over d \omega} \,,
\\
&& {1 \over \pi} \, {\rm Im}_{\omega^{\prime}} \, \int_0^{\infty} \, d \omega_1  \,
\int_0^{\infty} \, d \omega_2 \, \frac{1}{\omega-\omega^{\prime}-i0} \,
\ln^2{\omega-\omega^{\prime} \over  \omega_2 -\omega^{\prime}} \,\, \psi_4(\omega_1,\omega_2, \mu) \nonumber \\
&& = - \omega^{\prime} \, \int_{\omega^{\prime}}^{\infty}  \, d \omega \,
\left [ \ln^2 {\omega -\omega^{\prime} \over \omega^{\prime}}
+ 2\, \ln {\omega -\omega^{\prime} \over \omega^{\prime}}
-{\pi^2 \over 3} + 2 \right ]   \, {d \phi_4(\omega, \mu) \over d \omega} \,\,,
\\
&& {1 \over \pi} \, {\rm Im}_{\omega^{\prime}} \, \int_0^{\infty} \, d \omega_1  \,
\int_0^{\infty} \, d \omega_2 \, \frac{1}{\omega-\omega^{\prime}-i0} \,
{ \omega_2 - \omega^{\prime} \over \omega_1 } \,
\ln{\omega-\omega^{\prime} \over  \omega_2 -\omega^{\prime}} \,\, \psi_4(\omega_1,\omega_2, \mu) \nonumber \\
&& = \int_{\omega^{\prime}}^{\infty}  \, d \omega \,
\left [ \ln{\omega \over  \omega - \omega^{\prime}} + \omega^{\prime} \,
\left ( \ln{\omega - \omega^{\prime} \over \omega^{\prime}}  + 1 \right ) \,
{d \over d \omega }\right ] \, \phi_4(\omega, \mu) \,\,,
\\
&& {1 \over \pi} \, {\rm Im}_{\omega^{\prime}} \, \int_0^{\infty} \, d \omega_1  \,
\int_0^{\infty} \, d \omega_2 \, \frac{1}{\omega-\omega^{\prime}-i0} \,
\ln{\omega-\omega^{\prime} \over  \omega_2 -\omega^{\prime}} \,\, \psi_4(\omega_1,\omega_2, \mu) \nonumber \\
&& = - \omega^{\prime} \, \int_{\omega^{\prime}}^{\infty}  \, d \omega \,
\left [ \ln{\omega - \omega^{\prime} \over \omega^{\prime}}  + 1 \right ] \,
{d \phi_4(\omega, \mu) \over d \omega } \,\,,
\\
&& {1 \over \pi} \, {\rm Im}_{\omega^{\prime}} \, \int_0^{\infty} \, d \omega_1  \,
\int_0^{\infty} \, d \omega_2 \, \frac{1}{\omega-\omega^{\prime}-i0} \,
\ln{\mu^2 \over n \cdot p^{\prime} \, (\omega-\omega^{\prime})} \,\, \psi_4(\omega_1,\omega_2) \nonumber \\
&& = \int_{0}^{\omega^{\prime}}  \, d \omega \,
\ln{\mu^2 \over n \cdot p^{\prime} \, (\omega^{\prime}-\omega)} \,
{d \tilde{\psi}_4(\omega, \mu) \over d \omega } \,\,,
\end{eqnarray}
where we have defined
\begin{eqnarray}
 \phi_4(\omega, \mu) = \psi_4 \left (u\, \omega \,, (1-u) \, \omega, \mu \right )   \,, \qquad
\tilde{\psi}_4(\omega, \mu)= \omega \, \phi_4(\omega, \mu)\,.
\end{eqnarray}

\vspace{0.2 cm}

%%%%%%%%%%%%%%%%%%%%%%%%%%%%%%%%%%%%%%%%%%%%%%%%%%%%%%%%%%%%%%%%%%%%%%%%%%%
%\newpage

\end{document}